\documentclass[twocolumn]{aastex631}

\usepackage{amssymb}
\usepackage{amsthm}
\usepackage{newtxtext,newtxmath}
\usepackage{graphicx}

\newcommand{\dtpar}{\Delta t_\mathrm{par}}
\newcommand{\M}{\boldsymbol{\rm M}}
\newcommand{\R}{\boldsymbol{\rm R}_{s,\rm RKL}}
\newcommand{\f}{\boldsymbol{\rm f}}
\newcommand{\F}{\mathcal{F}}
\newcommand{\lambdaRKL}{\lambda_{k,\mathrm{RKL}}(s)}
\newcommand{\lambdaM}{\lambda_{k,\mathrm{M}}}
\newcommand{\skopt}{s_{k,\mathrm{opt}}}
\newcommand{\sopt}{s_\mathrm{opt}}
\newcommand{\lambdaapprx}{\lambda_{k,\mathrm{RKL}}^\mathrm{apprx}(s)}

\newcommand{\norm}[1]{|\!|#1|\!|}
\newcommand{\cs}{c_\mathrm{s}}
\newcommand{\tff}{t_\mathrm{ff}}
\newcommand{\kms}{\mathrm{km\,s^{-1}}}
\newcommand{\DS}{\displaystyle}
\newcommand{\HALF}{\frac{1}{2}}

\submitjournal{ApJS}

\shortauthors{Mandal et al.}

\begin{document}

\title{A self-gravity module for the PLUTO code}

\author[0000-0002-0912-8081]{Ankush Mandal}
\affiliation{Inter-University Centre for Astronomy and Astrophysics, Pune 411007, India}

\author[0000-0003-0632-1000]{Dipanjan Mukherjee}
\affiliation{Inter-University Centre for Astronomy and Astrophysics, Pune 411007, India}

\author[0000-0002-8352-6635]{Andrea Mignone}
\affiliation{Dipartimento di Fisica, Universitá di Torino via Pietro Giuria 1 (I-10125) Torino, Italy}

\correspondingauthor{Ankush Mandal}
\email{ankushm@iucaa.in}



\begin{abstract}

We present a novel implementation of an iterative solver for the solution of the Poisson equation in the PLUTO code for astrophysical fluid dynamics.
Our solver relies on a relaxation method in which convergence is sought as the steady-state solution of a parabolic equation, whose time-discretization is governed by the \textit{Runge-Kutta-Legendre} (RKL) method.
Our findings indicate that the RKL-based Poisson solver, which is both fully parallel and rapidly convergent, has the potential to serve as a practical alternative to conventional iterative solvers such as the \textit{Gauss-Seidel} (GS) and \textit{successive over-relaxation} (SOR) methods. 
Additionally, it can mitigate some of the drawbacks of these traditional techniques.
We incorporate our algorithm into a multigrid solver to provide a simple and efficient gravity solver that can be used to obtain the gravitational potentials in self-gravitational hydrodynamics.
We test our implementation against a broad range of standard self-gravitating astrophysical problems designed to examine different aspects of the code. We demonstrate that the results match excellently with the analytical predictions (when available), and the findings of similar previous studies.

\end{abstract}



\section{Introduction}
\label{sec:intro}
Elliptical partial differential equations (PDE) are crucial in numerous mathematical and physical problems. One prime example is Poisson's equation, which emerges naturally in many contexts such as gravity, incompressible flow, electromagnetism, thermal conduction, divergence-cleaning procedure in magneto-hydrodynamics (MHD), and several others.
In this sense, the development of efficient and fast solvers for such PDEs represents one of the primary goals in computational physics. 
In particular, solving the Poisson equation for the gravitational potential for a given mass distribution is of great interest in those astrophysical environments where self-gravitational forces play a crucial role, such as the formation of stars \citep{Ostriker_2001,McKee2007}, planet formation \citep{Boss_1997,Rice_2005}, supernova explosions \citep{Nordhaus_2010,Couch_2013}, the evolution of galaxies \citep{Mo_2010}, and the structure formation \citep{Bertschinge_1998} in the universe.
Thus, the accurate representation of self-gravity is critical for achieving realistic and reliable results in the numerical modeling of such systems.

The need for solving the Poisson equation has led to the development of various algorithms with different levels of complexity and applicability. 
One can rely on direct solvers such as the Fast Fourier Transform (FFT) method to achieve a solution accurate up-to round-off error. 
On the other hand, the traditional relaxation methods such as \textit{Gauss-Seidel} (GS) or \textit{successive over-relaxation} (SOR) are also of great interest due to their simplicity of implementations \cite[see][for details]{Press_1992}.
Over the past two decades, there has been an increasing interest in these iterative methods due to the advancement of massively parallel computers.
Since one generally does not need the solution to be accurate upto the round-off error (as provided by the FFT method), iterative methods are preferred for practical purposes as iterations can be stopped early when the iteration error in the solution drops below the truncation error.
Although these iterative solvers suffer from the slow convergence rate for the long wavelength modes \citep{Press_1992}, the usage of the multigrid scheme can improve the convergence rate drastically \citep{Brandt_1977,Briggs_2000}. 
The multigrid method uses a hierarchy of grids with decreasing spatial resolution to correct the error in the guess solution on the finest grid. 
The usage of coarser grids ensures that long-wavelength modes of the error on the finer grids appear short on the coarser ones, thus, can be damped down efficiently. 
Thus, the multigrid method consists of three primary elements: (i) a smoothing operator on the finer grids that damp small-wavelength (high-frequency) modes smoothing out the solution; (ii) some restriction and prolongation operators to map different quantities between grids; (iii) an efficient Poisson solver on the coarsest grid, which is necessary for solving the error equation \citep{Trottenberg_2000}.

While the GS and SOR methods are commonly used in several codes for the coarsest grid solver, both pose certain limitations. 
In particular, the GS algorithm converges very slowly, whereas the convergence rate of the SOR method heavily depends on the optimal choice of the over-relaxation parameter ($\omega$), which is typically problem-dependent and difficult to derive analytically \citep{Press_1992}. 
Furthermore, none of the methods achieves full parallelization, meaning that all grid points within the computational domain cannot be updated simultaneously during a single iteration step. 
Rather, the updates are accomplished in an alternating odd-even pattern, requiring twice the amount of parallel communication between processors compared to a fully parallel algorithm, which incurs additional computational overheads \citep{Trottenberg_2000}.

In this study, we present a new Poisson solver for the \textsc{Pluto} code \footnote{\textsc{Pluto} is freely available at \url{http://plutocode.ph.unito.it}.} \citep{Mignone_2007,Mignone_2012} for self-gravitational astrophysical fluids.
The method is implemented within a multigrid V-cycle algorithm structure and it takes advantage of the accelerated super-time-stepping technique based on the \textit{Runge-Kutta-Legendre} (RKL) method used for solving parabolic PDEs \citep{Meyer_2012,Meyer_2014, Vaidya_2017}, by viewing the Poisson equation as the steady-state solution of a parabolic equation.
RKL time-stepping methods are optimized for stability rather than accuracy, so that a single time-step, consisting
$s$ number of sub-stages, can be taken much larger compared to the time-step permitted by a standard explicit method, and it increases monotonically with $s$ \citep{Meyer_2014}. 
In theory, a sufficiently high $s$ can produce a steady-state solution in a single iteration, but the \textit{convergence rate} does not necessarily increase monotonically with increasing $s$. 
In fact, we demonstrate that there exists an optimal value of $s$ that produces the maximum convergence rate, which depends only on the domain size and boundary conditions and it can be analytically derived. 
An optimized RKL-based Poisson solver exhibits a much higher convergence rate than the GS method and can become competitive with the SOR scheme. 
Moreover, it eliminates the uncertainties associated with the optimal over-relaxation parameter ($\omega$) of the SOR algorithm. 
In addition, the RKL method is fully parallelizable, allowing all grid points to be updated simultaneously during a single iteration, which results in lower parallel communication overhead, making the scheme well-suited for massively parallel systems.

The article is organized as follows. 
In Sec.~\ref{sec:the_algorithm}, we describe the basic V-cycle multigrid algorithm, its individual components, and the boundary conditions. 
Sec.~\ref{sec:coupling} outlines how gravity is coupled with hydrodynamics. 
In Sec.~\ref{sec:results}, we present the accuracy of the Poisson solver and the gravity module as a whole for a wide variety of standard self-gravitating astrophysical problems. 
The chosen problems are designed to showcase the accuracy of various aspects of the code, including energy conservation, variation in the equation of state, momentum conservation, and boundary conditions. 
Finally, we summarize our findings in Sec.~\ref{sec:summary}.

\section{The Algorithm}\label{sec:the_algorithm}


\subsection{The iterative multigrid algorithm}\label{sec:multigrid_algorithm}
The core idea of the multigrid technique is to use a hierarchy of grids with geometrically decreasing mesh sizes.
The method solves a system of discrete equations on a given grid iteratively, through constant interactions with a series of coarser grids.
In this method, the coarser grids are used to find the error in the approximate solution which can be used to obtain a better guess solution for the next iteration, traditionally called \textit{the correction scheme} multigrid method \citep{Briggs_2000,Trottenberg_2000}. In the following, we briefly summarize the scheme.

Suppose we wish to solve the following elliptic equation,
\begin{equation}\label{eq:elliptic_problem}
    \mathcal{L}\Phi = f,
\end{equation}
where $\mathcal{L}$ is a linear elliptic operator and $f$ is the source function defined in some domain $\Omega$. Let us assume, our domain is to be discretized into $N_x$, $N_y$, and $N_z$ grid points with a grid spacing of $h$. This discretized domain is be denoted as $\Omega^h$, where the grid points are represented by $x_i = ih$, $y_j = jh$, and $z_k = kh$. Similarly, if we have a discretization of $\Omega$ with $N_x/2$, $N_y/2$, and $N_z/2$ points and a grid spacing of $2h$, we represent it as $\Omega^{2h}$. Then, the discretization of Eq.~\eqref{eq:elliptic_problem} on $\Omega^h$ can be written as,
\begin{equation}\label{eq:discretization}
    \mathcal{L}_h\Phi_h=f_h.
\end{equation}
Let us assume that the guess to the solution of Eq.~\eqref{eq:discretization} at the first iteration is $\Psi_h$. Then the error can be found as,
\begin{equation}\label{eq:error_definition}
    e_h = \Phi_h - \Psi_h,
\end{equation}
where $\Phi_h$ is the exact solution of Eq.~\eqref{eq:discretization}. Now, the residual can be defined as,
\begin{equation}\label{eq:residual}
    r_h = \mathcal{L}_h\Psi_h - f_h.
\end{equation}
Interestingly, since $\mathcal{L}_h$ is a linear operator, the error $e_h$ satisfies
\begin{equation}\label{eq:error_equation}
    \mathcal{L}_h e_h = -r_h.
\end{equation}
Thus, instead of relaxing the original solution $\Phi_h$ with an arbitrary guess solution, we can relax Eq.~\eqref{eq:error_equation} with a specific guess $e_h=0$.

\begin{figure}
    \centering
    \includegraphics[width=\linewidth]{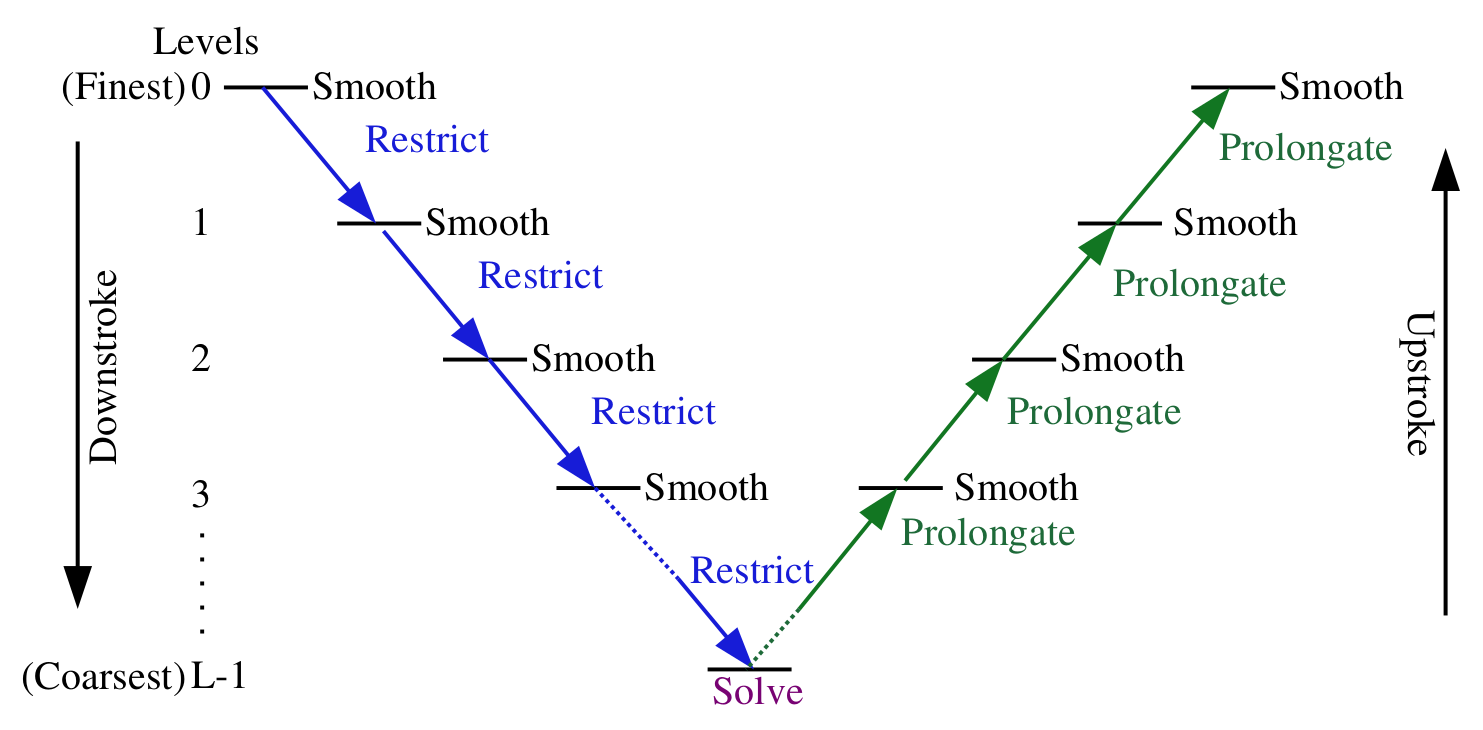}
    \caption{A schematic diagram of the multigrid V-cycle algorithm.}
    \label{fig:V-cycle}
\end{figure}

Now, if we expand $e_h$ into a discrete Fourier series, it can be shown that the long-wavelength modes of $e_h$ converge very slowly, whereas the oscillatory or short-wavelength modes of $e_h$ can be reduced more easily \citep{Press_1992,Briggs_2000,Trottenberg_2000}.
Thus, if we restrict the error to a coarser grid ($\Omega^{2h}$) from the original grid ($\Omega^h$), long-wavelength modes of the error will appear as short-wavelength modes on the coarser grid, which can be reduced by a large factor with each iteration \citep{Briggs_2000}. 
Once, we get the solution for $e_{2h}$ on $\Omega^{2h}$, it can be interpolated back to the fine grid to calculate the approximated solution $\Tilde{e}_h$ of the error. 
Then the guess to the solution is updated by,
\begin{equation}\label{eq:new_guess}
    \Psi_h^\mathrm{new} = \Psi_h + \Tilde{e}_h.
\end{equation}
Thus, we can utilize these concepts recursively across a hierarchy of grids with geometrically decreasing mesh sizes to carry out an entire multigrid cycle.
In order to transfer data between grids with  different mesh spacing, we need to define restriction and prolongation operators.
Fine-to-coarse grid interpolation is done through the \textit{restriction operator} ($\mathcal{R}$) while the inverse process is achieved using the coarse-to-fine \textit{prolongation operator} ($\mathcal{P}$). 
We will discuss this in detail in Sec.~\ref{sec:operators}.

One of the simplest forms of multigrid cycles is the V-cycle, which is schematically shown in Fig.~\ref{fig:V-cycle}.
The V-cycle algorithm initially approximates the solution by performing $N_{\rm pre}$ smoothing iterations on the finest grid level ($l=0$). 
The residual is then restricted to the next coarse level, where another $N_{\rm pre}$ smoothing iterations are carried out. 
This process of restriction followed by smoothing is repeated until the coarsest level is reached (downstroke), where we solve the correction equation with a fast iterative solver. 
Subsequently, the correction solution is prolongated to the next fine level, which is utilized to adjust the solution. 
Using the modified solution, $N_{\rm post}$ number of smoothing operations are performed. 
The sequence of prolongation and smoothing steps persists until the correction is incorporated into the solution at the finest grid level (upstroke).
The whole procedure forms a V-cycle, which is repeated until the desired convergence level is reached.
In the following, we give the V-cycle multigrid algorithm that contains $L$ number of levels \cite[see][for details]{Briggs_2000,Trottenberg_2000}. 
\begin{itemize}\label{v-cycle}
    \item do \{
    \begin{enumerate}
        \item At $l=0$, perform $N_{\rm pre}$ relaxation on $\mathcal{L}^{0}\Phi^{0} = f^{0}$.
        \item for ($l=1$ to $l < L$)\{
        \begin{enumerate}
            \item[2a.] Compute the residual: $r^{l-1}=\mathcal{L}^{l-1}\Phi^{l-1} - f^{l-1}$.

            \item[2b.] Calculate $f^{l}$ by performing restriction on $r^{l-1}$: $f^{l} = \mathcal{R} (r^{l-1})$.

            \item[2c.] Set the guess solution for this level to zero: $\delta\Phi^{l} = 0$.

            \item[2d.] If $l<L-1$
            
            \begin{enumerate} 
                \item  Perform $N_{\rm pre}$ relaxation on  $\mathcal{L}^{l}\delta\Phi^{l} = -f^{l},\; {\rm  with}\; \delta\Phi^{l} = 0 \; {\rm on} \; \partial\Omega^{l}$.

                \item $l \xleftarrow[]{} l+1$
            \end{enumerate}
            
            \item[2e.] Else if $l=L-1$, solve $\mathcal{L}^{l}\delta\Phi^{l} = -f^{l}$ with a fast iterative solver.
        \end{enumerate}
        \}
        
        \item for ($l=L-2$ to $l>0$)\{
        \begin{enumerate}
            \item[3a.] Perform prolongation of $\delta\Phi^{l+1}$ to correct $\delta\Phi^{l}$: $\delta\Phi^{l} \xleftarrow[]{} \delta\Phi^{l} + \mathcal{P}(\delta\Phi^{l+1})$.

            \item[3b.] Perform $N_{\rm  post}$ relaxations on $\mathcal{L}^{l}\delta\Phi^{l} = -f^{l},\; {\rm  with}\; \delta\Phi^{l} = 0 \; {\rm on} \; \partial\Omega^{l}$.
            
            \item[3c.] $l \xleftarrow[]{} l-1$.
        \end{enumerate}
        \}
        
        \item At $l=0$,
        \begin{enumerate}
            \item[4a.] Correct the finest grid solution: $\Phi^0 \xleftarrow[]{}\Phi^0 +\mathcal{P}(\delta\Phi^1)$
            \item[4b.] Perform $N_{\rm post}$ relaxation on $\mathcal{L}^{0}\Phi^{0} = f^{0}$. 
        \end{enumerate}
    \end{enumerate}
    \} while (residual is above the specified tolerance level)
\end{itemize}
From the above steps, we see that, at the coarsest level of the multigrid hierarchy, we need a very efficient iterative solver to estimate the solution of the error equation, which is sometimes called the `\textit{kernel}' of the iterative multigrid algorithm. We discuss this in detail in Sec.~\ref{sec:smooth_kernel}.


\subsection{Operator definitions}\label{sec:operators}
 
Spatial discretization can be based on either staggered or cell-centered meshes, but the structure of different operators and the treatment of boundary conditions change accordingly \citep{Trottenberg_2000}.
For application purposes, we adopt in what follows a multigrid algorithm that makes usage of a cell-centered discretization, essential in devising the iterative multigrid algorithm, as discussed in Sec.~\ref{sec:multigrid_algorithm}.

 
\subsubsection{Discretization of the Laplace operator}\label{sec:laplace}
In Sec.~\ref{sec:multigrid_algorithm}, we discussed how the iterative multigrid algorithm works for any general linear elliptic PDEs. 
However, in this paper, our primary goal is to implement the multigrid solver for Poisson's equation for gravitational potential, which has the following form,
\begin{equation}\label{eq:poisson_eqaution}
    \nabla^2\Phi = 4\pi G\rho
\end{equation}
where $\nabla^2$ is the Laplace operator. 
We use the second-order accurate five-point and (in 2D) or seven-point (in 3D) stencil for the discretization of the Laplace operator on a particular grid level $L$. 
For two dimensions, the discretization is given by
\begin{align}\label{eq:discrete_laplace_2D}
    \nabla^2\Phi^L_{i,j} \approx & \frac{1}{\Delta x^2_L}\left(\Phi_{i-1,j}^L-2\Phi_{i,j}^L+\Phi_{i+1,j}^L\right) \nonumber \\
    & + \frac{1}{\Delta y^2_L}\left(\Phi_{i,j-1}^L-2\Phi_{i,j}^L+\Phi_{i,j+1}^L\right)
\end{align}
and for three dimensions,
\begin{equation}\label{eq:discrete_laplace_3D}
\begin{split}
    \nabla^2\Phi^L_{i,j,k} &\approx \frac{1}{\Delta x^2_L}\left(\Phi_{i-1,j,k}^L-2\Phi_{i,j,k}^L+\Phi_{i+1,j,k}^L\right)\\
     &+ \frac{1}{\Delta y^2_L}\left(\Phi_{i,j-1,k}^L-2\Phi_{i,j,k}^L+\Phi_{i,j+1,k}^L\right)\\
     &+ \frac{1}{\Delta z^2_L}\left(\Phi_{i,j,k-1}^L-2\Phi_{i,j,k}^L+\Phi_{i,j,k+1}^L\right),
\end{split}
\end{equation}
where $\Delta x_L$, $\Delta y_L$ and $\Delta z_L$ are the grid spacing along $x$-, $y$- and $z$- directions on $L^\mathrm{th}$ level.

\begin{figure*}
     \centering
     \begin{minipage}[c]{0.31\textwidth}
         \centering
         \includegraphics[width=\textwidth]{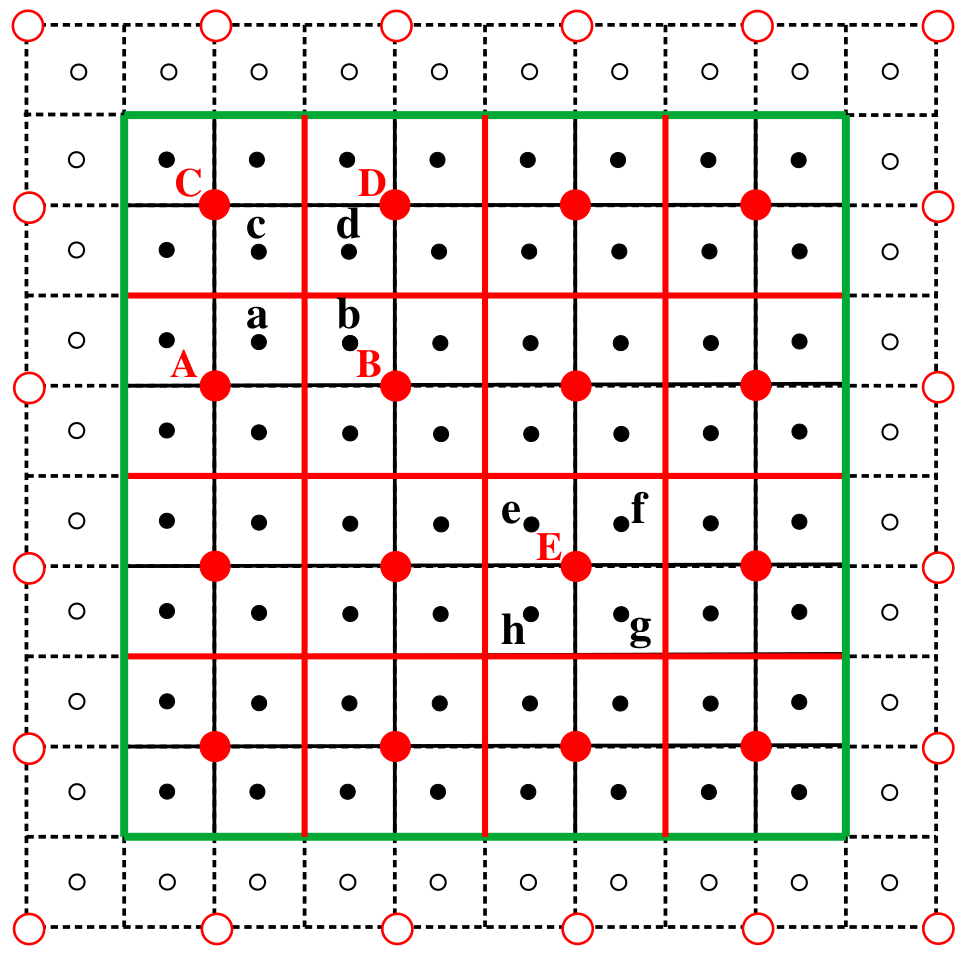}
         {(a)}
         \label{fig:stencil_2D}
     \end{minipage}
     \hfill
     \begin{minipage}[c]{0.31\textwidth}
         \centering
         \includegraphics[width=\textwidth]{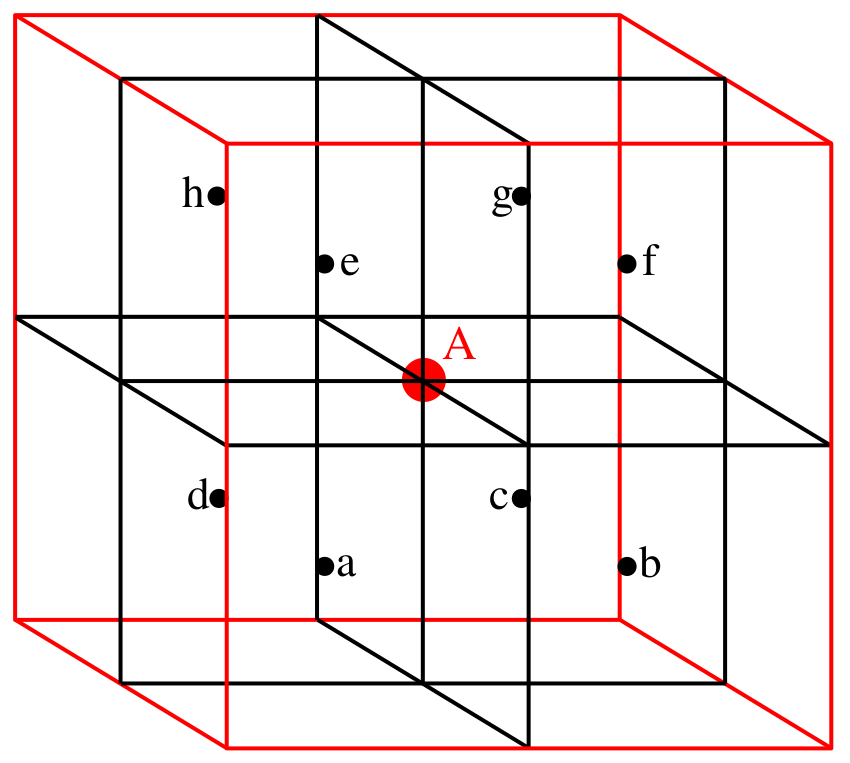}
         {(b)}
         \label{fig:Restriction3D}
     \end{minipage}
     \hfill
     \begin{minipage}[c]{0.33\textwidth}
         \centering
         \includegraphics[width=\textwidth]{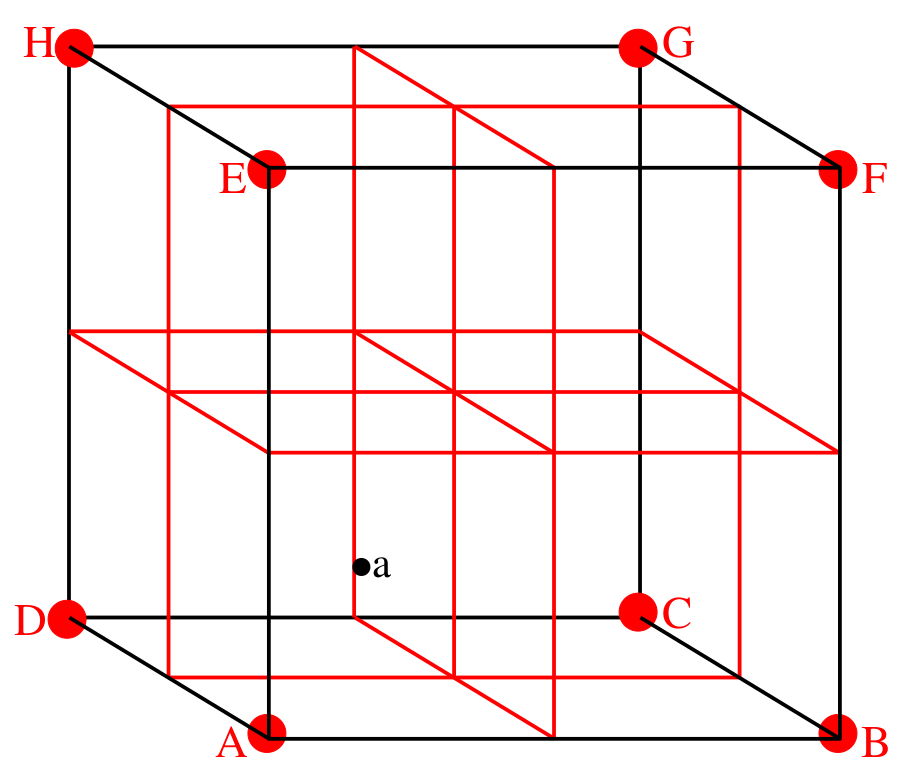}
         {(c)}
         \label{fig:Prolongation3D}
     \end{minipage}
        \caption{Visual representations of the of the restriction ($\mathcal{R}$) and prolongation ($\mathcal{P}$) operators in 2D (a) and 3D (b and c). The upper case letters in red represent the value cell-centered values of the coarse grid and the lower case black letters are the same for the fine grid. If we consider \ref{fig:stencil_2D}, in order to calculate the coarse grid value $E$, we simply take the average of the fine grid values $e$, $f$, $g$, and $h$ that are closest to $E$ (Eq.~\ref{eq:restriction_example_2D}). Conversely, the fine grid value $a$ can be estimated by taking a weighted sum of the four nearest coarse grid values $A$, $B$, $C$, and $D$ (Eq.~\ref{eq:prolongation_example_2D}). Likewise, the 3D restriction operator (depicted in \ref{fig:Restriction3D}) performs an averaging operation on the eight closest fine grid values (as expressed in Eq.~\ref{eq:restriction_example_3D}). On the other hand, the 3D prolongation operator (illustrated in \ref{fig:Prolongation3D}) calculates a weighted sum of the eight nearest coarse grid points (as indicated in Eq.~\ref{eq:prolongation_example_3D}).}
        \label{fig:stencils}
\end{figure*}


\subsubsection{Restriction operator}\label{sec:restriction}
As discussed in Sec.~\ref{sec:multigrid_algorithm}, a restriction operator ($\mathcal{R}^h_{2h}$) is needed that maps or restricts the error from the fine grid ($\Omega^h$) to the coarse grid ($\Omega^{2h}$). 
Here, we adopt the simple  $2^{\rm nd}$-order  accurate four-point average for two dimensions \citep{Trottenberg_2000}. Application of $\mathcal{R}^h_{2h}$ to a fine grid ($\Omega^h$) function $r_h(x,y)$ at coarse grid point $(x,y)\in \Omega^{2h}$ yields,
\begin{equation}\label{eq:restriction_equation_2D}
    \begin{split}
        r_{2h}(x,y) =& \mathcal{R}^h_{2h}r_h(x,y) \\
                    =& \frac{1}{4}\left[ r_h\left(x-\frac{h}{2}, y-\frac{h}{2}\right) + r_h\left(x+\frac{h}{2}, y-\frac{h}{2}\right) \right. \\
                    & \left.+ r_h\left(x-\frac{h}{2}, y+\frac{h}{2}\right)+ r_h\left(x+\frac{h}{2}, y+\frac{h}{2}\right) \right].
    \end{split}
\end{equation}
In Fig.~\ref{fig:stencil_2D}, we show a simple stencil of the fine (black lines) and coarse grids (red lines). 
The values are defined at the center of the cells. From Eq.~\eqref{eq:restriction_equation_2D}, the coarse grid value $E$ is simply given by,
\begin{equation}\label{eq:restriction_example_2D}
    E = \frac{1}{4}\left(h+g+e+f\right),
\end{equation}
where $g$, $h$, $e$ and $f$ are the fine grid values at specified location in Fig.~\ref{fig:stencil_2D}.

In three dimensions, the restriction operator can be easily extended from the two-dimensional counterpart by considering an eight-point average, which reads,
\begin{equation}\label{eq:restriction_equation_3D}
    \begin{split}
        r_{2h}(x,y,z) =& \mathcal{R}^h_{2h}r_h(x,y,z) \\
                      =& \frac{1}{8}\left[ r_h\left(x-\frac{h}{2}, y-\frac{h}{2}, z-\frac{h}{2}\right)+r_h\left(x+\frac{h}{2}, y-\frac{h}{2}, z-\frac{h}{2}\right)\right. \\
                      & + r_h\left(x-\frac{h}{2}, y+\frac{h}{2}, z-\frac{h}{2}\right) + r_h\left(x-\frac{h}{2}, y-\frac{h}{2}, z+\frac{h}{2}\right) \\
                      & + r_h\left(x+\frac{h}{2}, y-\frac{h}{2}, z+\frac{h}{2}\right) + r_h\left(x-\frac{h}{2}, y+\frac{h}{2}, z+\frac{h}{2}\right) \\
                      &\left. + r_h\left(x+\frac{h}{2}, y+\frac{h}{2}, z-\frac{h}{2}\right) + r_h\left(x+\frac{h}{2}, y+\frac{h}{2}, z+\frac{h}{2}\right)
                       \right]
    \end{split}
\end{equation}
From Fig.~\ref{fig:Restriction3D}, the coarse grid value $A$ can be obtained by,
\begin{equation}\label{eq:restriction_example_3D}
    A = \frac{1}{8}\left(a+b+d+c+e+f+h+g\right)
\end{equation}
%

\subsubsection{Prolongation operator}\label{sec:prolongation}
In order to transfer the coarse grid correction to the fine grid, we use the $2^{\rm nd}$-order accurate cell-centered prolongation operator as shown in Fig.~\ref{fig:stencils}. 
In two dimensions, with reference to  Fig.~\ref{fig:stencil_2D}, the values at fine grid points are calculated as,
\begin{equation}\label{eq:prolongation_example_2D}
  \begin{array}{lcl}
    a &=& \DS  \frac{1}{16}\left(9A + 3B + 3C + D \right) \\ \noalign{\medskip} 
    b &=& \DS  \frac{1}{16}\left(3A + 9B + C + 3D \right) \\ \noalign{\medskip}
    c &=& \DS  \frac{1}{16}\left(3A + B + 9C + 3D \right) \\ \noalign{\medskip}
    d &=& \DS  \frac{1}{16}\left(A + 3B + 3C + 9D \right)
  \end{array}
\end{equation}
where $a$, $b$, $c$ and $d$ are the fine grid values at the cell centres and $A$, $B$, $C$ and $D$ are the coarse grid cell-centred value as depicted in Fig.~\ref{fig:stencil_2D}. 
In three dimensions, we can easily extend the prolongation operator, by inspecting Fig.~\ref{fig:Prolongation3D}, the cell-centered fine grid value $a$ can be obtained as follows,
\begin{equation}\label{eq:prolongation_example_3D}
     a = \frac{1}{64}\left(27A + 9B + 9D + 9E + 3C + 3F + 3H + G\right),
\end{equation}
where $A$, $B$, $C$, $D$, $E$, $F$, $G$ and $H$ are the cell-centred coarse grid values.


\subsubsection{Smoothing operator and kernel solver}\label{sec:smooth_kernel}
The two most important components of the iterative multigrid algorithm are the smoothing operator which is used to smooth out the solution error on the finer grids, and the coarsest grid solver (or the kernel solver), which solves the error equation on the coarsest level for the given boundary conditions. 
For the smoothing operations, the \textit{Gauss-Seidel} relaxation method with red-black ordering (GS-RB) has been shown to be one of the best methods available in the literature \citep{Press_1992,Trottenberg_2000} and it will also be our method of choice. 
The GS-RB solver has been used in several the (magneto-) hydrodynamic codes where the multigrid algorithm is used for solving the Poisson equation \citep{Ziegler_2005,Ricker_2008,Almgren_2010,Guillet_2011,Wang_2020}. 
In the GS-RB scheme, cell values are updated in two passes, each pass covering half of the computational domain according to the colors of a checkerboard pattern (red or black, indeed).

For the coarsest grid direct solver, there are various algorithms ranging from iterative solvers to the conjugate gradient method and fast Fourier transform (FFT). 
Traditionally, iterative methods are employed as kernel solvers owing to their simplicity of implementation. 
Still, one important criterion for choosing a particular solver is the degree of parallelization as it can degrade the overall performance of a highly-parallelized code. 
The traditional relaxation $\omega$-Jacobi scheme \citep{Trottenberg_2000} is fully parallel (i.e., the operator can be applied to all the grid points simultaneously and the new value are mutually exclusive) but has a very poor convergence rate. 

The GS-RB method has a relatively better convergence rate than the $\omega$-Jacobi method, but it becomes inefficient as the domain becomes larger and it is also half-parallel\footnote{In GS-RB method, one iteration consists of two half-steps. 
Thus, only half of the total grid points inside the computational domain can be updated simultaneously. This implies the method is half-parallel \cite[see][for details]{Trottenberg_2000}.}. 
The $\omega$-GS-RB or the \textit{successive-overrelaxation} (SOR) method can be designed to improve the convergence rate by choosing an optimal value of $\omega$.
However, the efficiency of this method heavily depends on the overrelaxation parameter ($\omega$), which is typically problem dependent. 
A slight deviation from the true value of $\omega$ can degrade the convergence rate considerably.

In this work, we propose a novel approach for the solution of Poisson's equation based on the \textit{Runge-Kutta-Legendre} (RKL) method for solving parabolic differential equation \citep{Meyer_2012,Meyer_2014,Vaidya_2017}.
The RKL method is a class of super-time-stepping methods (STS) that have been developed to deal with the restrictive time-step for an explicit scheme for parabolic equations of the form $\partial u/\partial t = \boldsymbol{M}u$. 
This allows us to consider a larger time-step ($\tau$) than the maximum permissible explicit parabolic time-step ($\dtpar$) without affecting the stability of the solution.
For a $s$-stage RKL method, the maximum possible time-step ($\tau_\mathrm{max}$) can be shown to be
\begin{equation}\label{eq:tau_max}
    \tau_\mathrm{max} = \dtpar\frac{s^2+s}{2},
\end{equation}
where $\dtpar = 2/|\lambda_\mathrm{max}|$ is the explicit parabolic time-step ($\lambda_\mathrm{max}$ is the maximum eigenvalue of the elliptic operator $\boldsymbol{M}$). 
As Poisson's equation can be sought as the steady-state manifestation of the following non-homogeneous parabolic equation,
\begin{equation}\label{eq:parabolic_eq}
    \frac{\partial\Phi}{\partial t} = \nabla^2\Phi -\rho \,,
\end{equation}
we can easily take advantage of the RKL technique to solve for the gravitational potential $\Phi$ in Eq.~\eqref{eq:parabolic_eq} in the $t\xrightarrow{}\infty$ limit with much larger time-steps than allowed by explicit methods. 
While this could (in principle) be obtained within a single time step $\tau$ using a very large value of $s$, it does not ensure a monotonically increasing convergence rate with growing values of $s$.

In fact, although increasing the value of $s$ improves the convergence rate of the RKL operator, we find that there exists an optimal value of $s$ ($\sopt$) for which the convergence rate of an RKL-based Poisson solver is the highest, and further increment of $s$ degrades the convergence rate. 
This value of $\sopt$ depends only on the domain size and the boundary conditions, which removes the uncertainty of the optimal overrelaxation parameter $\omega$ in the SOR algorithm. 
In \ref{sec:RKL_convergence}, we derive the convergence property of the RKL-based Poisson solver in detail. 
Another advantage of the RKL scheme is that it is fully parallel, i.e., the calculation of the cell values at the current time-step entirely depends on the previous step, thus a single communication between processors is enough to update all the points during a single iteration.
Therefore, being fully parallel, highly optimized, and specifically not fine-tuned, the RKL-based Poisson solver can be an alternative to the traditional relaxation methods for solving Poisson's equation. 
For instance, in \ref{sec:performance_test}, we demonstrate the comparative performance of the GS, SOR, and RKL methods for a 3D problem. 
We find that the RKL scheme not only achieves a given residual level in the least amount of time but also exhibits significantly lower parallel communication overhead compared to the GS and SOR methods.
Hence, in the multigrid algorithm described in this paper, we use the RKL-based Poisson solver for the coarsest grid kernel solver. 


\subsection{Boundary conditions}\label{sec:boundary_conditions}
For the cell-centered multigrid method, the boundary points need special attention as for the cell-centered discretization, no grid points lie along the boundary (see the green line in Fig.~\ref{fig:stencil_2D}, which lies along the cell edges).
For a Dirichlet-type boundary condition, linear extrapolation is used to specify solution values at the centers of the ghost cells.
For example, for the two boundaries at the start and end points of $x$-axis, we have:
\begin{gather}\label{eq:solution_ghost}
    \Phi_{0,j,k} = 2\Phi_{\HALF,j,k} - \Phi_{1,j,k}\,, \\
    \Phi_{N_x+1,j,k} = 2\Phi_{N_x+\HALF,j,k} - \Phi_{N_x,j,k},
\end{gather}
where $\Phi_{\HALF,j,k}$ and $\Phi_{N_x+\HALF,j,k}$ are the potential values at the edges of the first and last active cells and correspond to the physical boundaries, which are specified by the user.
The ghost cells along the $y\mbox{-}$ and $z\mbox{-}$axis can be filled in a similar way.
Notice that boundary conditions must be specified at the beginning of the relaxation procedure only on the finest level ($l=0$).
As we are solving the error equations on levels other than the finest one, the error along the boundary should be zero for a Dirichlet-type boundary condition. 
Thus the ghost cell values at the levels with $l>0$ will be,
\begin{equation}\label{eq:error_ghost}
    e_{0,j,k} = -e_{1,j,k}\,,
    \qquad {\rm and}\qquad e_{N_x+1,j,k} = -e_{N_x,j,k}.
\end{equation}

In the case of periodic boundary conditions (let's say along $x$-axis), left ghost cells are filled by copying the last active cell values along the $x$-axis and vice-versa for the right ghost cells:
\begin{equation}\label{eq:periodic}
    \Phi_{0,j,k} = \Phi_{N_x,j,k}.
\end{equation}
The same periodic boundary conditions (Eq.~\ref{eq:periodic}) are also employed at  the coarser levels ($l>0$).

The periodic boundary condition for gravitational potential, however, requires some modifications of the Poisson equation (Eq.~\ref{eq:poisson_eqaution}).
Let $\Phi$ be the solution of Eq.~\eqref{eq:poisson_eqaution} on the domain $\Omega$ with periodic boundary conditions. 
Direct application of Gauss' theorem upon integrating yields
\begin{equation}\label{eq:volume_integral}
    \int_\Omega \nabla^2 \Phi dV = \int_{\partial \Omega} \frac{\partial \Phi}{\partial\boldsymbol{n}}dS.
\end{equation}
Now, for periodic boundary condition on $\Phi$, the right hand side of Eq.~\eqref{eq:volume_integral} vanishes, which implies
\begin{equation}
    \int_{\Omega}\nabla^2 \Phi dV = 0.
\end{equation}
Thus, if $\Phi$ satisfies Eq.~\eqref{eq:poisson_eqaution}, we must have
\begin{equation}\label{eq:condition_periodic}
    \int_{\Omega} \rho dV = 0.
\end{equation}
However, the mass density $\rho$ can never be less than $0$ physically\footnote{This is analogous to the well-known contradiction in linear stability analysis of such systems, popularly known as Jeans' Swindle \citep{Binney_1987}}. 
Thus, it is impossible to satisfy Eq.~\eqref{eq:condition_periodic} without modifying the source term.
The usual way the overcome this issue is to subtract the mean value of the density from the source term, as any infinite homogeneous density distribution does not affect the potential \citep{Binney_1987,Kiessling_1999,Falco_2013}. 
Hence, we can subtract the mean homogeneous density and calculate the potential due to the fluctuation. This leads to the following modification of the source term in Eq.~\eqref{eq:poisson_eqaution}:
\begin{equation}\label{eq:source_modification}
    \rho \xrightarrow{} \rho - \Bar{\rho}\,,
    \quad{\rm where}\quad
    \Bar{\rho} = \frac{\int_{\Omega}\rho dV}{\int_{\Omega} dV}.
\end{equation}
Eq.~\eqref{eq:source_modification} ensures that Eq.~\eqref{eq:condition_periodic} is satisfied, thus implying the convergence of the multigrid algorithm is guaranteed. 
We also subtract the average from the residual on all the coarser levels ($l>0$) explicitly \citep{Ricker_2008}.
This ensures that the convergence rate of the multigrid algorithm is not affected by the non-zero uniform component of the residuals for the periodic boundary condition.


\section{Coupling gravity with hydrodynamics}\label{sec:coupling}
The effect of self-gravity is generally treated by adding source terms in Euler's equations, which are given by,
\begin{gather}
      \frac{\partial\rho}{\partial t} 
    + \boldsymbol{\nabla}\cdot(\rho\boldsymbol{v}) = 0 \label{eq:mass_conservation} \\
    \frac{\partial(\rho\boldsymbol{v})}{\partial t} 
    + \boldsymbol{\nabla}\cdot\left[\rho\boldsymbol{v}\boldsymbol{v} + p\boldsymbol{I}\right] = \rho \boldsymbol{g} \label{eq:momentum_conservation} \\
    \frac{\partial E_t}{\partial t} 
    + \boldsymbol{\nabla}\cdot\left[\left(E_t + p \right)\boldsymbol{v} )\right] = \rho\boldsymbol{v}\cdot\boldsymbol{g} \label{eq:energy_conservation}\\
    \nabla^2\Phi = 4\pi G \rho \label{eq:poisson}
\end{gather}
where $\rho$ is the mass density, $\boldsymbol{v}$ is the velocity, $p$ is the thermal pressure, $\boldsymbol{g}=-\boldsymbol{\nabla}\Phi$ is the acceleration due to gravity and $E_t$ is the total energy density given by,
\begin{equation}
    E = \rho e + \frac{\rho\boldsymbol{v}^2}{2}.
\end{equation}
The above equations are combined with an equation of state (EOS) $\rho e = \rho e(p,\rho)$ to provide the closure. Thus, the change in momentum and energy due to self-gravity is contributed by the source terms in the momentum (Eq.~\ref{eq:momentum_conservation}) and energy (Eq.~\ref{eq:energy_conservation}) equations, respectively.
In order to compute gravitational acceleration $\boldsymbol{g}$ from the potential ($\Phi$), we use a $4^{\rm th}$-order finite difference approximation, e.g.,
\begin{equation}
    \begin{split}\label{eq:4thorder_acceleration}
        (g_x)_{i,j,k} = & -\nabla_x\Phi \\
                      = & -\frac{4}{3}\frac{\Phi_{i+1,j,k}-\Phi_{i-1,j,k}}{2\Delta x}\\
                        &+ \frac{1}{3}\frac{\Phi_{i+2,j,k}-\Phi_{i-2,j,k}}{4\Delta x} + \mathcal{O}((\Delta x)^4) \,.
    \end{split}
\end{equation}

There have been recent developments for not treating gravity as a source term but in a fully conservative fashion where the momentum and energy equations are evolved by the divergence of a gravitational stress tensor and a gravitational energy flux, as the traditional source term-based approach does not guarantee the explicit conservation of energy and momentum \citep[e.g.,][]{Jiang_2013}. 
This  approach explicitly conserves the energy and the momentum of the system to the machine accuracy. 
However, some studies \citep{Springel_2010,Katz_2016,Hanawa_2019,Mullen_2021} argue that if the equations are formulated in a conservative manner the source term-based approach can give a similar level of accuracy. 
Nonetheless, all of these schemes require that the error in the gravitational potential must be in the order of round-off error \citep{Mullen_2021}, which can be achieved by the FFT algorithm but generally not possible for an iterative Poisson solver like the multigrid method. 
The small amount of residual in the solution obtained by an iterative method can not ensure the round-off error level accuracy in the energy and momentum. 
Moreover, the conservative approach requires solving Poisson's equation in multiple stages along with the underlying time-stepping scheme (e.g. RK2 or RK3), which can become computationally expensive. 
Therefore, in this work, we follow the traditional source term-based approach with some level of approximation. 

If we consider Eq.~\eqref{eq:mass_conservation}-\eqref{eq:energy_conservation} in the following form,
\begin{equation}\label{eq:conservative}
    \frac{\partial\boldsymbol{U}}{\partial t} = \boldsymbol{\mathcal{L}}, \hspace{0.5cm}\text{where}\hspace{0.5cm} \boldsymbol{\mathcal{L}} = -\boldsymbol{\nabla}.\boldsymbol{T} + \boldsymbol{S},
\end{equation}
then the semi-discrete RK2 method advances Eq.~\eqref{eq:conservative} as,
\begin{align}
    \boldsymbol{U}^* =\; & \boldsymbol{U}^n + \Delta t \boldsymbol{\mathcal{L}}^n,\\
    \boldsymbol{U}^{n+1} =\; & \frac{1}{2}\left(\boldsymbol{U}^n + \boldsymbol{U}^* + \Delta t\boldsymbol{\mathcal{L}}^*\right) \\
    =\; & \boldsymbol{U}^n + \frac{\Delta t}{2}\left(\boldsymbol{\mathcal{L}}^* + \boldsymbol{\mathcal{L}}^n\right) \label{eq:finalstep_RK2},
\end{align}
where $\boldsymbol{U}^n$ is the solution of the conservative variables at the $n^{\rm th}$ step and $\boldsymbol{\mathcal{L}}^n$ is the corresponding right-hand side {\bf array}.
 
Note that the calculation of $\boldsymbol{\mathcal{L}}^*$ at the last stage requires the gravitational potential at the predictor stage ($*$). This demands solving the Poisson equation twice during a single RK2 step, which will be computationally expensive.
Some codes use an extrapolation technique to compute the gravitational potential at the intermediate stage by linearly extrapolating the value of the potential at $t^n$ and $t^{n-1}$ \citep[e.g.,][]{Bryan_1995,Fryxell_2000}.
However, in the \textsc{Pluto} implementation, we provide the user with the option to choose between two alternatives: (i) solving the Poisson equation at each intermediate step, as explained earlier; or (ii) assuming that the potential remains constant within a single time step and using the acceleration $\boldsymbol{g}^n$ at all intermediate stages to advance the solution to $(n+1)^\mathrm{th}$ step.
The later approximation is valid if the change in density during a time-step is not dramatic and may fail for a region undergoing gravitational collapse towards singularity. 
However, the benchmarks considered in Sec.~\ref{sec:results} demonstrate an excellent agreement with analytical predictions (when available) and with many prior studies by various authors, even for a cloud undergoing gravitational collapse to a singularity.

\begin{table*}
    \centering
    \caption{List of numerical tests considered here to examine the accuracy of the self-gravity module.}
    \begin{tabular}{|c|c|c|}
        \hline
        Test Name & Primary goal & Resolution\\
        \hline
        \hyperref[sec:accuracy]{Accuracy Test}  & Test for the numerical accuracy of the Poisson solver & $16^3\mbox{-}1024^3$ \\
        \hline
        \hyperref[sec:jean_instability]{Jeans Instability} & Test for self-gravity with periodic boundary condition & $128^3$ \\
        \hline
        \hyperref[sec:adiabatic_collapse]{Adiabatic Collapse of a Sphere} & \begin{tabular}{@{}c@{}} Check for energy conservation, \\ shock formation and virialization \end{tabular} & $1024^3$ \\
        \hline
        \hyperref[sec:non_rotating]{\begin{tabular}{@{}c@{}} Collapse of an isothermal \\ non-rotating uniform sphere\end{tabular}} & \begin{tabular}{@{}c@{}} Test for the accuracy of the self-gravity module for \\ a sphere undergoing gravitational collapse to singularity \end{tabular} & $512^3$ \\
        \hline
        \hyperref[sec:rotatingU]{\begin{tabular}{@{}c@{}}Collapse of an isothermal \\ rotating uniform sphere \end{tabular}} & Check for angular momentum conservation and disc formation & $512^3$ \\
        \hline
        \hyperref[sec:rotatingP]{\begin{tabular}{@{}c@{}c@{}}Collapse of an isothermal \\ non-rotating sphere with \\ azimuthal density perturbation \end{tabular}} & Test for cloud fragmentation and binary structure formation & $512^3$ \\
        \hline
    \end{tabular}
    \label{tab:test_list}
\end{table*}

\section{Test results}
\label{sec:results}
In this section, we verify our implementation against standard astrophysical problems involving self-gravity. A list of different tests with primary goals of each of the problems is summarized in Table~\ref{tab:test_list}.
While some of these tests have been investigated by means of adaptive mesh refinement techniques, here we focus the attention on the convergence property of the multigrid RKL solver on a static grid.


\begin{figure*}
\centerline{
\def\arraystretch{1.0}
\setlength{\tabcolsep}{0.0pt}
\begin{tabular}{lcr}
  \includegraphics[width=0.4\linewidth]{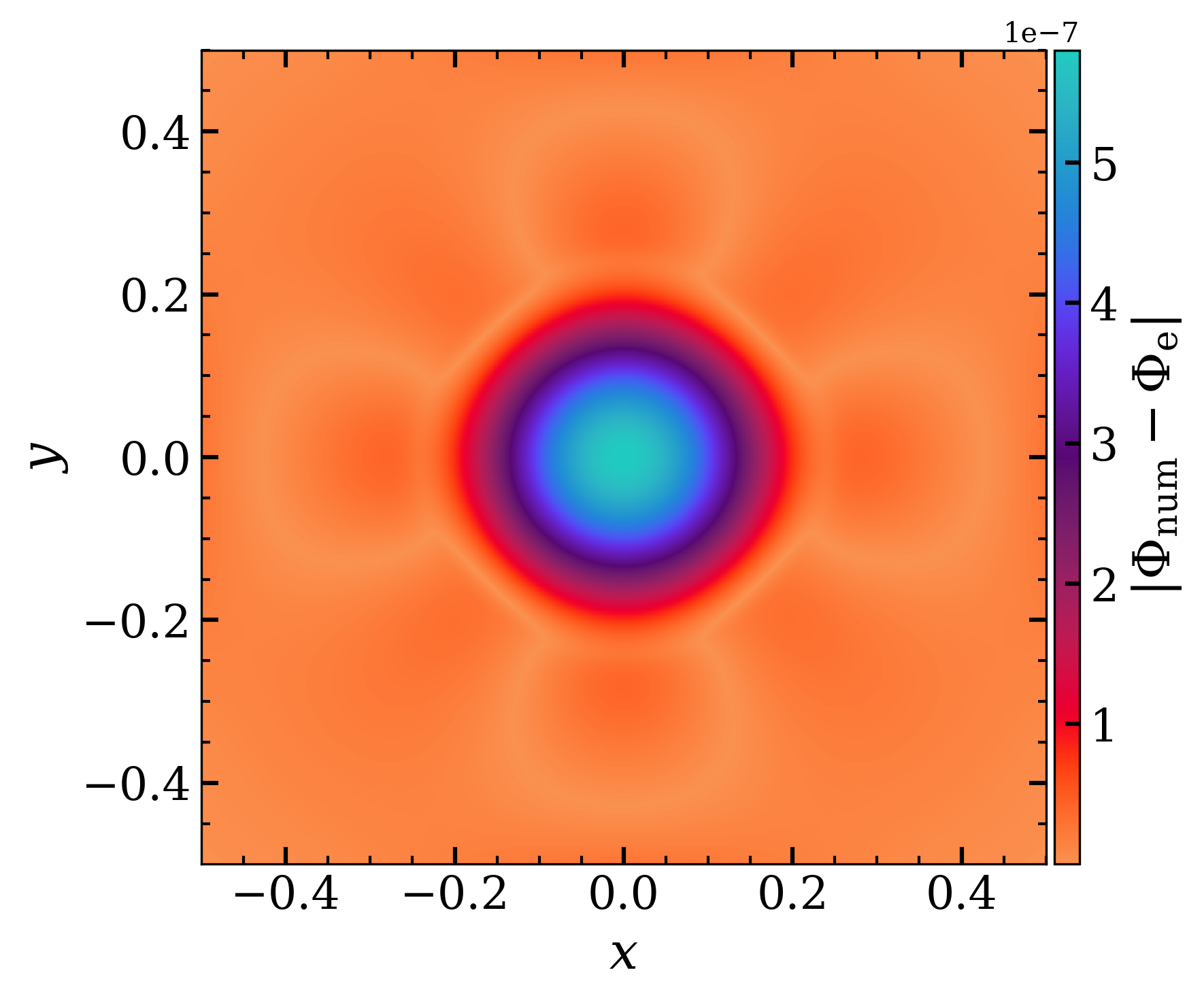} &
  \includegraphics[width=0.4\linewidth]{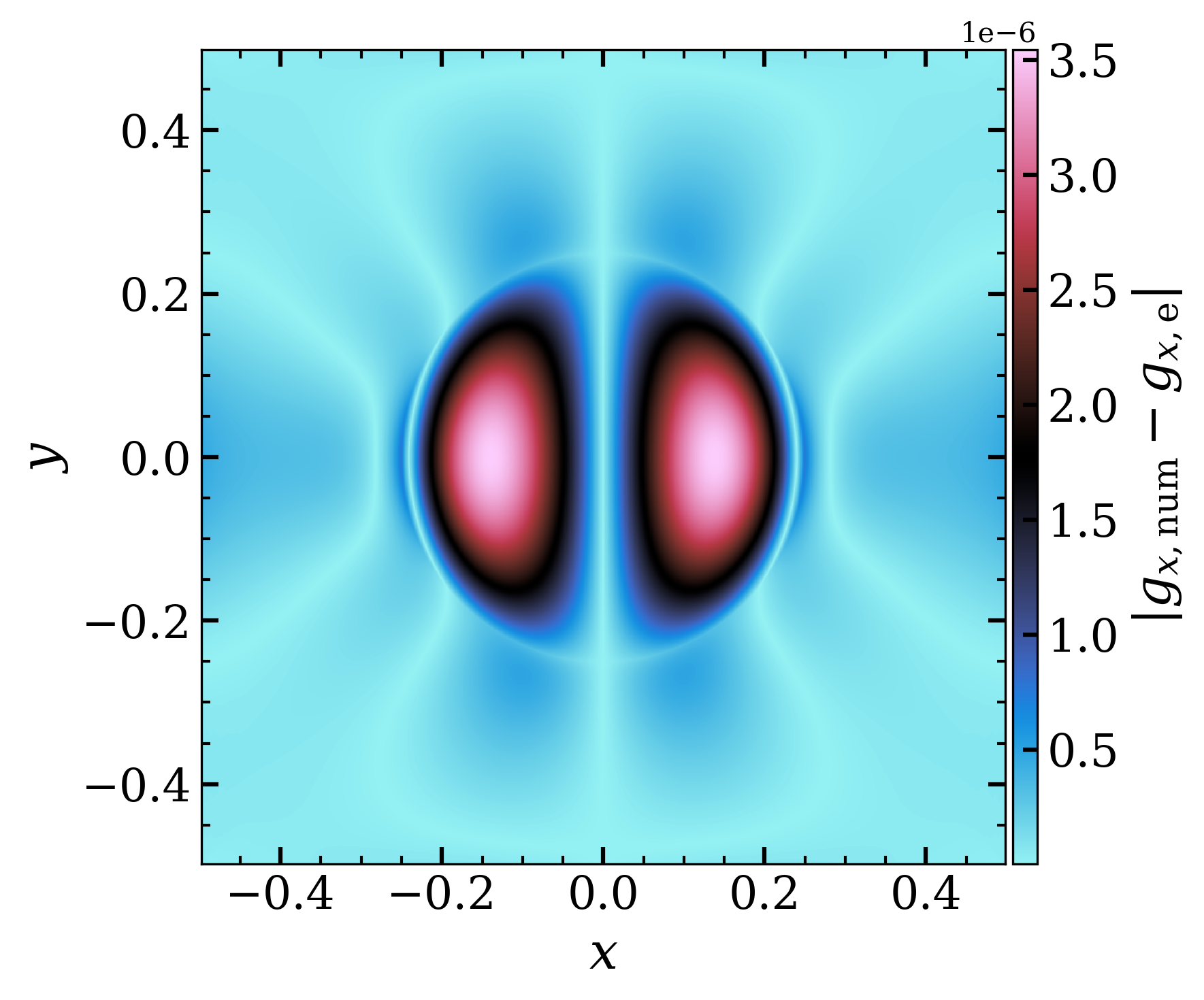}
\end{tabular}}
  \caption{Error in the numerical result of the potential (left) and $x$-component of the acceleration (right) for the 3D model problem (Sec.~\ref{sec:accuracy}). In both instances, we observe that the error magnitude is approximately $10^{-6}$ at the considered resolution of $1024^3$, which aligns with the discretization error ($\mathcal{O}(\Delta x^2)$).}
  \label{fig:error_map}
\end{figure*}

\begin{figure*}
\centerline{
\def\arraystretch{1.0}
\setlength{\tabcolsep}{0.0pt}
\begin{tabular}{lcr}
  \includegraphics[width=0.4\textwidth]{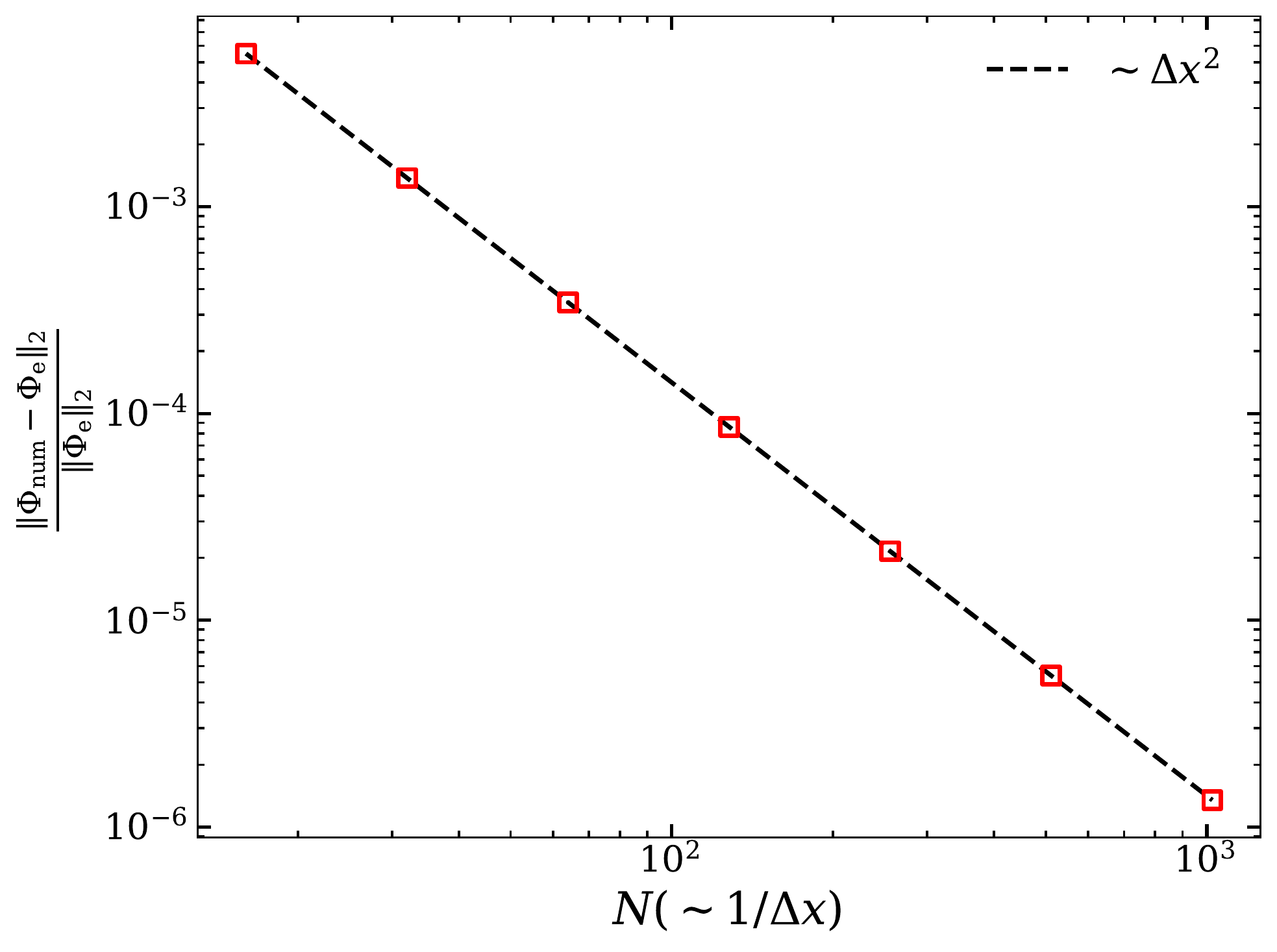} &
  \includegraphics[width=0.4\textwidth]{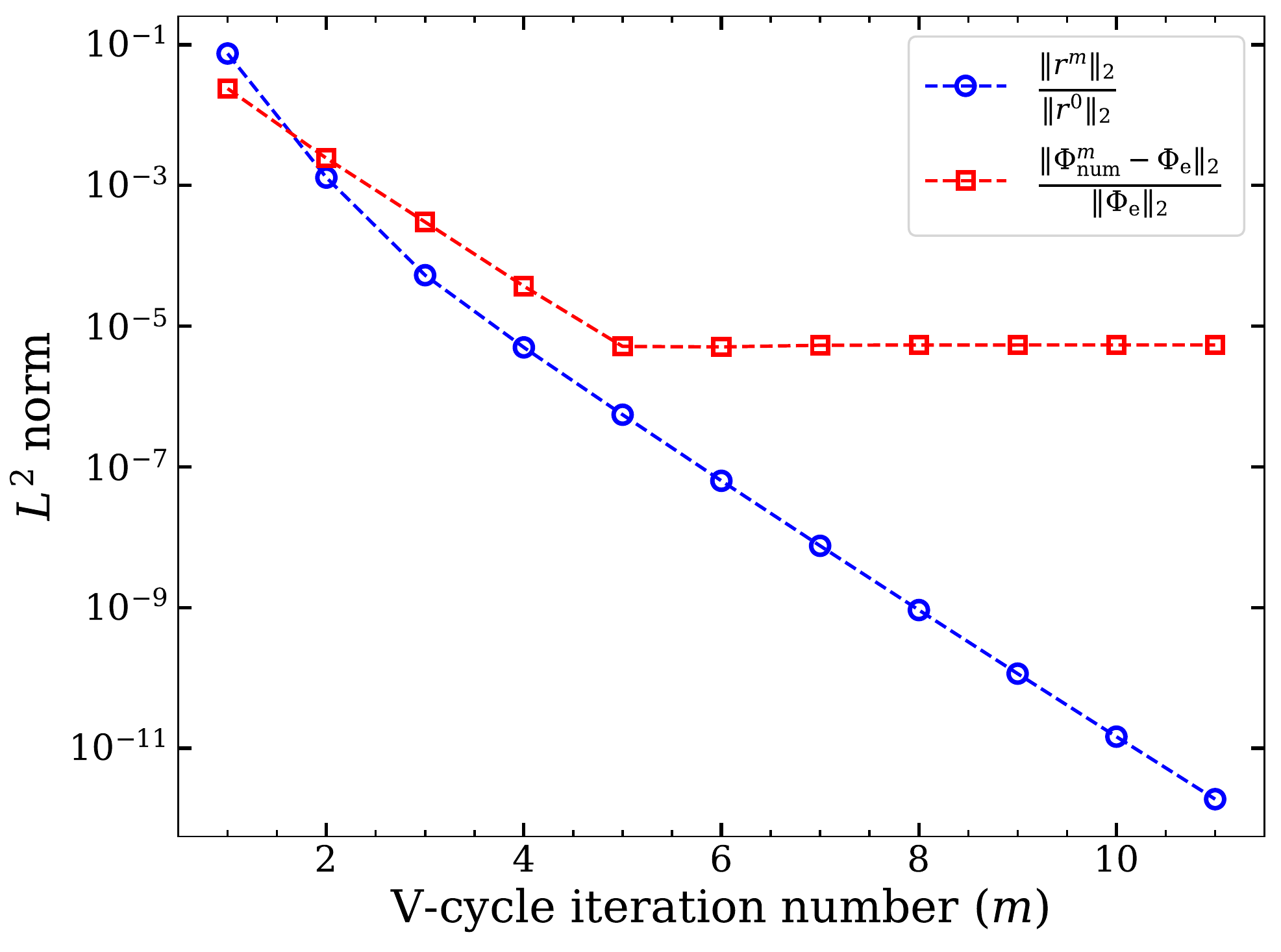}
\end{tabular}}
  \caption{\textit{Left}: The $L^2$ norm of the absolute error in the numerical solution of the potential as a function of the resolution of the finest grid level. The black dashed line is $\mathcal{O}((\Delta x)^2)$. \textit{Right}: The $L^2$ norm of the residual (blue circles) at $m^\mathrm{th}$ (scaled to $L^2$ norm of the initial residual assuming $\Phi=0$) V-cycle iteration and the $L^2$ norm of the absolute error at the $m^\mathrm{th}$ iteration step (red square) as a function of the number of V-cycle ($m$). The residual at different steps have been calculated using Eq.~\eqref{eq:residual}.}
  \label{fig:error}
\end{figure*}

\subsection{Accuracy tests for a 3D model problem}
\label{sec:accuracy}
We first begin with the accuracy of the Poisson solver described in Sec.~\ref{sec:the_algorithm}. 
To accomplish this, we consider a 3D model problem with a known analytical solution, namely,
\begin{equation}\label{eq:source_3D}
    \nabla^2\Phi = 4\pi\rho\,, \quad \text{where}\quad
    \rho(r) = 
    \begin{cases}
        \rho_0\left(1-\frac{r^2}{r_0^2}\right)^2, & \text{if $r\leq r_0$} 
        \\[4pt]
        0, & \text{if $r>r_0$}\,,
    \end{cases}
\end{equation}
where $\rho_0 = 1$ and $r_0 = 0.25$. For simplicity, we set $G=1$. 
The analytical solution of Eq~(\ref{eq:source_3D}) can be derived as
\begin{equation}\label{eq:phi_3D}
    \Phi_e(r) = 
    \begin{cases}
       \DS  -\frac{2}{3}\pi\rho_0r_0^2 + 4\pi\rho_0\left(\frac{r^2}{6} - 
        \frac{1}{10}\frac{r^4}{r_0^2} + \frac{1}{42}\frac{r^6}{r_0^4}\right), & \text{if $r\leq r_0$} 
        \\ \noalign{\medskip}
        \DS -M/r, & \text{if $r>r_0$}
    \end{cases}
\end{equation}
where $M=32\pi\rho_0r_0^3/105$ is the mass of the sphere. 
The gravitational acceleration ($\boldsymbol{g}$) corresponding to this potential (Eq.~\ref{eq:phi_3D}) is given by, 
\begin{equation}\label{eq:acceleration_3D}
    \boldsymbol{g}(\boldsymbol{r}) = 
    \begin{cases}
        -4\pi\rho_0\left(\frac{r}{3} - \frac{2}{5}\frac{r^3}{r_0^2} + \frac{1}{7}\frac{r^5}{r_0^4}\right) \hat{\boldsymbol{r}}, & \text{if $r\leq r_0$} \\
        -(M/r^2) \hat{\boldsymbol{r}}, & \text{if $r>r_0$}
    \end{cases}
\end{equation}
We consider the computational domain of $(x,y,z)\in [-0.5,0.5]$ with different resolutions ($N=16,\,32,..,1024$ along each dimension) at the finest level ($l=0$) and boundary conditions at the finest level specified using the exact solution  (Eq.~\ref{eq:phi_3D}). 
We employ the maximum possible number of coarse levels $L$ (including the finest level) such that $N/(2^{L-1})$ is an even number and each processor contains at least one active cell (excluding the ghost cells) of the coarsest level ($l=L-1$). 
For the best performance, we use $N_\mathrm{pre}=2$ and $N_\mathrm{post}=1$ in the multigrid algorithm presented in Sec.~\ref{v-cycle}.

In Fig.~\ref{fig:error_map}, we show the error in the numerical solution in potential (left) and the $x$-component of the acceleration vector for $1024^3$ box calculation of the problem described above. 
The acceleration has been calculated from the potential using Eq.~\eqref{eq:4thorder_acceleration}. 
In both cases, we observe that the error is in the order of the discretization error ($\mathcal{O}(\Delta x^2)\approx 10^{-6}$). 
To be more quantitative, we define the $L^2$ norm ($\norm{\cdot}_2$) of a quantity $\mathcal{F}_{i,j,k}$ as,
\begin{equation}
    \norm{\mathcal{F}}_2 = \left[\frac{1}{V}\sum_{i,j,k}(\mathcal{F}_{i,j,k})^2 dV_{i,j,k}\right]^{1/2},
\end{equation}
where $dV_{i,j,k}$ is volume of the cell with index ($i,j,k$) and $V$ is the total volume of the computational domain.
In the left panel of Fig.~\ref{fig:error}, we show the $L^2$-norm of the absolute error, defined as $(\norm{\Phi_{\rm num} - \Phi_{\rm e}})/\norm{\Phi_{\rm e}}$, in the numerical solution of the potential (red square) as a function of the finest grid resolution (the black dashed line represents the $2^{\rm nd}$-order predicted accuracy). 
As expected, our numerical results match the $2^{\rm nd}$-order convergence for the considered resolutions. 
The right panel of Fig.~\ref{fig:error} displays the $L^2$-norm of the residual of the numerical solution at the $m^{\rm th}$ V-cycle (blue circles) as a function of the number of V-cycle iteration ($m$) for the $512^3$ case. 
We also show the $L^2$-norm of the absolute error ($\norm{\Phi^m-\Phi_e}/\norm{\Phi_e}$) as a function of the iteration number (red squares).
It is worth noting that every multigrid V-cycle diminishes the error norm by approximately one order of magnitude, i.e., the convergence factor is about $10$. 
Likewise, the absolute error is also reduced by a similar factor initially and after $4-5$ V-cycles, it quickly attains the truncation error value. 
Thus, we only need just a few numbers of V-cycle for a problem to attain accuracy of the order of truncation. 
Furthermore, if the solution at the previous step is used as the guess for the current step, it just takes 2-3 V-cycle to converge to an error level $\max_\Omega|\Phi^m-\Phi^{m-1}|<10^{-6}$, where $\Phi^m$ is the solution at the $m^\mathrm{th}$ V-cycle.

\subsection{Jeans instability test}
\label{sec:jean_instability}

\begin{figure*}
\centerline{
\def\arraystretch{1.0}
\setlength{\tabcolsep}{0.0pt}
\begin{tabular}{lcr}
  \includegraphics[width=0.4\textwidth]{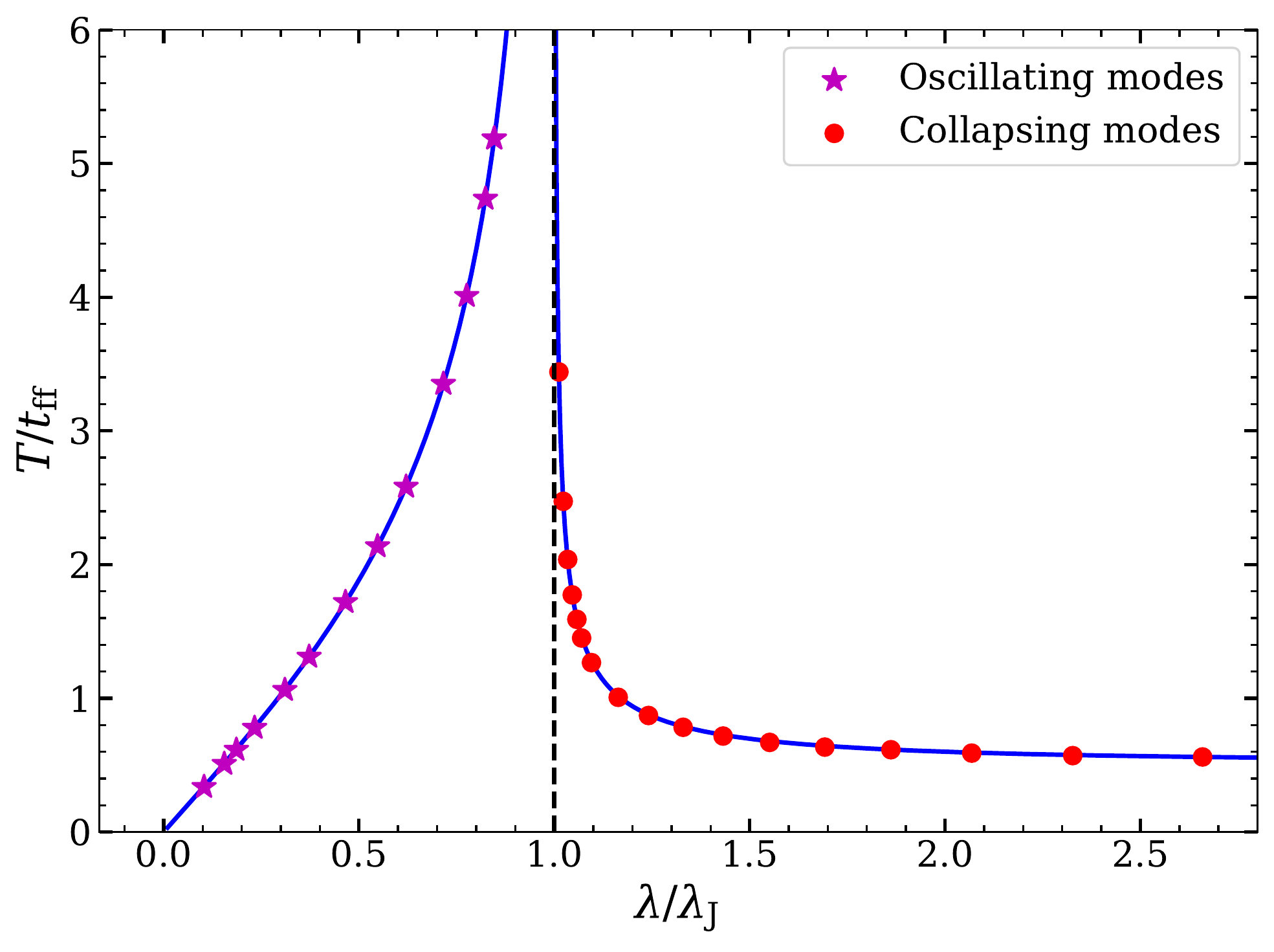} &
  \includegraphics[width=0.4\textwidth]{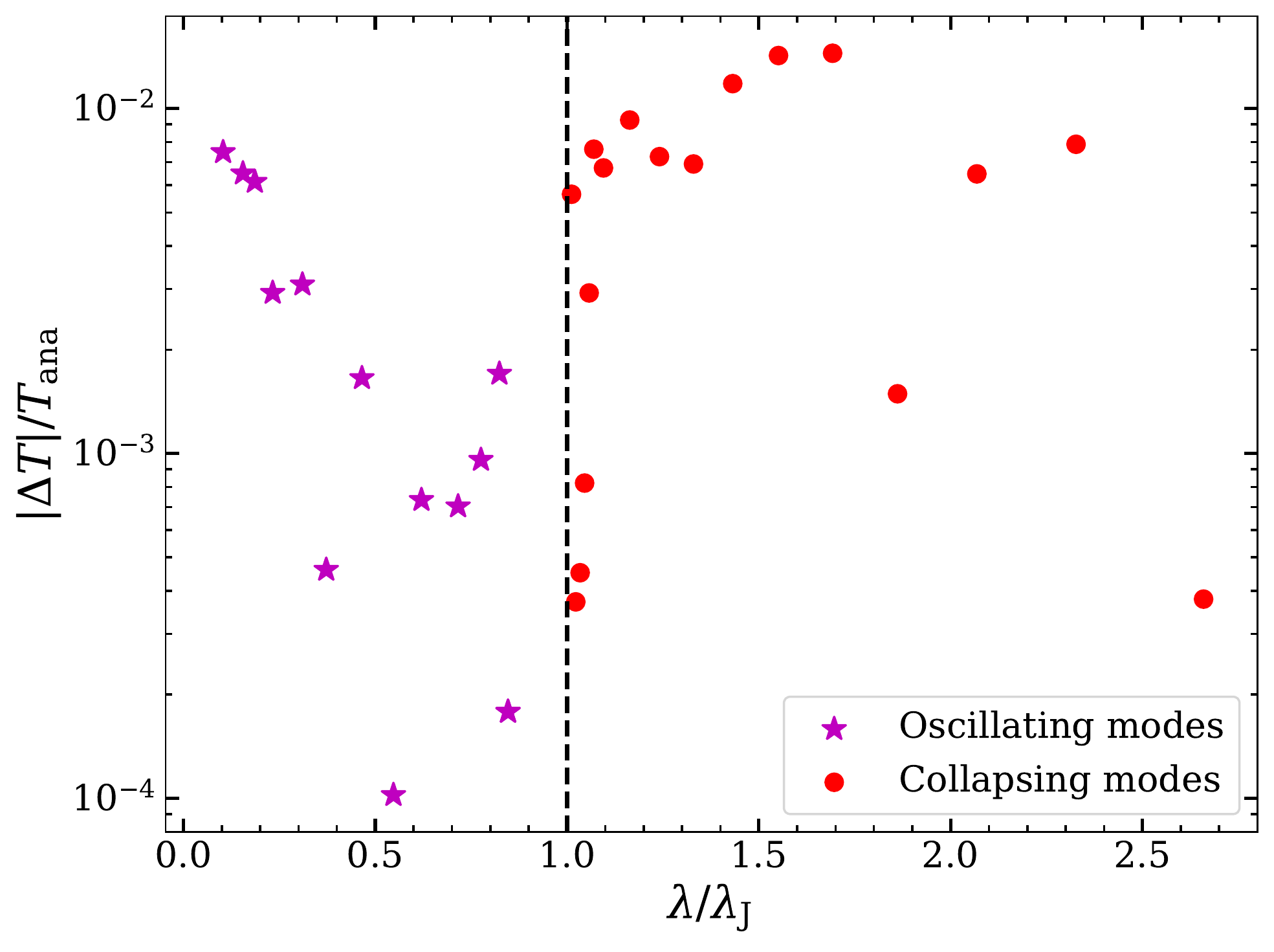}
\end{tabular}}
  \caption{\textit{Left}: The characteristic time-scales for the evolution of the sinusoidal perturbation in Jean's instability test. The $x$-axis represents the wavelength ($\lambda$) of the plane-wave perturbation in the unit of Jean's length ($\lambda_\mathrm{J}$) and the $y$-axis is the corresponding time-scale for oscillation (or collapse) in the unit of the free-fall time ($\tff$). The blue line in $\lambda<\lambda_\mathrm{J}$ region represents the oscillation time-scales (Eq.~\ref{eq:time_oscillation}) as a function of $\lambda$ of the oscillating modes and in the $\lambda>\lambda_\mathrm{J}$ it shows the characteristic collapse time-scale (Eq.~\ref{eq:time_collapse}) of the collapsing modes. The black dashed line marks the $\lambda=\lambda_\mathrm{J}$ point, where the oscillation/collapse timescale tends to infinity. The purple stars and the red circles are the corresponding results from the simulations, which agree very well with the analytical prediction. \textit{Right}: The error in numerical results. The quantity $\Delta T$ is defined as $\Delta T = T_{\rm num} - T_{\rm ana}$, where $T_{\rm ana}$ and $T_{\rm num}$ are the analytical and numerically calculated timescales, respectively. We find that the error in the numerical result for the considered range of $\lambda/\lambda_{\rm J}$ is less than $1\%$.}
  \label{fig:timescales_JeansInstability}
\end{figure*}

There exist very few problems with periodic gravity that have analytical solutions and the Jeans instability test is one of them \citep{Hubber_2006,Hubber_2018}. 
This problem can be used to test the accuracy of the periodic boundary conditions imposed for the gravitational potential. 
We briefly describe the problem below \cite[see][for a detailed description]{Hubber_2006}.

We start from an infinite uniform medium with density $\rho_0$ and isothermal sound speed $\cs$, initially at rest. By linearizing the isothermal Euler equations (i.e., Eq.~\ref{eq:mass_conservation}-\ref{eq:poisson}, and replacing Eq. \ref{eq:energy_conservation} with $p = c_s^2\rho$) with a small density perturbation $\rho_1(\boldsymbol{r},t) = A\rho_0e^{i(\boldsymbol{k}.\boldsymbol{r}\pm\omega t)}$, one finds that there exists a critical value of the wave-number ($k_\mathrm{J}$) below which $\omega$ becomes imaginary and the perturbations grow exponentially in time by the action of gravity. 
The corresponding wavelength ($\lambda_\mathrm{J}$) associated with $k_\mathrm{J}$ is called the Jeans length, given by,
\begin{equation}\label{eq:jeans_length}
    \lambda_\mathrm{J} = \sqrt{\frac{\pi \cs^2}{G\rho_0}} \,,
\end{equation}
which divides the oscillating short-wavelength perturbations from collapsing long-wavelength modes. 
The time period of the oscillating stable modes ($\lambda<\lambda_\mathrm{J}$) can be derived as \citep{Hubber_2006},
\begin{equation}\label{eq:time_oscillation}
    T^\mathrm{osc}_\lambda = \left(\frac{\pi}{G\rho_0}\right)^{1/2} \frac{\lambda}{(\lambda_\mathrm{J}-\lambda)^{1/2}}.
\end{equation}
Similarly, the collapse timescale (defined as the time taken by the perturbation to grow by a factor $\cosh(1)$) of the unstable wavelengths ($\lambda>\lambda_\mathrm{J}$) is given by \citep{Hubber_2006},
\begin{equation}\label{eq:time_collapse}
    T^\mathrm{col}_\lambda = \left(\frac{1}{4\pi G\rho_0}\right)^{1/2} \frac{\lambda}{(\lambda-\lambda_\mathrm{J})^{1/2}} \,.
\end{equation}
For the simulation setup, we consider a box with $128^3$ zones, extending from -1 to 1 along each direction with a uniform background density $\rho_0$ superimposed with a sinusoidal perturbation of amplitude $A\rho_0$ along $x$-direction:
\begin{equation}\label{eq:jeans_source}
    \rho(\boldsymbol{r}) = \; \rho_0\left[1+A\cos\left(\frac{2\pi x}{\lambda}\right)\right].
\end{equation}
The medium is set to be at rest initially with $\rho_0=1$ and $A=0.01$. 
With this numerical setup, we perform several simulations with different values of $\lambda/\lambda_{\rm J}$ covering both oscillatory and collapsing modes. 
We use the same value of the wavelength ($\lambda=2$) of the perturbation for all the simulations and adjust the sound speed which changes the Jeans length (Eq.~\ref{eq:jeans_length}), yielding an effective range of values of $\lambda/\lambda_\mathrm{J}$. 
For computing the oscillation timescale from the simulations, we first find the time evolution of the density at point $(x,y,z)=(0,0,0)$, which is a sinusoidal function in time. 
Then we calculate the period of oscillation by finding the consecutive local minima or maxima of the sinusoidal function. We consider three such periods of oscillation and take the average value, which gives the oscillation timescale. 
For the collapsing modes, the timescales are estimated by calculating the time it takes the density at $(x,y,z)=(0,0,0)$ to increase from the initial value $A\rho_0$ to a value of $A\rho_0\cosh{(1)}$.

In the left panel of Fig.~\ref{fig:timescales_JeansInstability}, we show the results from our simulation with different values of $\lambda/\lambda_\mathrm{J}$. 
The purple stars and red circles represent the oscillation timescales and characteristic collapse timescale for the oscillating modes ($\lambda<\lambda_\mathrm{J}$) and growing modes ($\lambda<\lambda_\mathrm{J}$), respectively. 
The blue lines are the corresponding analytical solutions (Eq.~\ref{eq:time_oscillation} and Eq.~\ref{eq:time_collapse}). 
The black dashed line marks the $\lambda=\lambda_\mathrm{J}$ threshold, where the oscillation/collapse timescale tends to infinity.
All the timescales are normalized to the freefall time ($\tff$), which is given by,
\begin{equation}\label{eq:tff}
    \tff = \sqrt{\frac{3\pi}{32\rho G}}
\end{equation}
It is evident that our simulation results exhibit a strong agreement with the analytical prediction across the entire range, with an error of less than $1\%$ (as shown in the right panel).
It is important to acknowledge that the analytical solution for this problem strictly holds for linearized Euler's equations.
Therefore, the amplitude of the perturbations (defined by the parameter $A$) must be significantly smaller than the background density. 
For a large amplitude perturbation, for instance, $A=0.1$ as considered by \citet{Hubber_2006,Hubber_2018}, some terms in Euler's equations can not be neglected.
Thus, the timescale calculated from the simulations will deviate from the analytical solutions (Eq.~\ref{eq:time_oscillation} and Eq.~\ref{eq:time_collapse}).


\begin{figure*}
\centerline{
\def\arraystretch{1.0}
\setlength{\tabcolsep}{0.0pt}
\begin{tabular}{lcr}
  \includegraphics[width=0.4\linewidth]{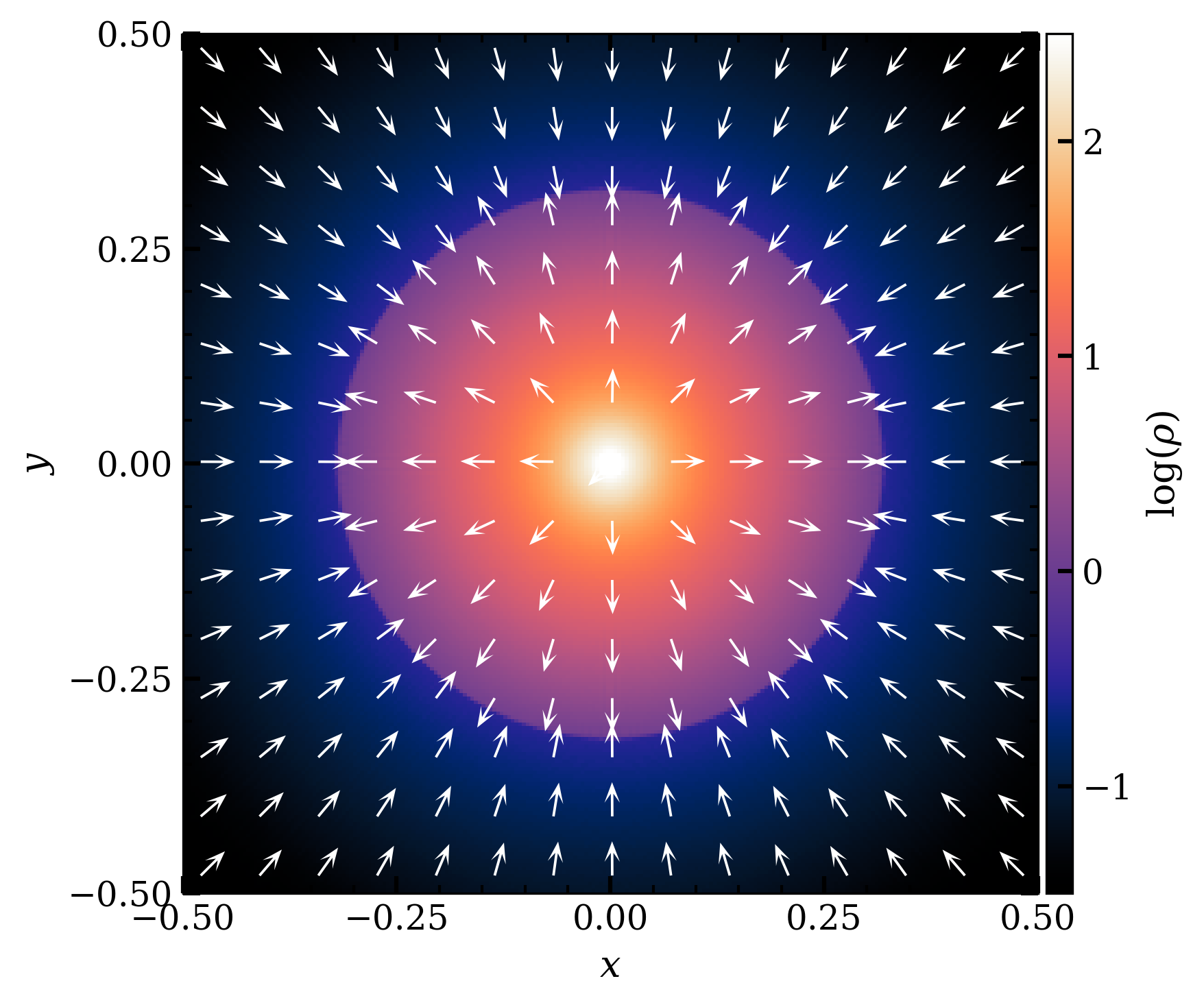} &
  \includegraphics[width=0.4\linewidth]{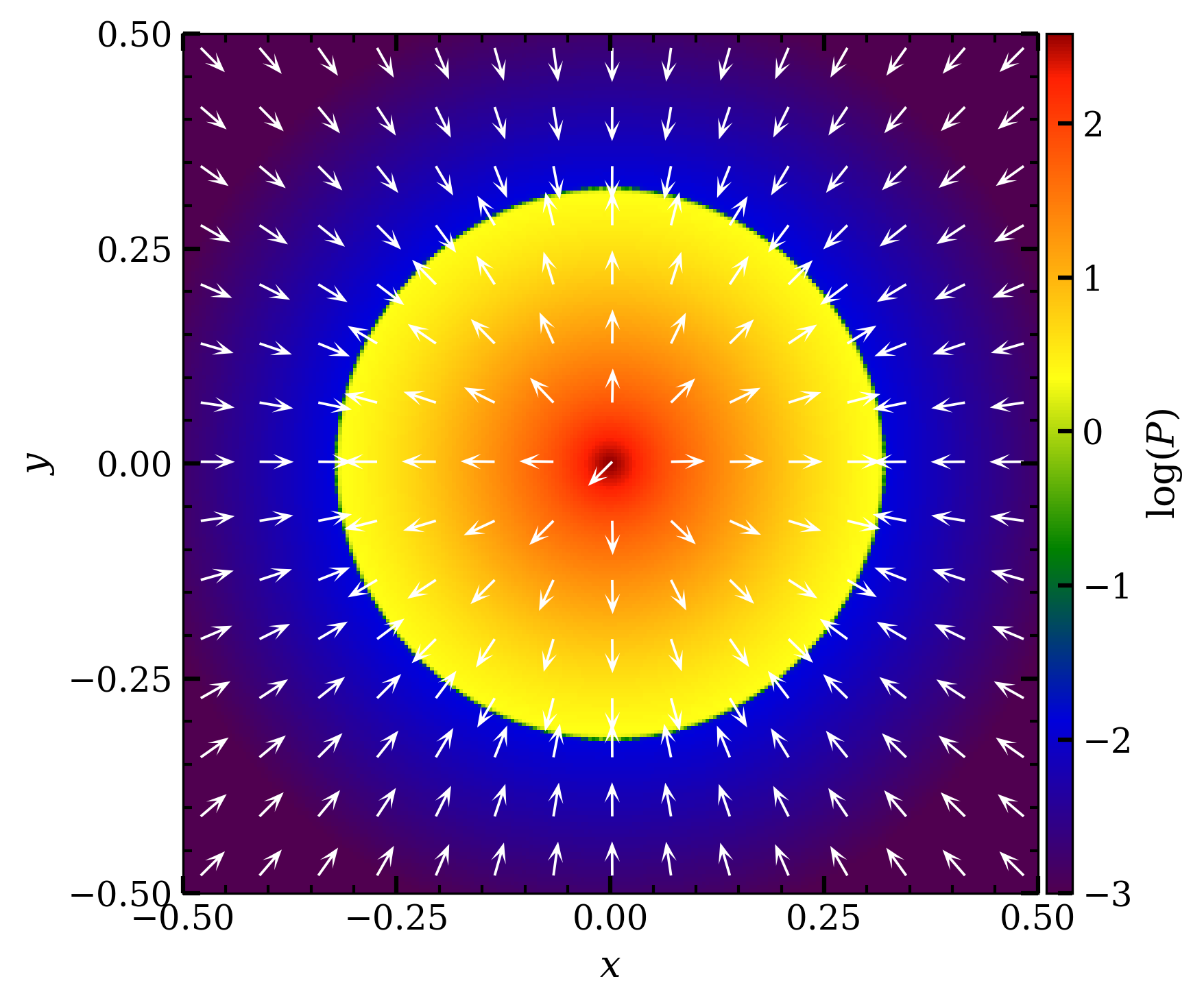}
\end{tabular}}
  \caption{The slice of the density (left) and pressure (right) distribution in the $x\mbox{-}y$ plane at $t=1.1$. The white arrows represent the local velocity field. We can clearly see the shock in both the density and pressure structure which is going outward through the still in-falling outer part.}
  \label{fig:Adiabatic_slice}
\end{figure*}

\begin{figure*}
    \centering
    \includegraphics[width=\linewidth]{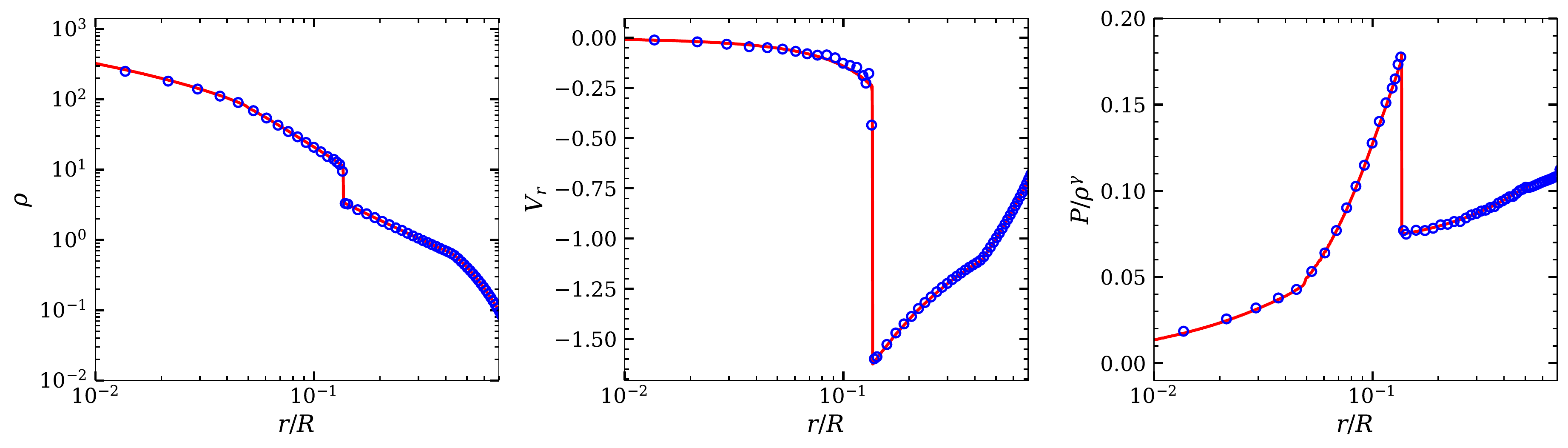}
    \caption{Radial profile of density (left), velocity (middle), and entropy function (right, defined as $P/\rho^\gamma$) for the adiabatic collapse problem (Evrard's collapse) at time $t=0.8$. The solid red lines in each plot are the results from a very high resolution (4096 grid points) spherically symmetric 1D PPM calculation of the same problem, similar to the calculation by \citet{Steinmetz_1993}, which can be taken as a reference solution of the problem. We see that the results from the 3D calculation in a cartesian geometry (blue circles) excellently match with the 1D PPM results in both the pre- and post-shock regions and also the shock profile is well recovered in the 3D simulation.}
    \label{fig:Adiabatic_profile}
\end{figure*}

\subsection{Collapse of an adiabatic gas sphere (Evrard's collapse)}
\label{sec:adiabatic_collapse}
One very interesting problem that includes self-gravity is the collapse of an initially cold gas cloud, originally proposed by \citet{Evrard_1988}. 
This test problem has been considered for validating the implementation of the self-gravity and energy conservation in SPH codes by many authors \cite[e.g.,][and many more]{Hernquist_1989,Springel_2001,Wadsley_2004,Springel_2005,Springel_2010,Hopkins_2015,Grudic_2021} but has been rarely used in grid codes. 
Here, we use our grid-based method to simulate the problem. 
The cloud is initially at rest and the density profile is given by,
\begin{equation}\label{eq:adiabatic_source}
    \rho = 
    \begin{cases}
         M/(2\pi R^2 r), & {\rm if} \quad r\leq R \,,\\
         0, & {\rm otherwise}
    \end{cases}
\end{equation}
where $M=1$ and $R=1$. 
Although in the original SPH formulation, the density of the background medium is set to $0$, we employ here a value of $10^{-4}$ to avoid the vacuum conditions in a grid-based code like \textsc{Pluto}.
The initial thermal energy per unit mass of the cloud is set to $u_{\rm th}=0.05$, which is very small compared to the gravitational binding energy ($u_{\rm g}$) of the cloud ($u_{g} = 0.6$, assuming $G=1$). The adiabatic index of the gas is set to be $\gamma=5/3$.

As the initial thermal pressure support is negligible compared to the gravitational force, the gas freely falls toward the center of the sphere. 
The adiabatic EOS results in a pressure build-up in the central region.
Eventually, a bounce back occurs, where a strong shock propagates outwards through the in-falling outer part of the sphere.
If a zero-gradient boundary condition is applied, this shock exits the computational domain, leading to a reduction in total energy due to the outflowing matter. 
However, as we are interested in the conservation of energy, we employ reflective boundary conditions at all the boundaries.
This choice ensures that the shock remains within the boundaries and gets reflected back whenever it reaches a boundary.
Furthermore, to safeguard the virialized sphere from the impact of the reflected shock, we confine our computational domain to the range of $(x,y,z)\in [-2,2]$.
We perform the simulation up to $t=3$ and confirm that the reflected shock does not reach the virialized sphere until the end of the simulation.

The formation of the shock is shown in Fig.~\ref{fig:Adiabatic_slice}, where a slice-cut of the density (left) and pressure (right) distributions in $x\mbox{-}y$ plane at $t=1.1$, overlayed with the local velocity field (white arrows) is represented.
As the shock sweeps throughout the entire domain, the system virializes and eventually settles into a hydrostatic equilibrium. 
The problem involves the conservation of gravitational to kinetic and thermal energy and it is thus particularly sensitive to the code's ability to conserve total energy.

In Fig.~\ref{fig:Adiabatic_profile}, we show the radial profile (blue circles) of density (left), radial velocity (middle), and entropy (right) at $t=0.8$ (after the formation of the outward shock).
For reference, we depict results from a very high resolution ($8192$ grid points) spherically symmetric 1D PPM calculation of the problem (red solid lines), similar to the computation done by \cite{Steinmetz_1993} which has been used as a common benchmark by numerous authors comparing. 
In our 1D calculation, we compute the gravitational potential ($\Phi$) by explicitly integrating the density field at each time-step and imposing the boundary condition: $\Phi(r) = 0$ as $r\xrightarrow{}\infty$. 
Thus, the high-resolution 1D PPM calculation can be considered as a reference solution to the problem as there is no analytical solution available. 
We see that the results obtained from the 3D simulation conducted in Cartesian geometry exhibit a very good agreement with the 1D results.
The shock structure in every plot is captured correctly as well as the pre- and post-shock regions.

\begin{figure}
    \centering
    \includegraphics[width=\linewidth]{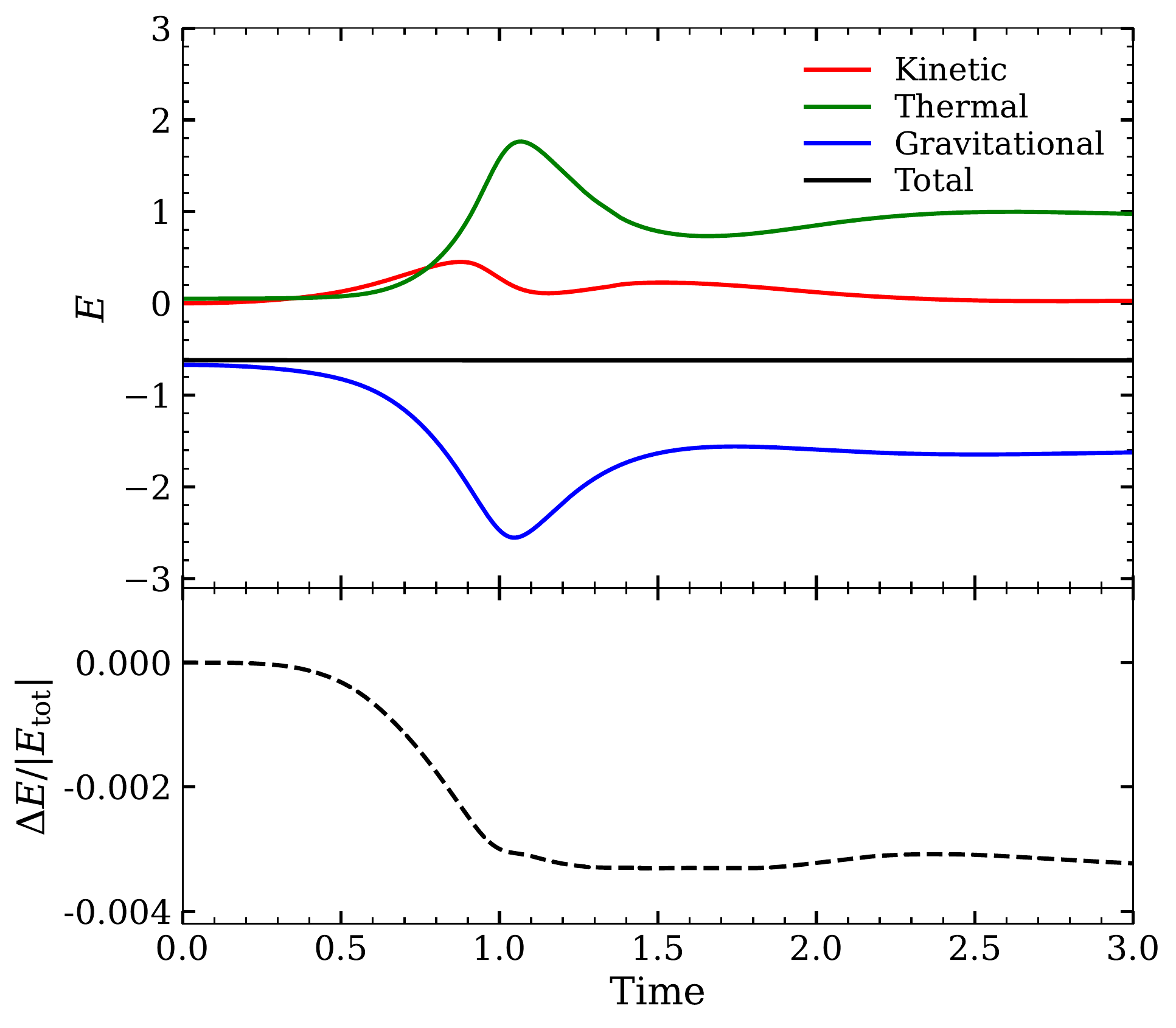}
    \caption{Top: The evolution of kinetic (red), gravitational (blue), thermal (green), and total (black) energies with time for the adiabatic collapse problem. Dimensionless units are considered for both energy and time. Bottom: The fractional error in total energy as a function of time. The conservation of total energy is quite well during the whole evolution with error $\lesssim 0.4 \%$.}
    \label{fig:Adiabatic_energy}
\end{figure}

Fig.~\ref{fig:Adiabatic_energy} presents the temporal evolution of thermal (green), kinetic (red), gravitational (blue), and total (black) energy densities as the sphere goes through different phases of evolution. 
We can readily observe the conversion of gravitational energy to kinetic energy (first) and then to the thermal energy of the gas.
Throughout the entire evolution, the total energy remains well conserved, with an error of less than $0.4\%$. 
Upon virialization of the sphere, the kinetic energy reaches zero, and the total energy is shared between the thermal and gravitational energy.


\subsection{Isothermal Collapse Tests}
\label{sec:isothermal_collapse}
We now consider the collapse of an isothermal gas sphere using various initial configurations, i.e., non-rotating uniform cloud (Sec.~\ref{sec:non_rotating}), rotating uniform cloud (Sec.~\ref{sec:rotatingU}) and rotating cloud with azimuthal perturbation (Sec.~\ref{sec:rotatingP}).
While the dynamical evolution of these systems is best described within the AMR framework, we still show that - at moderate resolution ($512^3$) - our gravity module successfully reproduces the main features of the results. 
We use a 3D cubic box for all the simulations and the gravitational potential at the boundary is calculated using the multipole expansion up to $l=4$ \citep{Mandal_2021} and supplied to the Poisson solver as a Dirichlet-type boundary condition. 
For the hydrodynamics, outflow boundary conditions are used at all the boundaries.

\begin{figure*}
\centerline{
\def\arraystretch{1.0}
\setlength{\tabcolsep}{0.0pt}
\begin{tabular}{lcr}
  \includegraphics[width=0.4\linewidth]{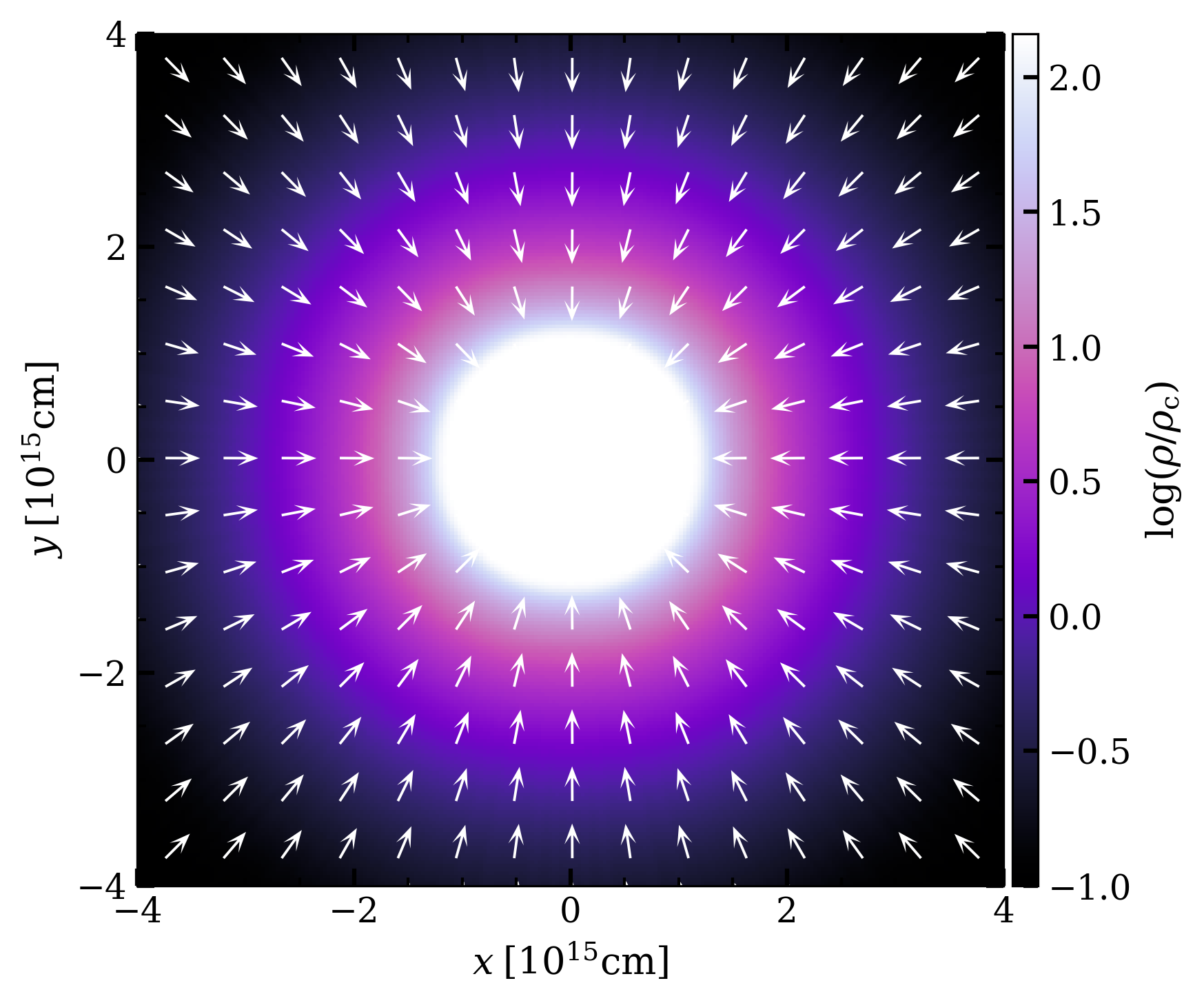} &
  \includegraphics[width=0.4\linewidth]{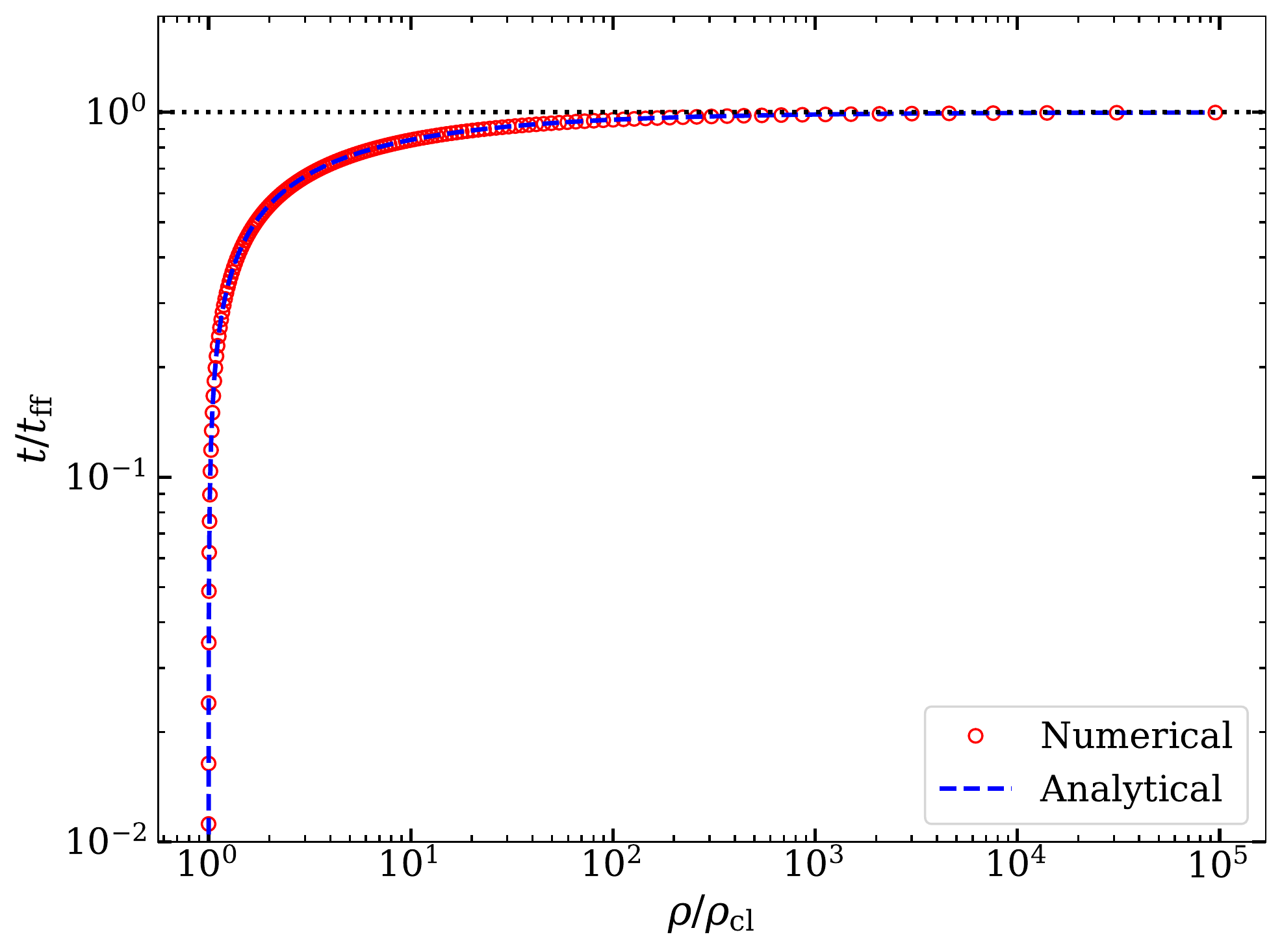}
\end{tabular}}
  \caption{\textit{Left}:Mid-plane slice of the density distribution in $x\mbox{-}y$ plane for the non-rotating isothermal collapse problem at $t=0.963\;\tff$. The white arrows show the local velocity field. The density values are normalized to the initial cloud density ($\rho_\mathrm{c}$). \textit{Right}: Time to reach a density $\rho$ for the non-rotating isothermal collapse test (red circles). The blue dashed line represents the analytical solution (Eq.~\ref{eq:isothermal_rho_t}).}
  \label{fig:isothermal_plot}
\end{figure*}

\subsubsection{Non-rotating uniform sphere}
\label{sec:non_rotating}
First, we consider the collapse of a uniform non-rotating cold gas sphere, originally proposed by \cite{Truelove_1998} and used to test the accuracy of the grid-based code \textsc{Nirvana} \citep{Ziegler_2005}. 
The initial condition consists of a gas sphere of radius $R = 7.8\times 10^{15}\,\mathrm{cm}$ with density $\rho_\mathrm{cl} = 10^{-15}\,\mathrm{g\,cm^{-3}}$ placed in a background medium of density $\rho_\mathrm{b} = 0.01\rho_\mathrm{cl}$. 
The isothermal sound speed inside the gas is set to $\cs = 0.167\,\kms$. 
This is an excellent problem as it provides an analytic solution in the pressureless ($T=0$) case which consists of a self-similar evolution. 
However, as we cannot model a pressureless and infinite medium, the solution consists of a rarefaction wave traveling radially inward from the cloud surface superimposed on a self-similar collapse. 
At any time, the region inside the rarefaction wave remains uniform and the density increases self-similarly which can be described by the analytical solution \citep{Truelove_1998}. This is illustrated in the left panel of Fig.~\ref{fig:isothermal_plot}, where we show the slice of the density distribution in the $x\mbox{-}y$ plane and the velocity field (white arrows) at $t=0.963\,\tff$. 
The central density plateau is clearly visible, which still follows the pressureless solution and is unaffected by the rarefaction front. 
The medium with radially decreasing density beyond the central plateau represents the outer part of the sphere which has been affected by the inward rarefaction wave that initially appeared at the cloud surface.
At $t=\tff$, it approaches a singular state of infinite density, and the temperature of the medium is chosen in such a way that the collapse occurs before the rarefaction wave reaches the cloud center. 
The time to  build-up a density $\rho$ inside the rarefaction wave can be derived analytically \cite[see][]{Truelove_1998}:
\begin{equation}\label{eq:isothermal_rho_t}
    \frac{t_\rho}{\tff} = \frac{2}{\pi}\left[\eta + \frac{1}{2}\sin{(2\eta)}\right],\;\;\text{where}\;\;\eta=\arccos{\left[\left(\frac{\rho}{\rho_\mathrm{cl}}\right)^{-1/6}\right]}
\end{equation}
In the right panel of Fig.~\ref{fig:isothermal_plot}, we compare the analytical expectation (Eq.~\ref{eq:isothermal_rho_t}, blue dashed line) with our numerical results (red circles).
It is evident that our numerical results align remarkably well with the analytical prediction across the entire range of density and time. 
In the simulation, the density increases by more than $5$ orders of magnitude at $0.9982\tff$ and the code breaks at $0.99994\tff$ as we can not numerically reach the singular state. 
The primary difference between our results and those of \cite{Truelove_1998} is that their solution features a time delay with respect to the analytic solution. 
This is due to the image boundary conditions they used for gravitational potential, which mimics the existence of neighboring clouds that oppose the collapse by the attracting force. 
As the image cloud separation is increased in their study, the time delay decreases. 
In our simulation, on the other hand, we use the Dirichlet boundary conditions for the gravitational potential, which is calculated from the multipole expansion considering $\Phi(r)\xrightarrow{}0$ as $r\xrightarrow{}\infty$. 
This better represents an isolated system and eventually leads to a better agreement with the analytical solution.
A similar result is also reported by \cite{Ziegler_2005}, where they have also used the multipole expansion to specify the boundary conditions and found no time delay compared to the analytical solution.

\begin{figure*}
    \centering
    \includegraphics[width=0.8\linewidth]{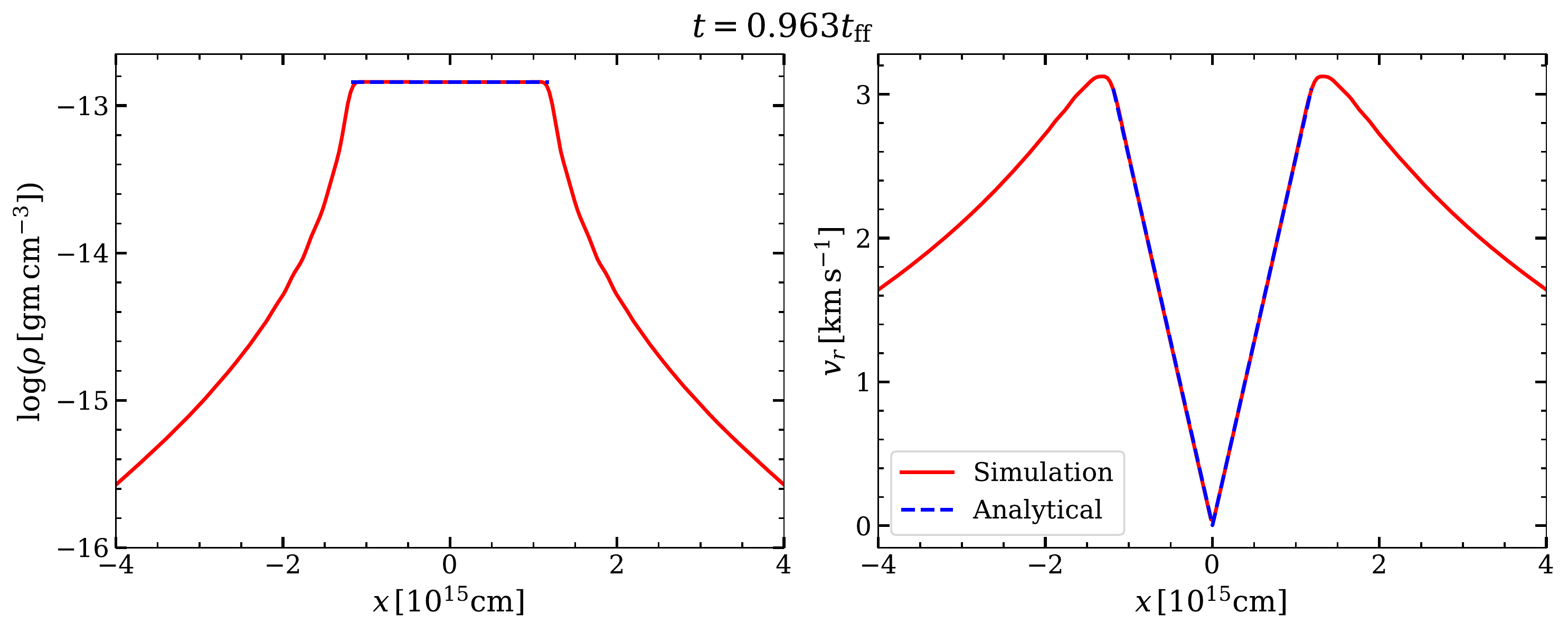}
    \caption{Cut of the density (left) and the radial velocity field (right) along $x$-axis of the left panel of Fig.~\ref{fig:isothermal_plot}. The red solid lines are the results of the simulation. The blue dashed lines show the analytical solutions which are only valid inside the rarefaction front.}
    \label{fig:isothermal_radial}
\end{figure*}

We can also derive the mass ($M_\mathrm{rf}$), size ($r_\mathrm{rf}$), and the radial velocity ($v(\boldsymbol{r},\rho))$) of the unaffected region inside the rarefaction wave as functions of central density \cite[see][for the derivation]{Truelove_1998}.
The mass inside the rarefaction wave at a density $\rho$ is given as,
\begin{equation}\label{eq:mass_rarefaction}
    M_\mathrm{rf} = M\left\{1-2\frac{\cs}{v_\mathrm{ff}}\arctan\left[\left(\frac{\rho}{\rho_\mathrm{cl}}\right)^{1/3}-1\right]^{1/2}\right\}^3,
\end{equation}
where $M$ is the mass of the cloud and $v_\mathrm{ff}=\sqrt{2GM/R}$ is the free-fall velocity. 
Thus, from Eq.~(\ref{eq:mass_rarefaction}), we can easily calculate the radius of the central region inside the rarefaction front ($r_\mathrm{rf}$) to be:
\begin{equation}\label{eq:radius_rarefaction}
    r_\mathrm{rf} =  R\left\{1-2\frac{\cs}{v_\mathrm{ff}}\arctan\left[\left(\frac{\rho}{\rho_\mathrm{cl}}\right)^{1/3}-1\right]^{1/2}\right\}.
\end{equation}
From the energy conservation, the flow velocity can be expressed as a function of the radius $r$ and of the density $\rho$ \citep{Truelove_1998}:
\begin{equation}\label{eq:flow_velocity}
    v(r,\rho) = -v_\mathrm{ff}\frac{r}{r_\mathrm{rf}},\left[\left(\frac{\rho}{\rho_\mathrm{cl}}\right)^{1/3}-1\right]^{1/2}\;\; \text{for}\;\; r\leq r_\mathrm{rf}.
\end{equation}
Fig.~\ref{fig:isothermal_radial} displays the cut of density (left) and radial velocity (right) along the $x$-axis at $t=0.963\,\tff$. 
The red solid lines in both panels represent the results from our simulation while the blue dashed lines are the corresponding analytical solutions, valid inside the rarefaction front. For both cases, the simulation results agree well with the analytical prediction inside the rarefaction front.

\begin{figure*}
    \centering
    \includegraphics[width=0.8\linewidth]{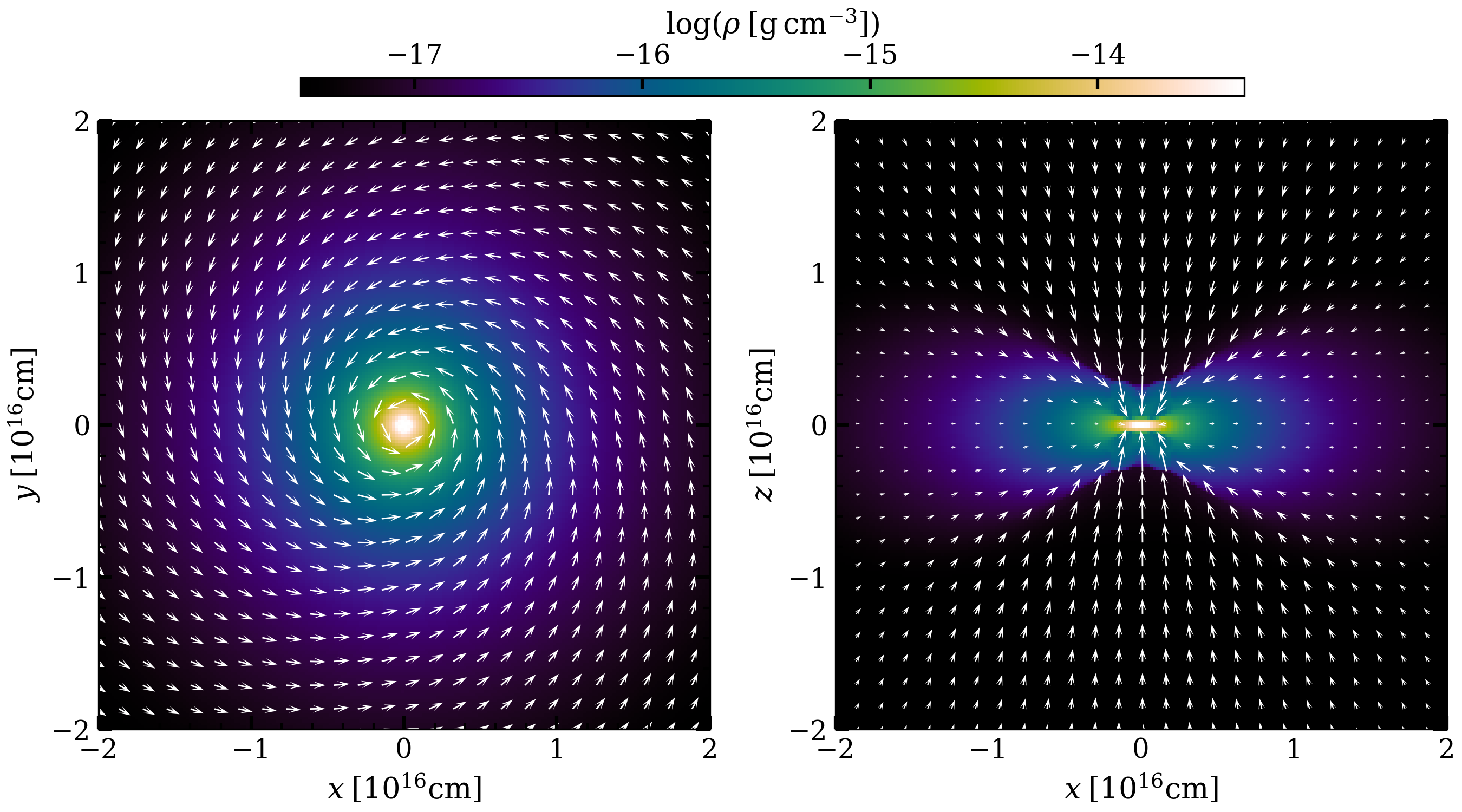}
    \caption{Slice of the density distribution in $x\mbox{-}y$ (left) and $x\mbox{-}z$ (right) plane for the rotating uniform sphere test at $t=1.26\tff$. Arrows in each plot represent the velocity field.}
    \label{fig:rotatingU_slice}
\end{figure*}

\subsubsection{Rotating uniform cloud}
\label{sec:rotatingU}
As the second test of the isothermal collapse problem, we consider the collapse of a rotating uniform cloud originally proposed by \cite{Norman_1980} and later investigated by various authors \citep{Boss_1992,Truelove_1998,Ziegler_2005}. 
This problem is particularly useful to verify the ability to conserve the angular momentum by numerical codes. 
The initial condition comprises of a uniform spherical gas sphere of mass $M=1\,M_\odot$ and radius $R = 7.01\times 10^{16}\,\mathrm{cm}$, giving a density $\rho_0 = 1.26\times10^{-18}\,\mathrm{g\,cm^{-3}}$. 
The cloud has a temperature of $T=5\,\mathrm{K}$ and rotates with an uniform angular velocity $\Omega = 3.04\times 10^{-13}\,\mathrm{rad\,s^{-1}}$. 
This yields an initial ratio of thermal to gravitational energy $\alpha=0.54$ and a ratio of rotational to gravitational energy  $\beta=0.08$. 
The surrounding medium is initialized with a constant density of $\rho_\mathrm{b}=0.01\rho_0$.

\begin{figure}
    \centering
    \includegraphics[width=\linewidth]{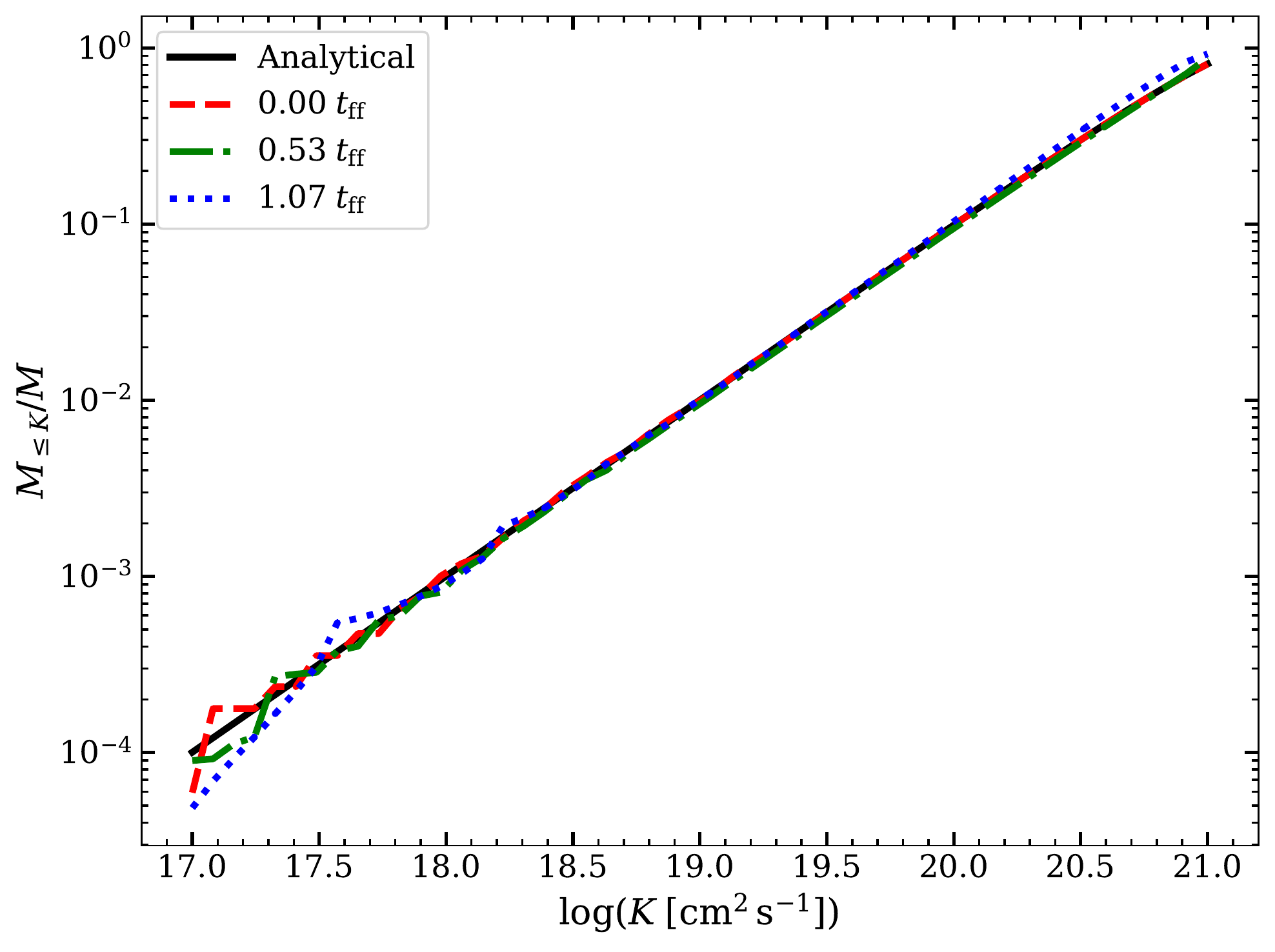}
    \caption{The specific angular momentum ($K$) spectrum at different times for the rotating uniform cloud collapse test. The $y$-axis represents the total mass inside the cloud with specific angular momentum less than or equal to $K$ normalized to the initial total mass of the cloud. The red, green, and blue lines are the spectrum calculated from the simulation at $t= 0,\,0.53.\,\text{and}\,1.07\,\tff$. The black line show the theoretical estimate (Eq.~\ref{eq:SAM}).}
    \label{fig:rotatingU_angular_momentum}
\end{figure}

As shown by \cite{Norman_1980}, poor conservation of angular momentum can produce an artificial ring-like structure instead of a disk in the process of collapse. 
On the other hand, if the conservation is adequate, the sphere forms a disk with an increasingly high central density as it evolves.
In Fig.~\ref{fig:rotatingU_slice}, we show the slice of density distribution in $x\mbox{-}y$ (left) and $x\mbox{-}z$ (right) plane at $t=1.26\,\tff$.
From a first inspection, our results are qualitatively comparable to those of \cite{Truelove_1998,Ziegler_2005}.
We can clearly see the disk has been formed and the velocity field in the $x\mbox{-}z$ plane shows that the collapse is progressing toward a singular state. 
No evidence of a ring structure has been found until the end of the simulation. 
Thus, the angular momentum is conserved quite, well although not explicitly designed to be in our scheme. 
A better method to quantify the conservation of angular momentum is the calculation of the spectrum of specific angular momentum ($K$) at different times as proposed by \cite{Norman_1980}, which is defined by the total mass inside the cloud less than or equal to a given value of $K$. 
For an ideal (inviscid) fluid, this distribution remains conserved and, for an initially rotating uniform cloud, the analytical form of the spectrum is given by \citep{Truelove_1998},
\begin{equation}\label{eq:SAM}
    M_{\leq K} = M\left[1-\left(1-\frac{K}{\Omega R^2}\right)^{3/2}\right],
\end{equation}
where $\Omega$ is the uniform angular velocity of the cloud. 
Ideally, the minimum and maximum extent of $K$ should be $K_{\min}=0$ (corresponds to the rotation axis) and $K_{\max}=\Omega R^2$ (corresponds to the surface of the sphere). 
However, due to finite resolution, any structure below the resolution of the simulation $\Delta x$ can not be resolved, which corresponds to a $K_\mathrm{min} \sim \Omega \Delta x^2 = 10^{18.2}\,\mathrm{cm^2\,s^{-1}}$ for our setup. 
Note that, due to the inability to resolve the density and velocity structures at this small scale, the numerical result may not agree with the analytical solution. 
However, if the angular momentum is conserved, then the shape of the distribution should not change with time irrespective of the initial shape. 
Thus, we compare the distributions at different times in Fig.~\ref{fig:rotatingU_angular_momentum}, where we show the specific angular momentum (SAM) spectrum from our simulation at $t=$ $0$ (red), $t=0.53\,\tff$ (green) and $t=1.07\,\tff$ (blue). 
For reference, the corresponding analytical estimate (Eq.~\ref{eq:SAM}) is also plotted in black. 
We notice the SAM distributions at different times agree satisfactorily at larger values of $K$ within $\lesssim 1\%$ error, except for the distribution at a later stage (blue), where the evolution is believed to be affected by the low-density surrounding medium of the cloud and the average fractional error in the numerical distribution is about $\sim 10\%$.

\begin{figure*}
    \centering
    \includegraphics[width=0.8\linewidth]{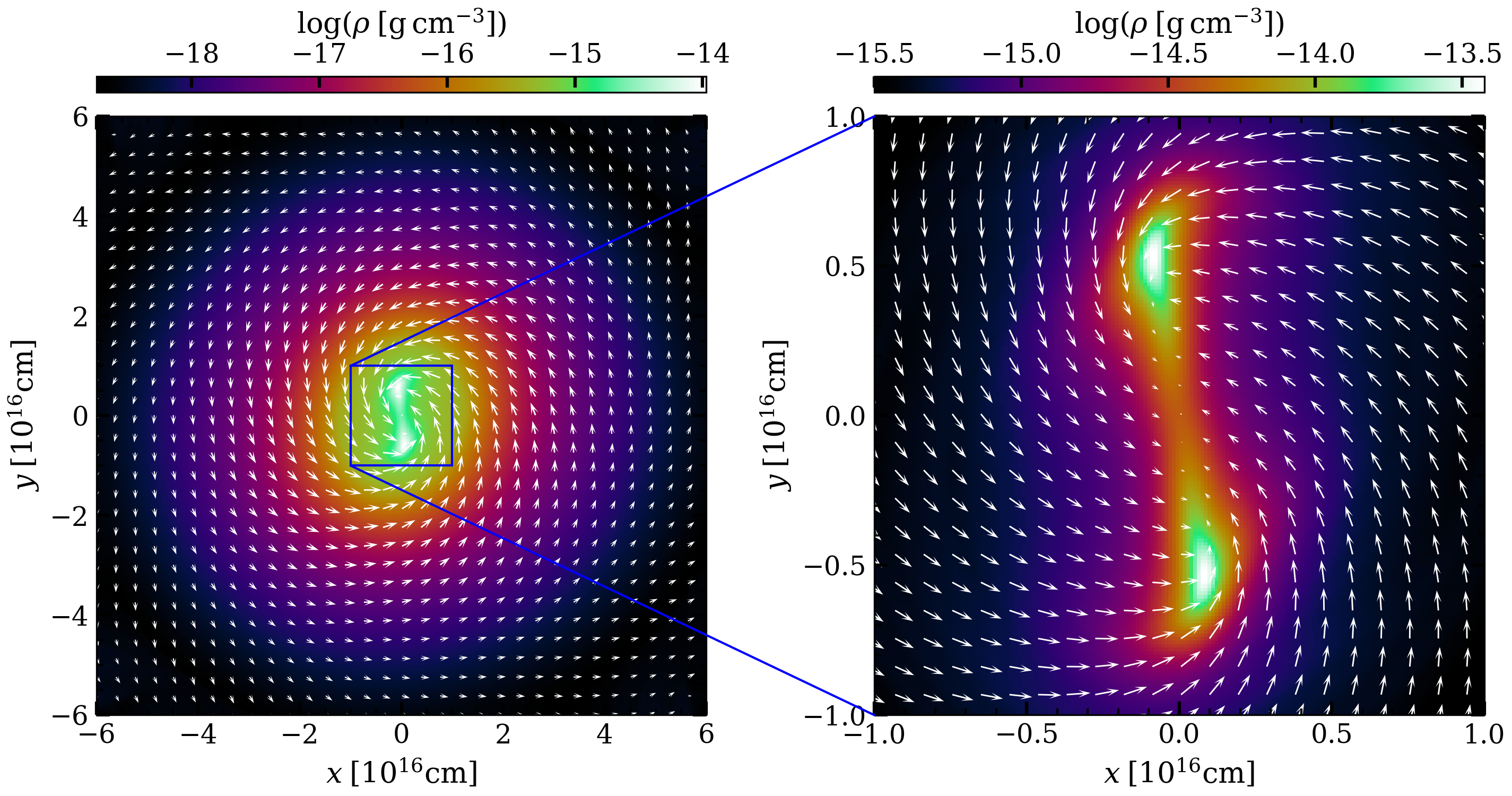}
    \caption{\textit{Left}: Slice of log density distribution in $x-y$ plane for the rotating perturbed cloud problem at $t=1.24\,\tff$ roughly when the two structures become gravitationally bound and undergo gravitational collapse independently. \textit{Right}: Only the central $2\times10^{16}\,\mathrm{cm}$ region of the left panel just to clearly show the structures.}
    \label{fig:rotatingP_slice}
\end{figure*}

\begin{figure}
    \centering
    \includegraphics[width=\linewidth]{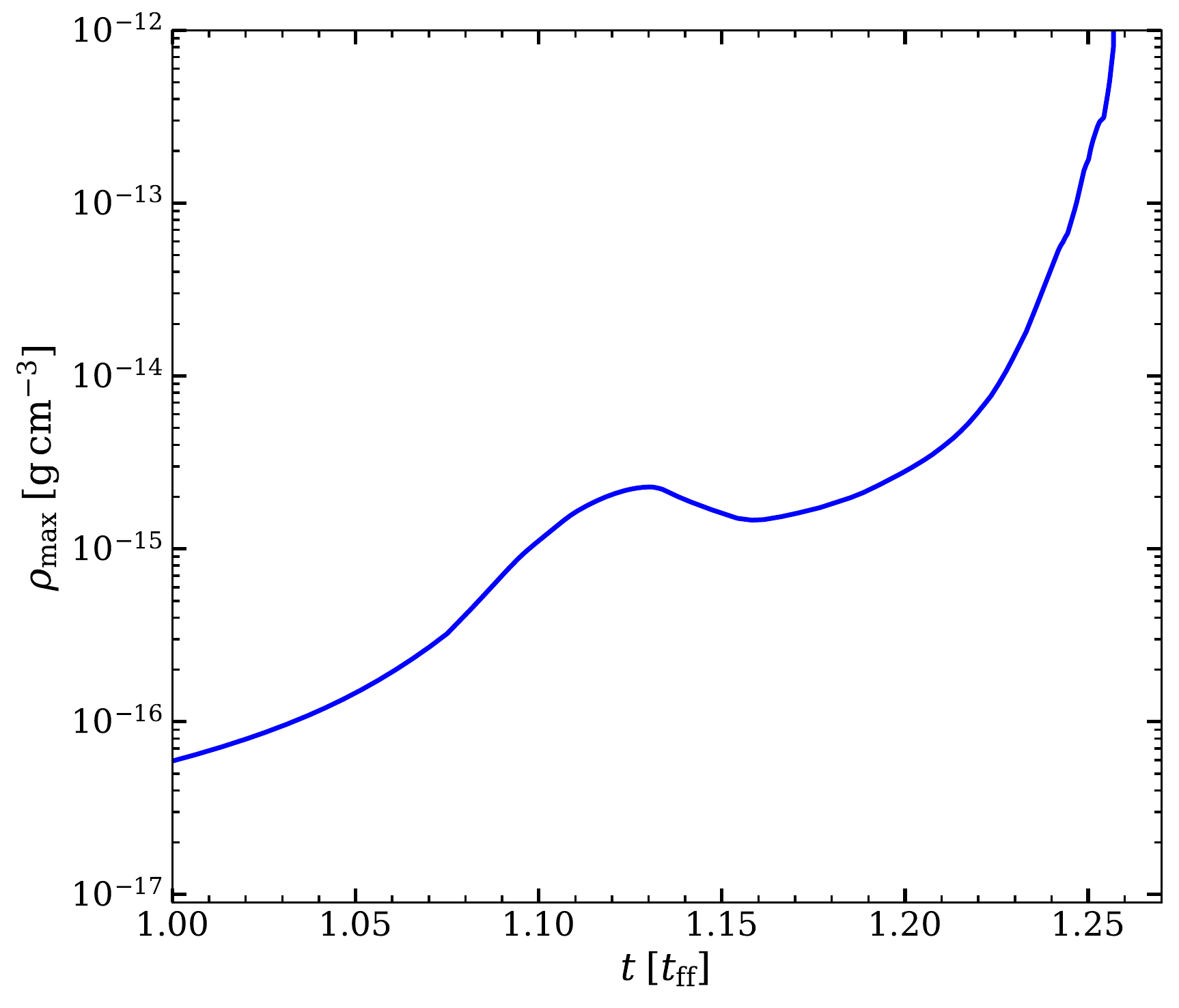}
    \caption{Maximum density in the simulation box as a function of time for the perturbed uniform cloud test.}
    \label{fig:rotatingP_rho_t}
\end{figure}

\subsubsection{Rotating cloud with azimuthal perturbation}
\label{sec:rotatingP}
Lastly, we consider a very demanding problem of fragmentation of an initially perturbed rotating cloud first proposed by \cite{Boss_1979} and employed by several authors \cite[e.g.,][]{Burkert_1993,Truelove_1997,Truelove_1998,Klein_1999,Ziegler_2005,Springel_2005,Hubber_2018,Grudic_2021} in different versions.
The original Boss \& Bonenheimer test consisted of an isothermal uniform cloud subjected to a $50\%$ initial azimuthal $m=2$ mode perturbation in density.
However, \cite{Burkert_1993} introduced the same problem with $10\%$ initial perturbation in density and showed that the evolution and subsequent fragmentation of such a small perturbation are more challenging to simulate and constitute a good benchmark for a numerical code.
This is the version we adopt here, which consists of an isothermal spherical cloud with sound speed $\cs = 0.167\,\kms$ and mass $M=1\,M_\odot$ with a radius of $R=5\times10^{16}\,\mathrm{cm}$.
The cloud rotates with an angular velocity of $\omega = 7.4\times10^{-13}\,\mathrm{rad\,s^{-1}}$, which gives the energy ratios of $\alpha=0.26$ and $\beta=0.16$. 
The uniform density of the cloud is modified by an azimuthal $m=2$ mode perturbation in the following way
\begin{equation}\label{eq:source_RotatingP}
    \rho = 
    \begin{cases}
        \rho_0\left[1+0.1\cos{(2\phi)}\right], & \text{if $r\leq R$}\\
        0, & \text{if $r>R$}
    \end{cases}
\end{equation}
where $\phi$ is the azimuthal angle around the rotation axis and $\rho_0 = 3.82\times10^{-18}\,\mathrm{g\,cm^{-3}}$ is the underlying constant density. 
The density of the medium outside the sphere is set to be $0.01\rho_0$.

Fig.~\ref{fig:rotatingP_slice} illustrates the equatorial slice of the density distribution (left panel) and a close-up of the central  structure (right panel) at $t=1.24\,\tff$. 
An elongated bar-like structure has formed connecting two high-density regions. 
The velocity field demonstrates that there are converging gas flows toward the center of mass of each structure as a result of gravitational collapse. 
This can be seen in Fig.~\ref{fig:rotatingP_rho_t}, where we show the maximum density inside the simulation box as a function of evolution time. 
Similar behavior has also been reported by \cite{Bate_1997,Springel_2005,Grudic_2021}. 
However, we find that in our case the collapse proceeds at a somewhat slower rate when compared to \cite{Springel_2005,Grudic_2021} owing to limited resolution near the collapsing regions. 
The low-resolution runs of the same test problem in \cite{Springel_2005} also exhibit similar behavior.

It is worth noting that the problem at hand exhibits a significant range of dynamics. 
This necessitates the use of high resolutions in the vicinity of the collapsing structure to witness the fragmentation collapse into a singular state, as illustrated in the works of \cite{Truelove_1998,Ziegler_2005}. 
Here, the employment of a uniform Cartesian grid prevents the computations from reproducing the elongation and subsequent gravitational collapse to a singular state.
From this perspective our results are fairly consistent with those of \cite{Bate_1997,Springel_2005} and adequately reproduce the major features of the test problem despite the limited resolution.


\section{Summary}
\label{sec:summary}
We have presented a novel implementation of an iterative method for solving the Poisson equation based on the \textit{Runge-Kutta-Legendre} time-stepping scheme developed for solving parabolic PDE. 
Our solver utilizes a relaxation scheme to achieve convergence on the solution of an elliptic problem sought as the steady-state solution of a parabolic equation, whose time-discretization is governed by the RKL scheme.
The algorithm is part of a V-cycle multigrid method and it has been implemented as a simple and efficient gravity solver in the \textsc{Pluto} code, to address self-gravitational hydrodynamics.
We demonstrate that our approach offers a potential substitute to conventional iterative techniques like Gauss-Seidel or SOR, given its significantly superior convergence rate when compared to the GS scheme, while also removing the uncertainties connected to the optimal over-relaxation parameter in the SOR method. 
Besides, being a fully parallel algorithm, our method incurs lower communication overhead between processors compared to GS or SOR, which are half parallel (see \ref{sec:performance_test} for the performance of these solvers for a model problem).

We have employed our methodology to perform a comprehensive set of astrophysical test problems, designed to assess different aspects of the code. 
In a 3D static problem, we have verified that the solver achieves $2^{\rm nd}$-order accuracy for both the potential and force, as expected. 
Our calculations of the timescales for collapsing and oscillating modes of the Jeans instability test agree well with the theoretical expectations, with errors of less than $1\%$. 
The shock structures observed in the density, velocity, and entropy functions of the adiabatic collapse problem match closely with those obtained from a highly resolved 1D PPM calculation, which serves as a reference solution. 
Moreover, our code conserves the total energy up to $1\%$, for the adiabatic collapse problem.
We have tracked the collapse of a uniform non-rotating sphere and observed that the central density increases by $\sim 5$ orders of magnitudes and the corresponding timescale matches very well with the analytical prediction. 
In the collapse of a rotating uniform sphere, which encompasses a density increase of $6$ decades, we have achieved angular momentum conservation up to $1\%$. However, at the later stage, the conservation degrades ($\sim 10\%$) due to the effect from the surrounding medium.
Lastly, we have reported the fragmentation and subsequent formation of binary structures of an initially perturbed rotating sphere where the density increases by more than $6$ orders of magnitude, and the results favorably compare to those obtained by previous investigators.
Although the test problems considered here can benefit even more from the employment of adaptive mesh refinement (AMR), we have demonstrated that our gravity module can reproduce the major features of the results using a uniform static grid already at moderate resolution.

Future extensions to AMR, require changes to the cycling strategy in the multigrid algorithm and to the treatment of the boundary layers for each multigrid level, which are well developed by many authors \citep{Ricker_2008,Guillet_2011,Wang_2020}. 
Smoothing, restriction, and prolongation operators as well as the coarsest grid solver should not be affected by the introduction of AMR.



We like to thank Christoph Federrath for the useful discussion about boundary conditions. 
Ankush Mandal thanks Moun Meenakshi for her valuable input and constant support throughout the project. We gratefully acknowledge the use of the high-performance computing facilities of IUCAA, Pune. Parts of the results in this work make use of the colormaps in the CMasher package.

%

\vspace{5mm}
\facilities{Pegasus, the high performance computing facility at IUCAA\footnote{\url{http://hpc.iucaa.in}}.}


\software{PLUTO\footnote{\url{http://plutocode.ph.unito.it}} \citep{Mignone_2007}, NumPy \citep{Harris_2020}, Matplotlib \citep{Matplotlib_2007}, CMasher \citep{Cmasher_2020}}



\appendix

\section{Convergence property of the RKL-based Poisson solver}
\label{sec:RKL_convergence}
The \textit{Runge-Kutta-Legendre} (RKL hereafter) method is a class of super-time-stepping methods (STS) that have been developed to deal with the restrictive time-step for an explicit scheme for parabolic equations \citep{Meyer_2012,Meyer_2014}. This allows us to consider a much longer time-step ($\tau$) than the maximum permissible explicit parabolic time-step ($\dtpar$) without affecting the solution. The parabolic diffusion equation can be written in the following form,
\begin{equation}\label{eq:diffusion}
    \frac{d u}{dt} = \M u(t),
\end{equation}
where $\M$ is a symmetric, constant coefficient matrix resulting from the space discretization of the parabolic operator. 
The STS methods are a class of stabilized \textit{Runge-Kutta} methods where one time-step ($\tau$) consists of additional sub-steps ($s$) to ensure stability. The solution after one $\tau$ step can be expressed in terms of the amplification factor $R_s$ as
\begin{equation}\label{eq:tau_step}
    u(t+\tau) = R_s(\tau\M)u(t).
\end{equation}
The stability of the solution is guaranteed is $|R_s(\tau\lambda)|\leq 1$ for all $\lambda$, where $\lambda$ is the eigenvalues of $\M$. Now the amplification factor $R_s(\tau\lambda)$, also known as the stability polynomial for a $s$-stage STS scheme can be defined as,
\begin{equation}\label{eq:stability_poly}
    R_s(\tau\lambda) = \prod_{i=1}^s (1+\lambda\Delta t_i),
\end{equation}
with $\sum_{i=1}^s \Delta t_i=\tau$. Matching the coefficients in Eq.~\eqref{eq:stability_poly} with the analytic expansion of Eq.~\eqref{eq:diffusion}, as given by
\begin{equation}\label{eq:analtic_expansion}
    u(t+\tau)=e^{\tau\M}u(t)\approx\left[1 + (\tau\M) +\frac{1}{2}(\tau\M)^2 + \cdot\cdot\cdot \right]u(t)
\end{equation}
gives the temporal accuracy of the solution.

In the RKL methods, the shifted Legendre polynomials are used as the stability polynomials. For a general $s$-stage RKL scheme, the stability polynomials are chosen as,
\begin{equation}\label{eq:RKL stability poly}
    R_s(\tau\lambda) = a_s +b_s P_s(w_0 + w_1(\tau\lambda)),
\end{equation}
where $w_0=1$ for all RKL methods. For first-order RKL scheme, $a_s=0$, $b_s=1$ and $w_1=2/(s^2+s)$. In the rest of the derivation, we consider only the first-order RKL technique. Thus, for simplicity, whenever ``RKL" is mentioned it implies the first-order RKL method. 
Therefore, the stability polynomial of a $s$-stage RKL scheme takes the following form,
\begin{equation}\label{RKL1 stability poly}
    R_s(\tau\lambda)=P_s\left(1+\frac{2}{s^2+s}\tau\lambda\right).
\end{equation}
Thus, the $j^\mathrm{th}$ stage of an $s$-stage RKL scheme can be in terms of the corresponding stability polynomial as,
\begin{equation}
    \begin{aligned}
        & u_j = P_j\left(1+\frac{2}{s^2+s}\tau\lambda\right)u^{(n)},\\
        & u^{(n+1)} = u_s = P_s\left(1+\frac{2}{s^2+s}\tau\lambda\right)u^{(n)},
    \end{aligned}
\end{equation}
where $u^{(n)}$ represents the solution at the beginning of the $\tau$-step, $u_j$ denotes the solution at the $j^\mathrm{th}$ intermediate step, and $u^{(n+1)}$ corresponds to the solution after advancing one $\tau$-step. Throughout the text, we use subscripts to represent the intermediate steps within one $\tau$-step and the superscript refers to each $\tau$-step.

In order to calculate the stability polynomials, we can use the three-point recursion relation of the Legendre polynomials:
\begin{equation}\label{eq:LP_recursion}
    P_j(x)=\frac{2j-1}{j}xP_{j-1}(x)-\frac{j-1}{j}P_{j-2}(x).
\end{equation}
Replacing $x=1+2\tau\lambda/(s^2+s)$, we can easily show that the stability polynomials of the RKL methods obey the following recursion relation
\begin{equation}\label{eq:RKL_recursion}
    P_j\left(1+\frac{2}{s^2+s}\tau\lambda\right)=\frac{2j-1}{j}\left(1+\frac{2}{s^2+s}\tau\lambda\right)P_{j-1}\left(1+\frac{2}{s^2+s}\tau\lambda\right)-\frac{j-1}{j}P_{j-2}\left(1+\frac{2}{s^2+s}\tau\lambda\right).
\end{equation}
Using Eq.~\eqref{eq:RKL_recursion}, one can write the $s$-stage RKL scheme for $\M$ as follows,
\begin{equation}\label{eq:RKL_intermediate}
    \begin{aligned}
        & Y_0 = u^{(n)} \\
        & Y_1 = Y_0 + \Tilde{\mu}_1\tau\M Y_0 \\
        & Y_j = \mu_j Y_{j-1} + \nu_j Y_{j-2} + \Tilde{\mu}_j\tau\M Y_{j-1};  \hspace{0.5cm} 2\leq j  \leq s\\
        & u^{(n+1)} = Y_s
    \end{aligned}
\end{equation}
where the coefficients $\mu_j, \nu_j$ and $\Tilde{\mu_j}$ can be obtained from the recursion relation (Eq.~\ref{eq:RKL_recursion}) and given by,
\begin{equation}
    \begin{aligned}
        & \mu_j = \frac{2j-1}{j} \\
        & \nu_j = \frac{1-j}{j}  \\
        & \Tilde{\mu}_j = \mu_j w_1 = \frac{2j-1}{j}\frac{2}{s^2+s}
    \end{aligned}
\end{equation}
The stability condition of Legendre polynomials provides the maximum permissible time-step $\tau$ for an $s$-stage RKL method. Legendre polynomials are bounded within the range $(-1,1)$, and therefore, to ensure solution stability, the condition $(1-w_1\tau|\lambda_{\max})|\geq -1$ must hold true. Here, $\lambda_{\max}$ represents the maximum (negative) eigenvalue of the operator $\M$. This condition determines the maximum value of $\tau$ applicable to an $s$-stage RKL scheme, yielding,
\begin{equation}
    \tau_{\max} = \frac{2}{w_1|\lambda_{\max}|} = \dtpar \frac{s^2+s}{2}
\end{equation}
where $\dtpar=2/|\lambda_{\max}|$ is the explicit parabolic time step.

\subsection{Poisson equation as a limiting case of the diffusion equation}
A simple extension of the diffusion equation (Eq.~\ref{eq:diffusion}) can be made to accommodate a time-independent source term ($\f$) on the right-hand side:
\begin{equation}\label{eq:diffusion_with_source}
    \frac{d u}{d t} = \M u(t) -\f
\end{equation}
The initial solution $u$ reaches an equilibrium as $t\rightarrow\infty$ and $du/dt\rightarrow 0$, which essentially becomes the solution of the Poisson equation, i.e.,
\begin{equation}\label{eq:Poisson}
    \M u-\f=0     
\end{equation}
Thus, successively applying one $\tau$-step to the previous step of Eq.~\eqref{eq:diffusion_with_source} until $t\rightarrow\infty$, gives the solution of the Poisson equation for a given source term $\f$.

To fit the source term ($\f$) in the intermediate stages of an $s$-stage RKL, we replace $\M Y_j\rightarrow \M Y_j -\f$ in Eq.~\eqref{eq:RKL_intermediate}, which takes the following form,
\begin{equation}\label{eq:RKL_intermediate_source}
    \begin{aligned}
        & Y_0 = u^{(n)} \\
        & Y_1 = Y_0 + \Tilde{\mu}_1\tau\M Y_0 -\Tilde{\mu}_1\tau\f \\
        & Y_j = \mu_j Y_{j-1} +\nu_j Y_{j-2} +\Tilde{\mu}_j\tau\M Y_{j-1} -\Tilde{\mu}_j\tau\f; \hspace{0.5cm} 2\leq j \leq s \\
        & u^{(n+1)} = Y_s
    \end{aligned}
\end{equation}
Using Eq.~\eqref{eq:RKL_intermediate_source}, it can be shown that the solution of Eq.~\eqref{eq:diffusion_with_source} after one $\tau$ step in an $s$-stage RKL scheme can be written as,
\begin{equation}\label{eq:Poisson_iteration}
    u^{(n+1)} = P_s(1+w_1\tau\M)u^{(n)} - \F_{(s-1)}(w_1\tau\M)w_1\tau\f,
\end{equation}
where $\F$ is a polynomial of degree $(s-1)$.
Therefore, Eq.~\eqref{eq:Poisson_iteration} can be thought of one iteration from $u^{(n)}$ to $u^{(n+1)}$ with the RKL iteration matrix
\begin{equation}\label{eq:Iteration_matrix}
    \R = P_s(1+w_1\tau\M).
\end{equation}
Thus, starting from a guess solution $u^{(0)}$, we can successively apply Eq.~\eqref{eq:Poisson_iteration} until $m$-steps for which $|u^{(m)}-u^{(m-1)}|<10^{-d}$, where $10^{-d}$ is some specified tolerance level.

\subsection{Convergence criteria}\label{sec:convergence}
Consider the initial guess solution of Eq.~\eqref{eq:Poisson_iteration} is $u^{(0)}$, and we apply one iteration step. Then, the solution after one $\tau$ step is,
\begin{equation}\label{eq:approx}
    u^{(1)}=\R u^{(0)} - \F_{(s-1)}(w_1\tau\M)w_1\tau\f
\end{equation}
Now, if $v$ is the exact solution of Eq.~\eqref{eq:Poisson}, then the iteration steps do not change the solution, i.e.,
\begin{equation}\label{eq:exact}
    v = \R v  - \F_{(s-1)}(w_1\tau\M)w_1\tau\f
\end{equation}
Subtracting Eq.~\eqref{eq:approx} from Eq.~\eqref{eq:exact}, we get
\begin{equation}
    v-u^{(1)} = \R (v - u^{(0)}) \Rightarrow e^{(1)} = \R e^{(0)},
\end{equation}
where $e^{(j)} = v-u^{(j)}$ is the error in the approximated solution at $j^\mathrm{th}$ step. Thus, it is possible to reduce the error from the previous step, if and only if
\begin{equation}\label{eq:convergence_cond_norm}
    \norm{\R}_p<1; \hspace{0.5cm} \forall p\in(1,\infty)
\end{equation}
where $\norm{\cdot}_p$ refers to the $p$-norm of the matrix. Now, if we consider $p=2$ (the Euclidean norm), then for the symmetric matrix $\R$, we have
\begin{equation}\label{eq:spectral_radius}
    \norm{\R}_2 = \sqrt{\rho ((\R)^T\R)} = \sqrt{\rho(\R^2)} = \rho(\R),
\end{equation}
where $\rho(\R)$ is the spectral radius of a matrix $\R$ which is defined as,
\begin{equation}
    \rho(\R) = \max|\lambda(\R)|
\end{equation}
where $\lambda(\R)$ are the eigenvalues of $\R$. 
Thus, from Eq.~\eqref{eq:convergence_cond_norm} and Eq.~\eqref{eq:spectral_radius}, we find that the convergence to the exact solution for any initial guess solution is possible through successive iterations, if and only if
\begin{equation}
    \rho(\R)<1
\end{equation}
So, after $m$ iteration, the error in the $m^\mathrm{th}$ approximation is given by,
\begin{equation}\label{eq:error}
    e^{(m)} = \R^m e^{(0)}
\end{equation}
Therefore, if we want to reduce the initial error by an order of $d$, we must need,
\begin{equation}\label{eq:minimum_iteration}
    \frac{\norm{e^{(m)}}_2}{\norm{e^{(0)}}_2}=\norm{\R}_2^m \leq 10^{-d} \Rightarrow \left[\rho(\R)\right]^m \leq 10^{-d}
\end{equation}
The minimum number of iterations ($m$) needed to reduce the initial error by $d$ order of magnitude in an $s$-stage RKL method is given by,
\begin{equation}\label{eq:iteration}
    m \geq - \frac{d}{\log_{10}\left[\rho(\R)\right]} = \frac{d}{\log_{10}\left[\rho(P_s(1+w_1\tau\M))\right]}
\end{equation}
%
\begin{figure}
    \centering
    \includegraphics[scale=0.44]{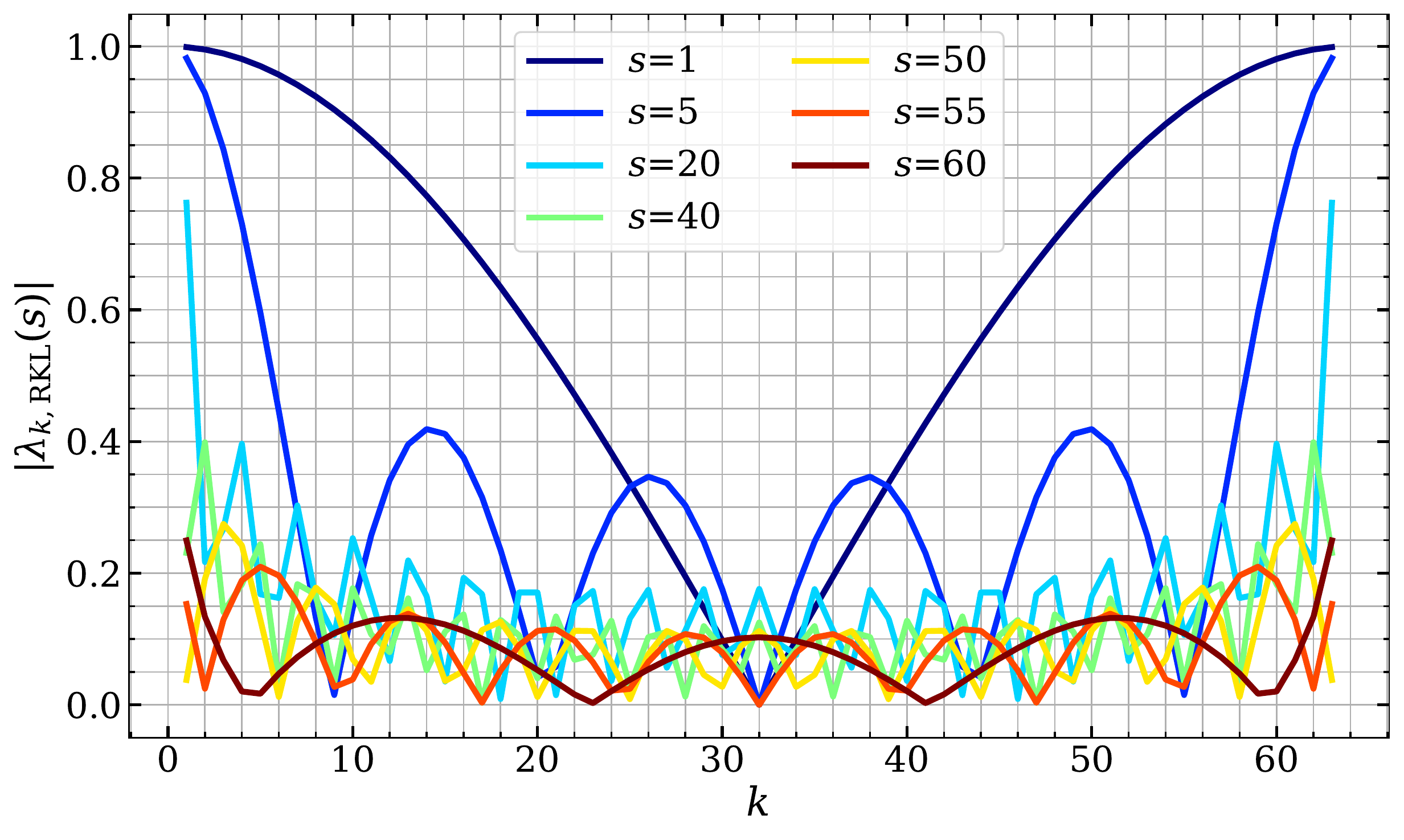}
    \caption{The absolute of the eigenvalues ($\lambdaRKL$) of the RKL iteration matrix ($\R$) as a function of $k$ for different values of $s$. }
    \label{fig:eigen_values}
\end{figure}

\subsection{Spectral analysis}
In order to find the spectral radius of the RKL iteration matrix ($\R$), we have to find how the eigenvalues ($\lambdaRKL$) behave. From Eq.~\eqref{eq:Iteration_matrix}, we note that $\R$ is a polynomial of $\M$. Thus, if $\lambdaM$ is the $k^\mathrm{th}$ eigenvalue of $\M$, then the eigenvalues of $\R$ is given by,
\begin{equation}
    \lambdaRKL = P_s(1+w_1\tau\lambdaM)    
\end{equation}

We first consider a 1-D problem with grid points at $j=0,1,...N$ with a grid spacing of $\Delta x$. The $\M$ matrix for this problem is given by,
\begin{equation}\label{eq:M_matrix}
    \M = \frac{1}{(\Delta x)^2}
    \begin{bmatrix}
        -2 &  1 & & & \\
        1  & -2 & 1 & & & \\
           & \cdot & \cdot & \cdot & &\\
           & & \cdot & \cdot & \cdot & \\
           & & & 1 & -2 & 1 \\
           & & & & 1 & -2
    \end{bmatrix}
    _{(N-1)\times (N-1)}
\end{equation}
The eigenvalues of $\M$ are,
\begin{equation}\label{eq:eigenvalue_M}
    \lambdaM = -\frac{4}{(\Delta x)^2}\sin^2\left(\frac{k\pi}{2N}\right), \hspace{0.5cm} k\in \mathbb{Z}^+ \,\, \mathrm{and}\,\, 1\leq k \leq N-1
\end{equation}
Also, it can be easily shown that the eigenvectors of $\M$ are given by,
\begin{equation}\label{eq:eigenvector_M}
    \boldsymbol{\Lambda}_{k,j,\rm M}=\sin\left(\frac{jk\pi}{N}\right), \hspace{0.5cm} 1\leq k\leq N-1, \hspace{0.5cm} 0\leq j \leq N
\end{equation}
where $\boldsymbol{\Lambda}_{k,j,M}$ is the $j^\mathrm{th}$ component of $k^\mathrm{th}$ eigenvector of the $\M$ operator. Interestingly, the eigenvectors of $\M$ are simply the Fourier modes in 1-D.

Now, assuming $\tau = \tau_{\max}=2/(w_1 \max\{\lambdaM\})=(\Delta x)^2/2w_1$ (where $\max\{\lambdaM\}=4/(\Delta x)^2$), the eigenvalues of $\R$ are given by,
\begin{equation}\label{eq:eigenvalue_RKL}
    \lambdaRKL = P_s\left(1-2\sin^2\left[\frac{k\pi}{2N}\right]\right) = P_s\left(\cos\left[\frac{k\pi}{N}\right]\right)
\end{equation}
From the properties of the Legendre polynomials, we know that $P_s(x)$ are symmetric and anti-symmetric functions when $s$ is even or odd. Thus, $\lambdaRKL$ have the following property,
\begin{equation}\label{eq:RKL_symmetry}
    \lambdaRKL = (-1)^s\lambda_{N-k,\mathrm{RKL}}\; \mathrm{or}\; |\lambdaRKL| = |\lambda_{N-k,\mathrm{RKL}}(s)|
\end{equation}
Therefore, for calculating the spectral radius for $\R$, we can only consider $1\leq k \leq N/2$ eigen-modes and the spectral radius of $\R$ can be expressed as,
\begin{equation}\label{eq:spectral_rad_RKL}
    \rho(\R) = \max_{1\leq k\leq N/2}\left\{ \left |P_s\left(\cos\left[\frac{k\pi}{N}\right]\right) \right|\right\}
\end{equation}
In Fig.~\ref{fig:eigen_values}, the absolute values of the eigenvalues of $\R$, denoted by $\lambdaRKL$, are plotted as a function of $k$ for various values of $s$. It can be observed that for small values of $s$, the maximum eigenvalue (which is essentially the spectral radius) of $\R$ is associated with the longest wavelength mode or the $k=1$ mode. However, as $s$ increases beyond a certain threshold, the maximum eigenvalue shifts to the first few $k$-modes, and this phenomenon repeats itself due to the highly oscillatory nature of the Legendre polynomials at large $s$. Therefore, it can be inferred that the first few $k$-modes determine the maximum eigenvalue of the $\R$ operator for different values of $s$.

\begin{figure}
\centerline{
\def\arraystretch{1.0}
\setlength{\tabcolsep}{0.0pt}
\begin{tabular}{lcr}
  \includegraphics[width=0.495\linewidth]{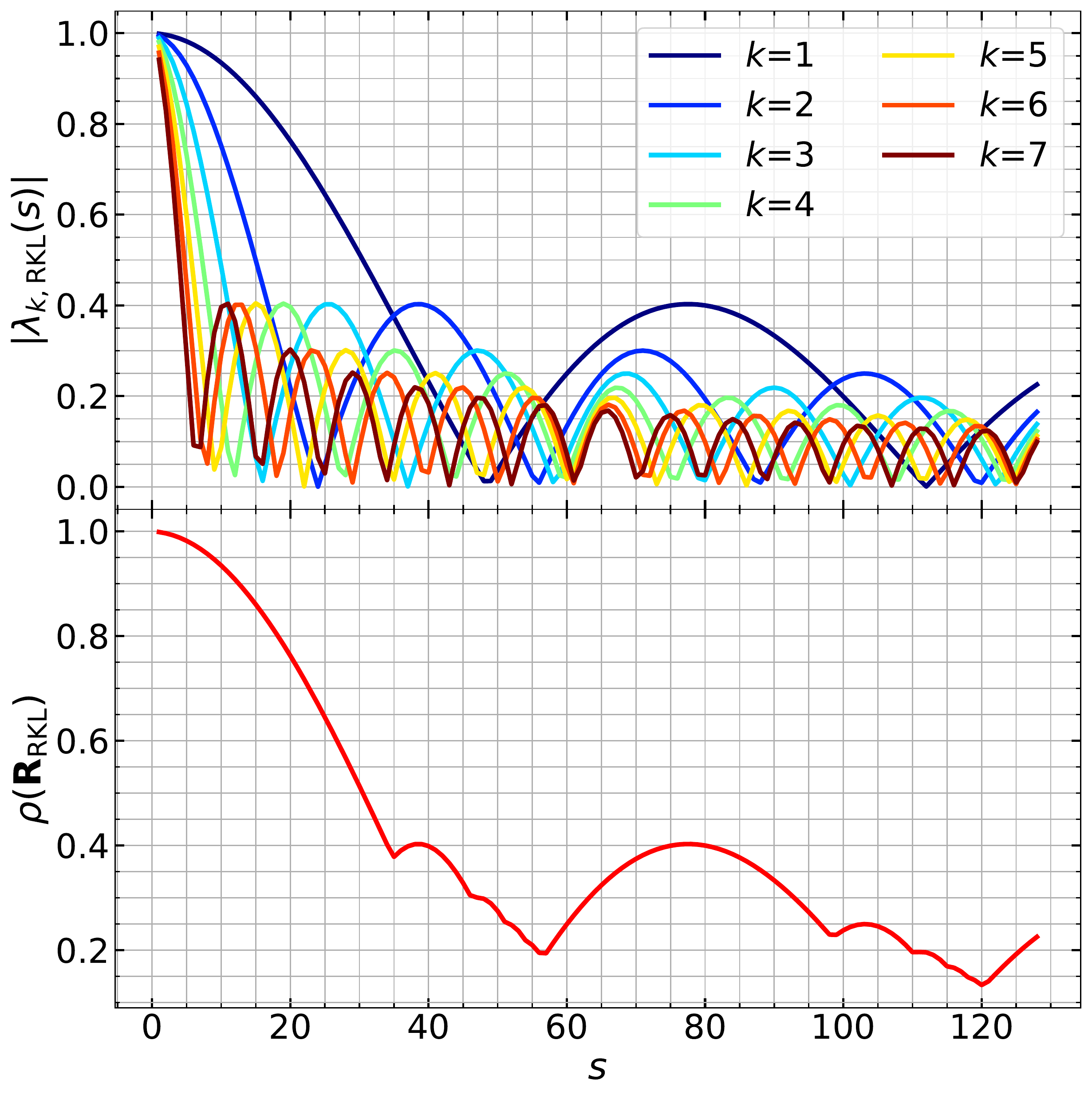} &
  \includegraphics[width=0.5\linewidth]{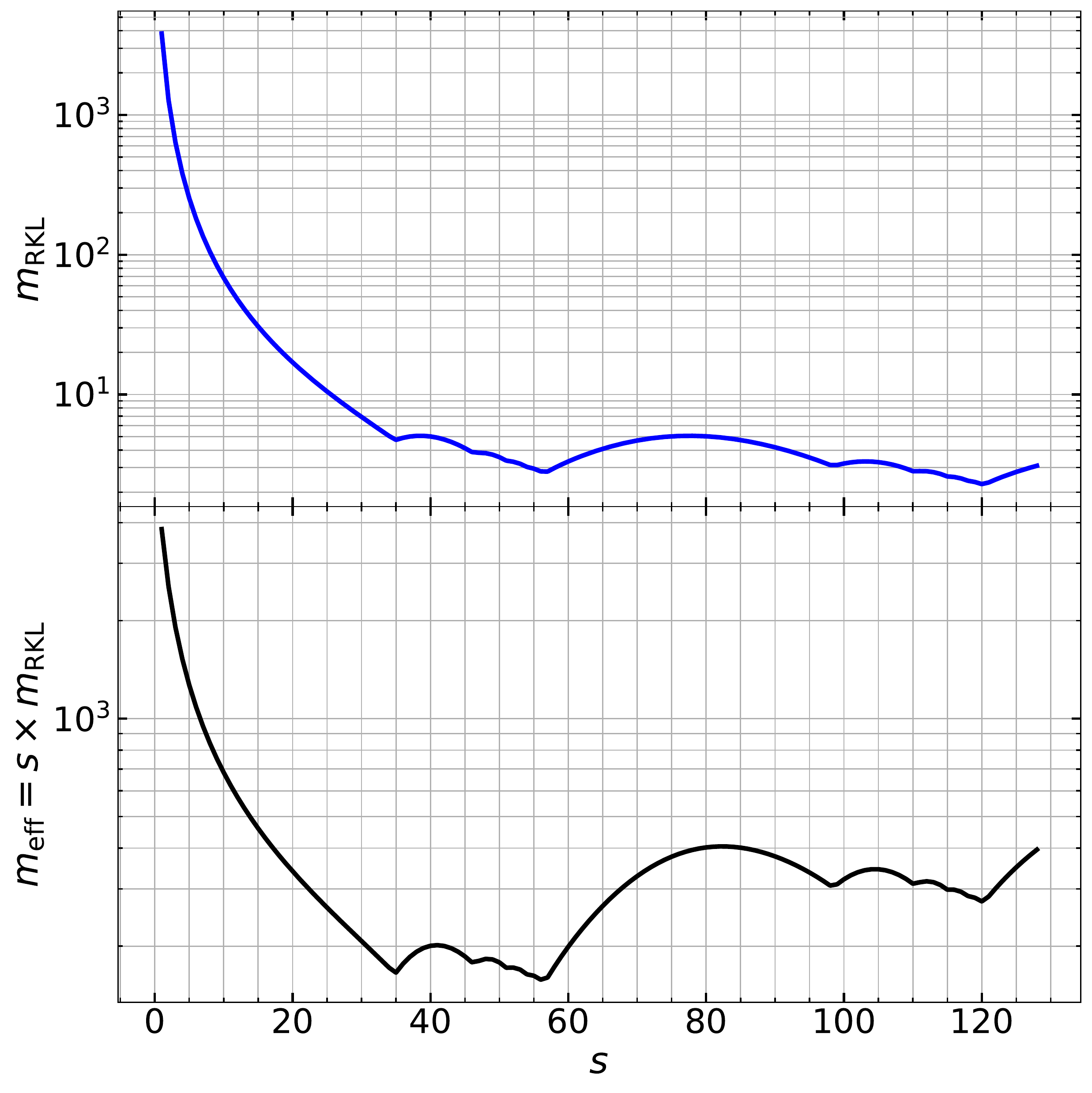}
\end{tabular}}
  \caption{\textit{Top left:} $|\lambdaRKL|$ as a function of $s$ for first seven eigen-modes. \textit{Bottom left}: The spectral radius ($\rho(\R)$) of the $\R$ operator. \textit{Top right}: Minimum number of $\tau$ steps required to reduce the initial error by an order of two in an $s$-stage RKL scheme as a function of $s$. \textit{Bottom right}: Total number of single loop iterations ($m_\mathrm{eff}=s\times m_\mathrm{RKL}$) required for the same convergence as a function of $s$.}
  \label{fig:spectral_iteration}
\end{figure}

In the top-left panel of Fig.\ref{fig:spectral_iteration}, we plot the variation of $\lambdaRKL$ as a function of $s$ for the first seven eigen-modes of $\R$. It is apparent that the maximum eigenvalue is not solely determined by a single eigen-mode. Rather, different eigen-modes dominate at different values of $s$, and this behavior oscillates. Additionally, we find that the first six modes, i.e., $1\leq k \leq 6$, are sufficient for calculating the maximum eigenvalue for 1-D problems. Therefore, the spectral radius does not exhibit a monotonically decreasing trend with increasing $s$, as evidenced by the bottom-left panel of Fig.\ref{fig:spectral_iteration}. Nevertheless, $\rho(\R)$ approaches zero asymptotically as $s\rightarrow\infty$.

In order to reduce the initial error to any value, a single $s$-stage RKL iteration is required by selecting a suitably high value that results in a spectral radius close to zero. The top-right panel of Fig.~\ref{fig:spectral_iteration} demonstrates this, displaying the minimum number of $\tau$ steps, $m_\mathrm{RKL}$ (Eq.~\ref{eq:minimum_iteration}), needed to decrease the initial error by a factor of two. It is evident that $m_\mathrm{RKL}$ decreases with an increase in $s$ and exhibits an oscillatory behavior similar to the spectral radius, which approaches 1 as $s\rightarrow\infty$. While it may seem that increasing $s$ indefinitely would result in faster convergence, it is important to note that each $\tau$ step in an $s$-stage RKL iteration consists of $s$ sub-stages. As a result, if a problem achieves a certain tolerance level in $m_\mathrm{RKL}$ $\tau$-steps, it actually requires $m_\mathrm{eff}=s\times m_\mathrm{RKL}$ single-loop iterations. Therefore, if the rate of decrease in $m_\mathrm{RKL}$ is slower than the rate of increase in $s$, then after a certain value of $s$, $m_\mathrm{eff}$ begins to increase as $s$ increases. As a result, there exists an optimal value of $s$ ($s_\mathrm{opt}$) at which the RKL iteration converges the fastest. This is depicted in the bottom-right panel of Fig.~\ref{fig:spectral_iteration}, where it is clear that $m_\mathrm{eff}$ initially decreases with increasing $s$, but begins to oscillate after $s=35$. The value of $m_\mathrm{eff}$ reaches its global minimum value at $s=56$.

Ideally, the theoretical value of $s_\mathrm{opt}$ can be found by the following equation,
\begin{equation}\label{eq:sopt_theory}
    \begin{aligned}
         \frac{d m_\mathrm{eff}}{ds}\Big|_{s=s_\mathrm{opt}} = 0 \\
         \frac{d}{ds} \left( \frac{s\times d}{\log_{10}[\rho(\R)]} \right)\Big|_{s=s_\mathrm{opt}} = 0.
    \end{aligned}
\end{equation}
Finding an analytical expression for $s_\mathrm{opt}$ from Eq.\eqref{eq:sopt_theory} is highly challenging in the absence of a closed-form expression for $\rho(\R)$. Nevertheless, $s_\mathrm{opt}$ can be computed numerically for a given problem prior to commencing the iteration process. As can be seen from the bottom panels of Fig.\ref{fig:spectral_iteration}, the optimal value of $s$ is where the value of $\rho(\R)$ is at its minimum within the range $1 \leq s \leq N$. Additionally, the top-left panel of Fig.~\ref{fig:spectral_iteration} demonstrates that the first six $k$-modes determine the spectral radius of the $\R$ operator. Thus, we find the value of $s_\mathrm{opt}$ in the following two steps:
\begin{enumerate}
    \item First we compute the spectral radius $\rho(\R)$ as a function of $s$ in the range $1\leq s \leq N$ using Eq.~\eqref{eq:spectral_rad_RKL} where $k$ is restricted in $1\leq k\leq 6$.
    
    \item Then we find for what value $s$ the above computed spectral radius has the minimum value. This will gives the $s_\mathrm{opt}$
\end{enumerate}
The process described above is a general approach for determining the value of $s_\mathrm{opt}$ for a 1-D problem. However, it can be further optimized by limiting the range of $s$ used in the computation of the spectral radius in Step~1. As can be observed from the top-left panel of Fig.\ref{fig:spectral_iteration}, the first few values of $s$ are redundant. We know that $s_\mathrm{opt}$ lies between the first zero of $\lambda_{1,\mathrm{RKL}}(s)$ and $N$. Thus, restricting the $s$-range to (first zero of $\lambda_{1,\mathrm{RKL}}(s)$ $\leq s \leq N$) will reduce the number of operations required to calculate the value of $s_\mathrm{opt}$. The procedure for identifying the first zero of $\lambdaRKL$ will be outlined in Sec.\ref{sec:acc_RKL}.  


\subsection{An accelerated RKL scheme}\label{sec:acc_RKL}
From the top-left panel of Fig.~\ref{fig:spectral_iteration}, we notice that each Fourier mode or $k$-mode of the error has a unique optimal value of $s$ denoted by $\skopt$. At $\skopt$, the $k^\mathrm{th}$ eigenvalue or $\lambdaRKL$ has a value near zero, and this occurs at the first zero of $\lambdaRKL$. Therefore, instead of selecting a universal $s$ for the entire spectrum, we can choose $\skopt$ for each $k$-mode and decrease the corresponding error individually. The approach is to commence the $\tau$ iteration with $s=\skopt$, where we choose $k=N/2$ at the initial iteration. Subsequently, after each $\tau$-step, decrease $k$ by one, i.e., $k=k_\mathrm{prev}-1$, and use the corresponding $\skopt$ value. This progression continues until we reach $k=1$. If we still have not achieved the specified tolerance, we use $s=s_{1,\mathrm{opt}}$ for the remaining iterations until the desired tolerance level is reached. This is because, during these accelerated iterations, starting from $k=N/2$ to $k=1$, all the short wavelength modes of the error are reduced below the tolerance level. Consequently, the remaining residual corresponds to the longest wavelength ($k=1$) of the error. Therefore, we use $s_{1,\mathrm{opt}}$ for the rest of the iterations, which yields the fastest convergence. We show the simplified algorithm in the following:

\begin{verbatim}
k=N/2
while (res>tol)
    s=s_opt(k)
    Iterate(s)
    k=k-1
    if k<1
        k=1
\end{verbatim}

\begin{figure}
    \centering
    \includegraphics[scale=0.44]{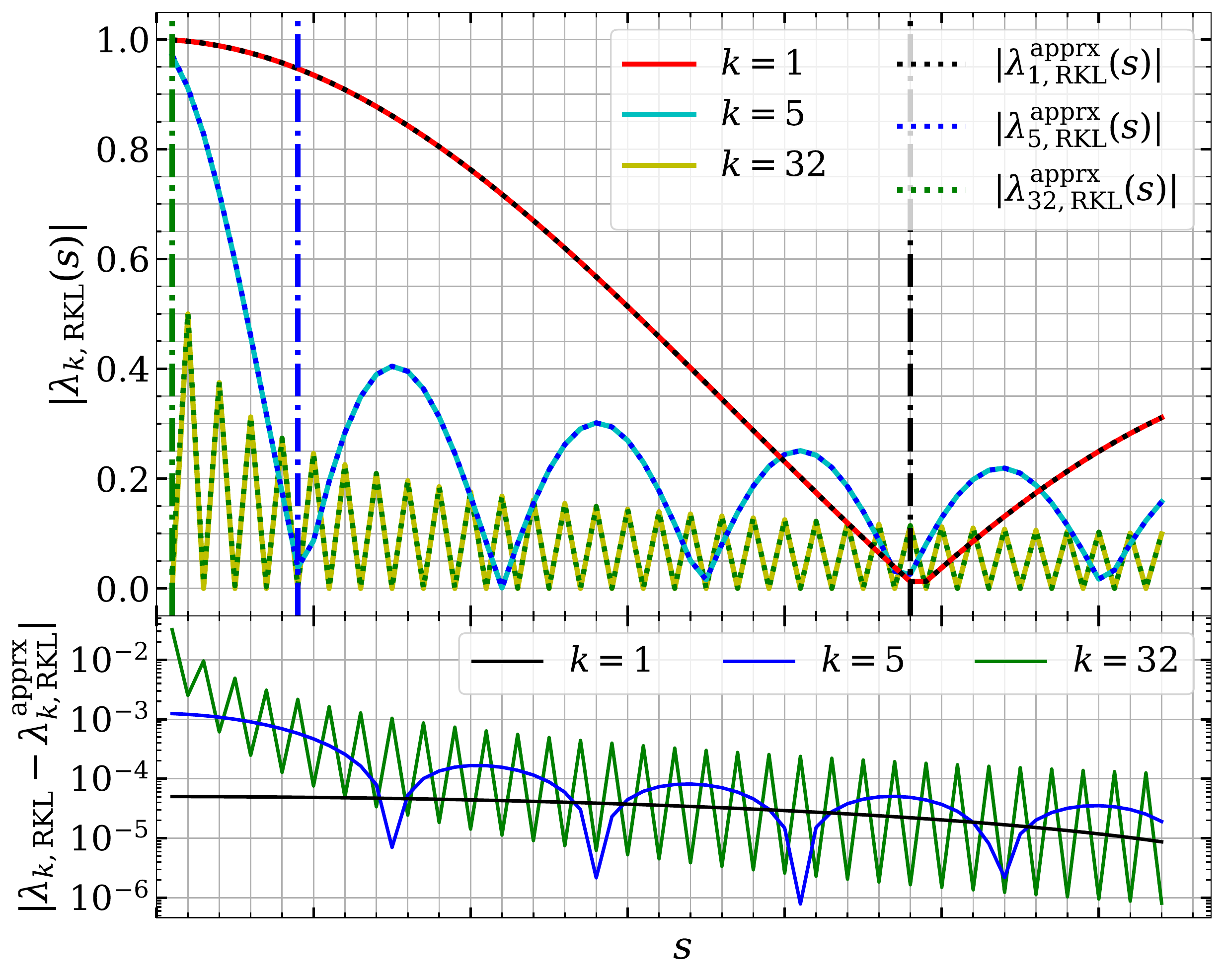}
    \caption{\textit{Top}: The analytic (solid) and approximated (dotted) values of $\lambdaRKL$ as a function of $s$ for $k=1, 5$ and $32$. To calculate the approximation we use Eq.~\eqref{eq:eigen_approx}. The dashed-dotted lines represent the value of $\skopt$ (Eq.~\ref{eq:sk_optimal}) for $k=1$ (red), 5 (blue), and 32 (green), i.e, the value of $s$ where the first zero of corresponding $\lambdaRKL$ occurs. \textit{Bottom}: The error in approximation, i.e., $|\lambdaRKL-\lambdaapprx|$ as a function of $s$ for $k=$1 (black), 5 (blue) and 32 (green). We notice that the error in approximation Eq.~\eqref{eq:eigen_approx} is $\lesssim 1\%$ even for $k=32$.}
    \label{fig:approximation}
\end{figure}

To determine $\skopt$ for a specific $k$-mode, we can numerically compute the first zero of $P_s(\cos[k\pi/N])$ using Eq.~\eqref{eq:eigenvalue_RKL}. However, we can also use the asymptotic form of the Legendre polynomials to numerically obtain the value of $\skopt$. For $-1+\delta<\cos{\theta}<1$ (with $\delta$ a small positive ﬁxed number), the asymptotic form of $P_{s}(\cos{\theta})$ is given by \citep{Temme_2014},
\begin{equation}\label{eq:legendre_approx}
    P_s(\cos{\theta})\sim \sqrt{\frac{\theta}{\sin{\theta}}} J_0\left(\left[s+\frac{1}{2}\right]\theta\right) + \mathcal{O}(s^{-1}),\,\,\,\theta\in (0,\pi)
\end{equation}
where $J_0(x)$ is the zeroth Bessel function of the first kind. Thus, from Eq.~\eqref{eq:eigenvalue_RKL} the eigenvalues of $\R$ can be approximated as, 
\begin{equation}\label{eq:eigen_approx}
    \lambda_{k,\mathrm{RKL}}^\mathrm{apprx}(s) = \sqrt{\frac{\theta}{\sin{\theta}}} J_0\left(\left[s+\frac{1}{2}\right]\theta\right),\hspace{0.5cm} \mathrm{where}\,\,\theta=\frac{k\pi}{N}
\end{equation}
Although the validity of Eq.\eqref{eq:eigen_approx} is limited to large values of $s$, we observe in Fig.\ref{fig:approximation} that it agrees well with the analytical values for $1\leq k\leq N/2$. The top panel of Fig.~\ref{fig:approximation} displays the eigenvalues ($\lambdaRKL$) for $k=1$ (red), 5 (cyan), and 32 (yellow) as a function of $s$, represented by solid lines. The dotted lines in the same panel correspond to the approximation ($\lambda_{k,\mathrm{RKL}}^\mathrm{apprx}(s)$) for the respective $k$-modes. The bottom panel shows the difference between $\lambdaRKL$ and $\lambdaapprx$ for $k=$1 (black), 5 (blue), and 32 (green).

For a given $k$ value, $\theta$ is constant in Eq.~\eqref{eq:eigen_approx}. Therefore, the first zero of $\lambdaapprx$ corresponds to the first zero of $J_0(x)$. It is known that the first zero of $J_0(x)$ occurs at $x\approx 2.4048$. Hence, we can obtain the optimal value of $s$ for the $k^\mathrm{th}$ mode as follows:
\begin{equation}\label{eq:sk_optimal}
    \begin{aligned}
        & \left(\skopt+\frac{1}{2}\right)\frac{k\pi}{N}  = 2.4048 \\
        & \skopt =\mathrm{Round} \left( \frac{2.4048}{\pi}\left[\frac{N}{k}\right]-\frac{1}{2} \right)
    \end{aligned}
\end{equation}
where $\mathrm{Round}(p)$ is defined as the nearest integer value of the real number $p$. In the top panel of Fig.~\ref{fig:approximation}, we show $\skopt$ (horizontal dashed-dotted line) for $k=$1 (black), 5 (blue), and 32 (green). We see that Eq.~\eqref{eq:sk_optimal} represents the first zero of the corresponding $\lambdaRKL$ very well.

\begin{figure}
    \centering
    \includegraphics[scale=0.5]{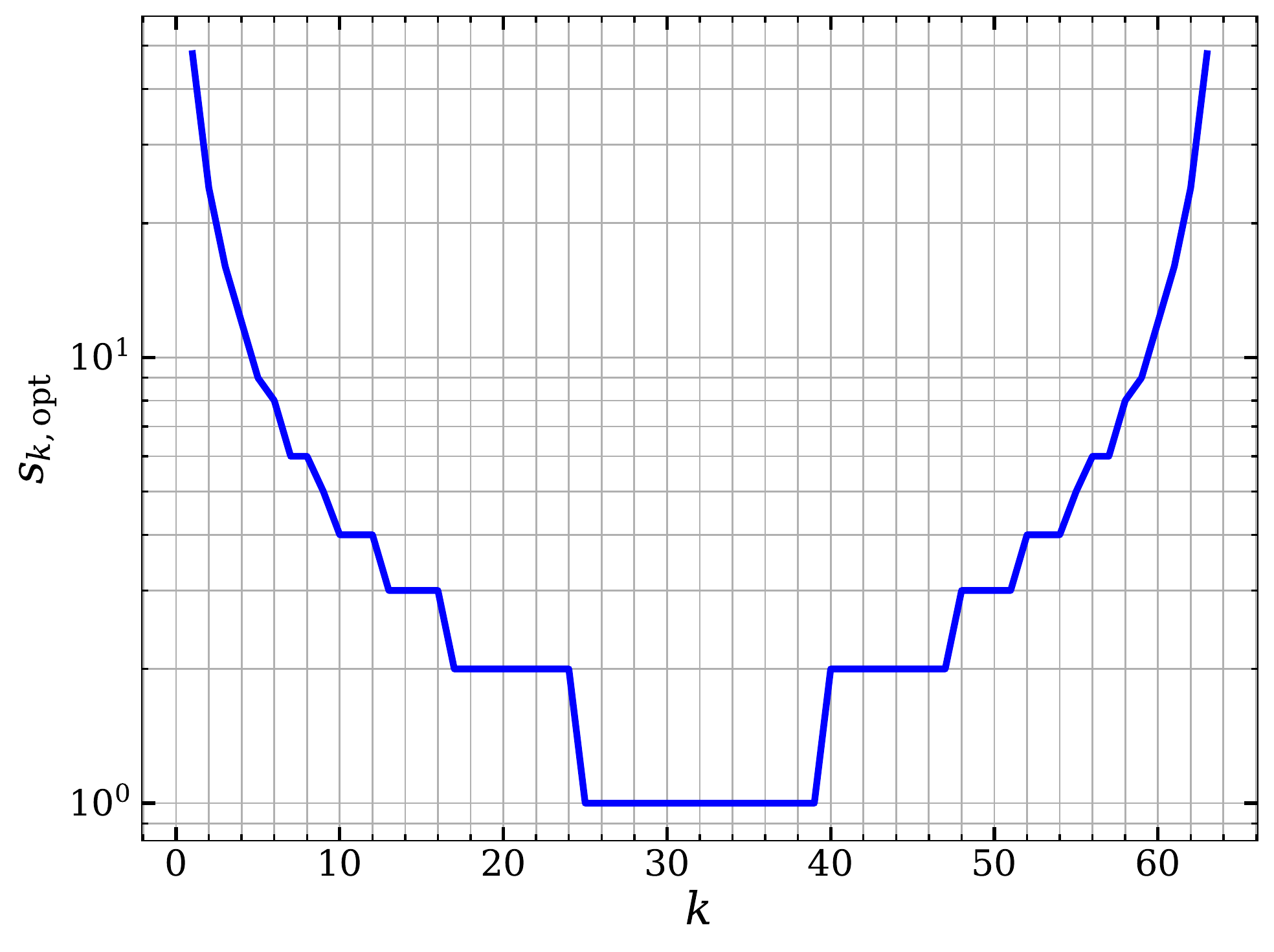}
    \caption{$\skopt$ as a function of wave number (k)}
    \label{fig:sk_opt}
\end{figure}

Once we have obtained the optimal value of $s$ ($\skopt$) for each $k$ using Eq.\eqref{eq:sk_optimal}, we can use it to perform the accelerated RKL iterations discussed earlier in this section. Fig.\ref{fig:sk_opt} displays the optimal value of $s$ as a function of the wavenumber $k$. We observe that $\skopt$ is symmetric about $k=N/2$ as expected. Additionally, a single $\skopt$ value corresponds to multiple wavenumbers, particularly for high $k$-modes. Consequently, one $\tau$-iteration with such an $\skopt$ value can simultaneously reduce the error for the corresponding $k$-values. Thus, there is no need to iterate the solution for each $k$-mode separately in the accelerated scheme discussed above. If $\skopt=s_{k+1,\mathrm{opt}}$ for some $k$, we can skip the iteration for that $k$ and proceed to the next $k$ value. An example of the algorithm is presented below.
\begin{verbatim}
k=N/2
s_old=0
while (res>tol)
    s=s_opt(k)
    if s==s_old and k!=1
        pass
    else
        Iterate(s)
        s_old=s
    k=k-1
    if k<1
        k=1
\end{verbatim}


\begin{figure*}
    \centering
    \includegraphics[width=0.5\linewidth]{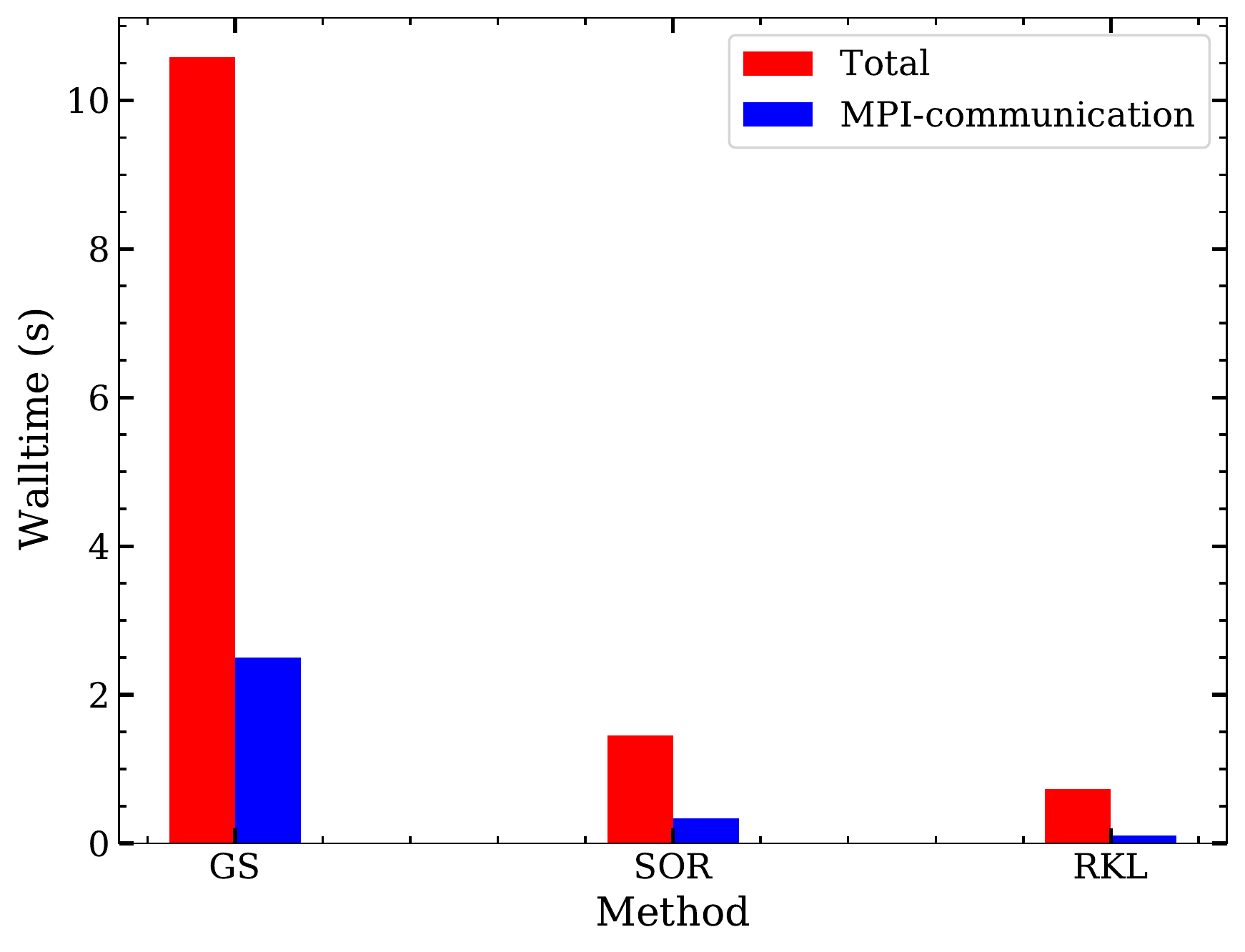}
    \caption{Performance test for different iterative solvers (excluding multigrid acceleration). The red and blue bars correspond to the time taken for the whole execution and MPI-communications, respectively.}
    \label{fig:performance}
\end{figure*}

\section{Performance test for different iterative solvers}\label{sec:performance_test}
Within this section, we present the comparative performance of the GS, SOR, and RKL methods (without the application of multigrid acceleration) in solving the 3D problem discussed in Sec.~\ref{sec:accuracy}. The optimal overrelaxation parameter ($\omega$) in the SOR algorithm is obtained by the following expression \citep{Press_1992}:
\begin{equation}
    \omega = \frac{2}{1+\sqrt{1-\rho_{\rm J}^2}}
\end{equation}
where, $\rho_{\rm J}$ is the spectral radius of the Jacobi iteration matrix. For a 3D grid of size $J\times L\times K$ with grid spacing ($\Delta_x$, $\Delta_y$, $\Delta_z$) and Dirichlet boundary conditions, $\rho_{\rm J}$ is given by,
\begin{equation}
    \rho_{\rm J} = 2\Delta_t\left[\frac{1}{\Delta_x^2}\cos\left(\frac{\pi}{J}\right)+\frac{1}{\Delta_y^2}\cos\left(\frac{\pi}{L}\right)+\frac{1}{\Delta_z^2}\cos\left(\frac{\pi}{K}\right)\right],
\end{equation}
where,
\begin{equation}
    \Delta_t = 2\left[\frac{1}{\Delta_x^2}+\frac{1}{\Delta_y^2}+\frac{1}{\Delta_z^2}\right]^{-1}
\end{equation}

We consider a box with $128^3$ zones for the calculation which was distributed between $128$ CPUs. The CPUs used were the \texttt{Intel(R) Xeon(R) Gold 6326} whose base clock frequency is $2.90\,{\rm GHz}$. 
The parallel communications between the CPUs were performed with the message passing interface (MPI) using the publicly available \texttt{Open MPI v4.1.5} library with \texttt{Infiniband EDR} interconnect.

Figure~\ref{fig:performance} displays the total time taken by various solvers to bring the maximum residual error below $10^{-10}$, as indicated by the red bars. 
The corresponding time spent in MPI-communications throughout the entire execution is also illustrated by the blue bars. 
The results clearly demonstrate that, in terms of the total execution time, the RKL method outperforms the SOR and GS techniques by a factor of roughly $2$ and $15$, respectively. 
As previously mentioned, the communication overhead for the SOR and GS solvers is considerably higher, comprising approximately $23\%$ of the total execution time. 
On the other hand, the time spent during parallel communications in the RKL scheme amounts to only around $13\%$ of the total time.


\bibliography{manuscript}{}

\begin{thebibliography}{}
\expandafter\ifx\csname natexlab\endcsname\relax\def\natexlab#1{#1}\fi
\providecommand{\url}[1]{\href{#1}{#1}}
\providecommand{\dodoi}[1]{doi:~\href{http://doi.org/#1}{\nolinkurl{#1}}}
\providecommand{\doeprint}[1]{\href{http://ascl.net/#1}{\nolinkurl{http://ascl.net/#1}}}
\providecommand{\doarXiv}[1]{\href{https://arxiv.org/abs/#1}{\nolinkurl{https://arxiv.org/abs/#1}}}

\bibitem[{{Almgren} {et~al.}(2010){Almgren}, {Beckner}, {Bell}, {Day},
  {Howell}, {Joggerst}, {Lijewski}, {Nonaka}, {Singer}, \&
  {Zingale}}]{Almgren_2010}
{Almgren}, A.~S., {Beckner}, V.~E., {Bell}, J.~B., {et~al.} 2010, ApJ, 715,
  1221, \dodoi{10.1088/0004-637X/715/2/1221}

\bibitem[{{Bate} \& {Burkert}(1997)}]{Bate_1997}
{Bate}, M.~R., \& {Burkert}, A. 1997, MNRAS, 288, 1060,
  \dodoi{10.1093/mnras/288.4.1060}

\bibitem[{{Bertschinger}(1998)}]{Bertschinge_1998}
{Bertschinger}, E. 1998, ARA\&A, 36, 599,
  \dodoi{10.1146/annurev.astro.36.1.599}

\bibitem[{{Binney} \& {Tremaine}(1987)}]{Binney_1987}
{Binney}, J., \& {Tremaine}, S. 1987, {Galactic dynamics} (Princeton University
  Press).
\newblock
  \url{https://press.princeton.edu/books/paperback/9780691130279/galactic-dynamics}

\bibitem[{{Boss}(1997)}]{Boss_1997}
{Boss}, A.~P. 1997, Science, 276, 1836, \dodoi{10.1126/science.276.5320.1836}

\bibitem[{{Boss} \& {Bodenheimer}(1979)}]{Boss_1979}
{Boss}, A.~P., \& {Bodenheimer}, P. 1979, ApJ, 234, 289, \dodoi{10.1086/157497}

\bibitem[{{Boss} \& {Myhill}(1992)}]{Boss_1992}
{Boss}, A.~P., \& {Myhill}, E.~A. 1992, ApJS, 83, 311, \dodoi{10.1086/191739}

\bibitem[{Brandt(1977)}]{Brandt_1977}
Brandt, A. 1977, Mathematics of Computation, 31, 333.
\newblock \url{http://www.jstor.org/stable/2006422}

\bibitem[{Briggs {et~al.}(2000)Briggs, Henson, \& McCormick}]{Briggs_2000}
Briggs, W.~L., Henson, V.~E., \& McCormick, S.~F. 2000, A Multigrid Tutorial,
  Second Edition, 2nd edn. (Society for Industrial and Applied Mathematics),
  \dodoi{10.1137/1.9780898719505}

\bibitem[{{Bryan} {et~al.}(1995){Bryan}, {Norman}, {Stone}, {Cen}, \&
  {Ostriker}}]{Bryan_1995}
{Bryan}, G.~L., {Norman}, M.~L., {Stone}, J.~M., {Cen}, R., \& {Ostriker},
  J.~P. 1995, Computer Physics Communications, 89, 149,
  \dodoi{10.1016/0010-4655(94)00191-4}

\bibitem[{{Burkert} \& {Bodenheimer}(1993)}]{Burkert_1993}
{Burkert}, A., \& {Bodenheimer}, P. 1993, MNRAS, 264, 798,
  \dodoi{10.1093/mnras/264.4.798}

\bibitem[{{Couch} {et~al.}(2013){Couch}, {Graziani}, \& {Flocke}}]{Couch_2013}
{Couch}, S.~M., {Graziani}, C., \& {Flocke}, N. 2013, ApJ, 778, 181,
  \dodoi{10.1088/0004-637X/778/2/181}

\bibitem[{{Evrard}(1988)}]{Evrard_1988}
{Evrard}, A.~E. 1988, MNRAS, 235, 911, \dodoi{10.1093/mnras/235.3.911}

\bibitem[{{Falco} {et~al.}(2013){Falco}, {Hansen}, {Wojtak}, \&
  {Mamon}}]{Falco_2013}
{Falco}, M., {Hansen}, S.~H., {Wojtak}, R., \& {Mamon}, G.~A. 2013, MNRAS, 431,
  L6, \dodoi{10.1093/mnrasl/sls051}

\bibitem[{{Fryxell} {et~al.}(2000){Fryxell}, {Olson}, {Ricker}, {Timmes},
  {Zingale}, {Lamb}, {MacNeice}, {Rosner}, {Truran}, \& {Tufo}}]{Fryxell_2000}
{Fryxell}, B., {Olson}, K., {Ricker}, P., {et~al.} 2000, ApJS, 131, 273,
  \dodoi{10.1086/317361}

\bibitem[{{Grudi{\'c}}(2021)}]{Grudic_2021}
{Grudi{\'c}}, M.~Y. 2021, MNRAS, 507, 1064, \dodoi{10.1093/mnras/stab2208}

\bibitem[{{Guillet} \& {Teyssier}(2011)}]{Guillet_2011}
{Guillet}, T., \& {Teyssier}, R. 2011, Journal of Computational Physics, 230,
  4756, \dodoi{10.1016/j.jcp.2011.02.044}

\bibitem[{Hanawa(2019)}]{Hanawa_2019}
Hanawa, T. 2019, Journal of Physics: Conference Series, 1225, 012015,
  \dodoi{10.1088/1742-6596/1225/1/012015}

\bibitem[{Harris {et~al.}(2020)Harris, Millman, van~der Walt, Gommers,
  Virtanen, Cournapeau, Wieser, Taylor, Berg, Smith, Kern, Picus, Hoyer, van
  Kerkwijk, Brett, Haldane, del R{\'{i}}o, Wiebe, Peterson,
  G{\'{e}}rard-Marchant, Sheppard, Reddy, Weckesser, Abbasi, Gohlke, \&
  Oliphant}]{Harris_2020}
Harris, C.~R., Millman, K.~J., van~der Walt, S.~J., {et~al.} 2020, Nature, 585,
  357, \dodoi{10.1038/s41586-020-2649-2}

\bibitem[{{Hernquist} \& {Katz}(1989)}]{Hernquist_1989}
{Hernquist}, L., \& {Katz}, N. 1989, ApJS, 70, 419, \dodoi{10.1086/191344}

\bibitem[{{Hopkins}(2015)}]{Hopkins_2015}
{Hopkins}, P.~F. 2015, MNRAS, 450, 53, \dodoi{10.1093/mnras/stv195}

\bibitem[{{Hubber} {et~al.}(2006){Hubber}, {Goodwin}, \&
  {Whitworth}}]{Hubber_2006}
{Hubber}, D.~A., {Goodwin}, S.~P., \& {Whitworth}, A.~P. 2006, A\&A, 450, 881,
  \dodoi{10.1051/0004-6361:20054100}

\bibitem[{{Hubber} {et~al.}(2018){Hubber}, {Rosotti}, \& {Booth}}]{Hubber_2018}
{Hubber}, D.~A., {Rosotti}, G.~P., \& {Booth}, R.~A. 2018, MNRAS, 473, 1603,
  \dodoi{10.1093/mnras/stx2405}

\bibitem[{Hunter(2007)}]{Matplotlib_2007}
Hunter, J.~D. 2007, Computing in Science \& Engineering, 9, 90,
  \dodoi{10.1109/MCSE.2007.55}

\bibitem[{{Jiang} {et~al.}(2013){Jiang}, {Belyaev}, {Goodman}, \&
  {Stone}}]{Jiang_2013}
{Jiang}, Y.-F., {Belyaev}, M., {Goodman}, J., \& {Stone}, J.~M. 2013, NewA, 19,
  48, \dodoi{10.1016/j.newast.2012.08.002}

\bibitem[{{Katz} {et~al.}(2016){Katz}, {Zingale}, {Calder}, {Swesty},
  {Almgren}, \& {Zhang}}]{Katz_2016}
{Katz}, M.~P., {Zingale}, M., {Calder}, A.~C., {et~al.} 2016, ApJ, 819, 94,
  \dodoi{10.3847/0004-637X/819/2/94}

\bibitem[{{Kiessling}(1999)}]{Kiessling_1999}
{Kiessling}, M. K.~H. 1999, arXiv e-prints, astro.
\newblock \doarXiv{astro-ph/9910247}

\bibitem[{{Klein}(1999)}]{Klein_1999}
{Klein}, R.~I. 1999, Journal of Computational and Applied Mathematics, 109, 123

\bibitem[{{Mandal} {et~al.}(2021){Mandal}, {Mukherjee}, {Federrath},
  {Nesvadba}, {Bicknell}, {Wagner}, \& {Meenakshi}}]{Mandal_2021}
{Mandal}, A., {Mukherjee}, D., {Federrath}, C., {et~al.} 2021, MNRAS, 508,
  4738, \dodoi{10.1093/mnras/stab2822}

\bibitem[{{McKee} \& {Ostriker}(2007)}]{McKee2007}
{McKee}, C.~F., \& {Ostriker}, E.~C. 2007, ARA\&A, 45, 565,
  \dodoi{10.1146/annurev.astro.45.051806.110602}

\bibitem[{{Meyer} {et~al.}(2012){Meyer}, {Balsara}, \& {Aslam}}]{Meyer_2012}
{Meyer}, C.~D., {Balsara}, D.~S., \& {Aslam}, T.~D. 2012, MNRAS, 422, 2102,
  \dodoi{10.1111/j.1365-2966.2012.20744.x}

\bibitem[{{Meyer} {et~al.}(2014){Meyer}, {Balsara}, \& {Aslam}}]{Meyer_2014}
---. 2014, Journal of Computational Physics, 257, 594,
  \dodoi{10.1016/j.jcp.2013.08.021}

\bibitem[{{Mignone} {et~al.}(2007){Mignone}, {Bodo}, {Massaglia}, {Matsakos},
  {Tesileanu}, {Zanni}, \& {Ferrari}}]{Mignone_2007}
{Mignone}, A., {Bodo}, G., {Massaglia}, S., {et~al.} 2007, ApJS, 170, 228,
  \dodoi{10.1086/513316}

\bibitem[{{Mignone} {et~al.}(2012){Mignone}, {Zanni}, {Tzeferacos}, {van
  Straalen}, {Colella}, \& {Bodo}}]{Mignone_2012}
{Mignone}, A., {Zanni}, C., {Tzeferacos}, P., {et~al.} 2012, ApJS, 198, 7,
  \dodoi{10.1088/0067-0049/198/1/7}

\bibitem[{{Mo} {et~al.}(2010){Mo}, {van den Bosch}, \& {White}}]{Mo_2010}
{Mo}, H., {van den Bosch}, F.~C., \& {White}, S. 2010, {Galaxy Formation and
  Evolution}

\bibitem[{{Mullen} {et~al.}(2021){Mullen}, {Hanawa}, \& {Gammie}}]{Mullen_2021}
{Mullen}, P.~D., {Hanawa}, T., \& {Gammie}, C.~F. 2021, ApJS, 252, 30,
  \dodoi{10.3847/1538-4365/abcfbd}

\bibitem[{{Nordhaus} {et~al.}(2010){Nordhaus}, {Burrows}, {Almgren}, \&
  {Bell}}]{Nordhaus_2010}
{Nordhaus}, J., {Burrows}, A., {Almgren}, A., \& {Bell}, J. 2010, ApJ, 720,
  694, \dodoi{10.1088/0004-637X/720/1/694}

\bibitem[{{Norman} {et~al.}(1980){Norman}, {Wilson}, \& {Barton}}]{Norman_1980}
{Norman}, M.~L., {Wilson}, J.~R., \& {Barton}, R.~T. 1980, ApJ, 239, 968,
  \dodoi{10.1086/158185}

\bibitem[{{Ostriker} {et~al.}(2001){Ostriker}, {Stone}, \&
  {Gammie}}]{Ostriker_2001}
{Ostriker}, E.~C., {Stone}, J.~M., \& {Gammie}, C.~F. 2001, ApJ, 546, 980,
  \dodoi{10.1086/318290}

\bibitem[{Press {et~al.}(1992)Press, Teukolsky, Vetterling, \&
  Flannery}]{Press_1992}
Press, W.~H., Teukolsky, S.~A., Vetterling, W.~T., \& Flannery, B.~P. 1992,
  Numerical Recipes in C (2nd Ed.): The Art of Scientific Computing (USA:
  Cambridge University Press)

\bibitem[{{Rice} {et~al.}(2005){Rice}, {Lodato}, \& {Armitage}}]{Rice_2005}
{Rice}, W.~K.~M., {Lodato}, G., \& {Armitage}, P.~J. 2005, MNRAS, 364, L56,
  \dodoi{10.1111/j.1745-3933.2005.00105.x}

\bibitem[{{Ricker}(2008)}]{Ricker_2008}
{Ricker}, P.~M. 2008, ApJS, 176, 293, \dodoi{10.1086/526425}

\bibitem[{{Springel}(2005)}]{Springel_2005}
{Springel}, V. 2005, MNRAS, 364, 1105, \dodoi{10.1111/j.1365-2966.2005.09655.x}

\bibitem[{{Springel}(2010)}]{Springel_2010}
---. 2010, MNRAS, 401, 791, \dodoi{10.1111/j.1365-2966.2009.15715.x}

\bibitem[{{Springel} {et~al.}(2001){Springel}, {Yoshida}, \&
  {White}}]{Springel_2001}
{Springel}, V., {Yoshida}, N., \& {White}, S. D.~M. 2001, NewA, 6, 79,
  \dodoi{10.1016/S1384-1076(01)00042-2}

\bibitem[{{Steinmetz} \& {Mueller}(1993)}]{Steinmetz_1993}
{Steinmetz}, M., \& {Mueller}, E. 1993, A\&A, 268, 391

\bibitem[{Temme(2014)}]{Temme_2014}
Temme, N.~M. 2014, Asymptotic Methods for Integrals (WORLD SCIENTIFIC),
  \dodoi{10.1142/9195}

\bibitem[{Trottenberg {et~al.}(2000)Trottenberg, Oosterlee, \&
  Schuller}]{Trottenberg_2000}
Trottenberg, U., Oosterlee, C., \& Schuller, A. 2000, Multigrid, 1st edn.
  (Academic Press).
\newblock
  \url{https://www.elsevier.com/books/multigrid/trottenberg/978-0-08-047956-9}

\bibitem[{{Truelove} {et~al.}(1997){Truelove}, {Klein}, {McKee}, {Holliman},
  {Howell}, \& {Greenough}}]{Truelove_1997}
{Truelove}, J.~K., {Klein}, R.~I., {McKee}, C.~F., {et~al.} 1997, ApJL, 489,
  L179, \dodoi{10.1086/310975}

\bibitem[{{Truelove} {et~al.}(1998){Truelove}, {Klein}, {McKee}, {Holliman},
  {Howell}, {Greenough}, \& {Woods}}]{Truelove_1998}
---. 1998, ApJ, 495, 821, \dodoi{10.1086/305329}

\bibitem[{{Vaidya} {et~al.}(2017){Vaidya}, {Prasad}, {Mignone}, {Sharma}, \&
  {Rickler}}]{Vaidya_2017}
{Vaidya}, B., {Prasad}, D., {Mignone}, A., {Sharma}, P., \& {Rickler}, L. 2017,
  MNRAS, 472, 3147, \dodoi{10.1093/mnras/stx2176}

\bibitem[{{van der Velden}(2020)}]{Cmasher_2020}
{van der Velden}, E. 2020, The Journal of Open Source Software, 5, 2004,
  \dodoi{10.21105/joss.02004}

\bibitem[{{Wadsley} {et~al.}(2004){Wadsley}, {Stadel}, \&
  {Quinn}}]{Wadsley_2004}
{Wadsley}, J.~W., {Stadel}, J., \& {Quinn}, T. 2004, NewA, 9, 137,
  \dodoi{10.1016/j.newast.2003.08.004}

\bibitem[{{Wang} \& {Yen}(2020)}]{Wang_2020}
{Wang}, H.-H., \& {Yen}, C.-C. 2020, ApJS, 247, 2,
  \dodoi{10.3847/1538-4365/ab66ba}

\bibitem[{{Ziegler}(2005)}]{Ziegler_2005}
{Ziegler}, U. 2005, A\&A, 435, 385, \dodoi{10.1051/0004-6361:20042451}

\end{thebibliography}
\bibliographystyle{aasjournal}



\end{document}